\documentclass[10pt]{wlscirep}
\usepackage[utf8]{inputenc}
\usepackage[T1]{fontenc}
\usepackage{amsmath, amsfonts, amssymb}
\usepackage{newtxmath}
\usepackage{subcaption}
\usepackage[font=small,labelfont=bf]{caption}

\usepackage{tikz}
\usetikzlibrary{arrows}
\usepackage{tcolorbox} 
\usepackage{listings}
\usepackage{xspace}
\usepackage{upgreek}
\makeatletter
\newcommand\setcurrentname[1]{\def\@currentlabelname{#1}}
\makeatother
\usepackage{nameref}
\usepackage[colorlinks=true, allcolors=blue]{hyperref}
\usepackage{cleveref}
\usepackage{siunitx}
\DeclareSIUnit\molar{M}
\DeclareSIUnit\molecule{}
\usepackage{longtable}
\usepackage{algorithm}
\usepackage{algpseudocode}
\usepackage{placeins}
\usepackage{setspace}
\usepackage{chemformula}  
\usepackage[version=4]{mhchem}
\newcommand{\caion}{\ce{Ca^{2+}}\xspace}
\newcommand{\naion}{\ce{Na^+}\xspace}

\newcommand{\R}{\mathbb{R}}
\newcommand{\inner}[2]{\langle #1, #2 \rangle}

\newcommand{\um}{$\upmu$m\xspace}
\newcommand{\uM}{$\upmu$M\xspace}
\newcommand{\thinsim}{{\raise.17ex\hbox{\(\scriptstyle\mathtt{\sim}\)}}}

\usepackage{float}
\floatstyle{plain}
\newfloat{video}{thp}{lop}
\floatname{video}{Video}
\crefname{video}{video}{videos}

\newcommand{\fixme}[1]{\textcolor{red}{#1}}

\newcommand{\beginsupplement}{%
    \setcounter{table}{0}
    \renewcommand{\thetable}{S\arabic{table}}%
    \setcounter{figure}{0}
    \renewcommand{\thefigure}{S\arabic{figure}}%
    \setcounter{section}{0}
    \renewcommand{\thesection}{S\arabic{section}}%
    \setcounter{equation}{0}
    \renewcommand{\theequation}{S\arabic{equation}}%
 }

\title{Spatial modeling algorithms for reactions and transport (SMART) in biological cells}

\author[1,+]{Emmet A.\@ Francis}
\author[1,2,+]{Justin G.\@ Laughlin}
\author[3]{J\o{}rgen S.\@ Dokken}
\author[4]{Henrik N.\@ T.\@ Finsberg}
\author[1]{Christopher T.\@ Lee}
\author[3,*]{Marie E.\@ Rognes}
\author[1,*]{Padmini Rangamani}
\affil[1]{Department of Mechanical and Aerospace Engineering, University of California San Diego, La Jolla, CA, USA}
\affil[2]{Computational Engineering Division, Lawrence Livermore National Laboratory, Livermore, CA, USA}
\affil[3]{Department of Numerical Analysis and Scientific Computing, Simula Research Laboratory, Oslo, Norway}
\affil[4]{Department of Computational Physiology, Simula Research Laboratory, Oslo, Norway}

\affil[*]{meg@simula.no, prangamani@ucsd.edu}
\affil[+]{these authors contributed equally to this work}

\begin{abstract}
Biological cells rely on precise spatiotemporal coordination of biochemical reactions to control their many functions.
Such cell signaling networks have been a common focus for mathematical models, but they remain challenging to simulate, particularly in realistic cell geometries.
Herein, we present our software, Spatial Modeling Algorithms for Reaction and Transport (SMART), a package that takes in high-level user specifications about cell signaling networks and molecular transport, and then assembles and solves the associated mathematical and computational systems.
SMART uses state-of-the-art finite element analysis, via the FEniCS Project software, to efficiently and accurately resolve cell signaling events over discretized cellular and subcellular geometries.
We demonstrate its application to several different biological systems, including YAP/TAZ mechanotransduction, calcium signaling in neurons and cardiomyocytes, and ATP generation in mitochondria.
Throughout, we utilize experimentally-derived realistic cellular geometries represented by well-conditioned tetrahedral meshes.
These scenarios demonstrate the applicability, flexibility, accuracy and efficiency of SMART across a range of temporal and spatial scales.
\end{abstract}
\begin{document}
\flushbottom
\maketitle

\thispagestyle{empty}

\section*{Introduction}

In the past decade, computational modeling has become an integral part of the discovery toolkit in biology along with advances in experimental technologies \cite{yueComputationalSystemsBiology2022, mogilnerQuantitativeModelingCell2006, kitanoComputationalSystemsBiology2002}.
One of the fundamental tenets of biology is that structure and function are very closely related \cite{wainwrightFormFunctionOrganisms1988, hermanUnifyingFrameworkUnderstanding2022}.
In single cell biology, this is reflected by the spatial compartmentalization of cellular signaling in different subcellular locations and organelles. 
Advances in microscopy in recent decades have provided data to support this notion from two approaches: electron microscopy for a detailed characterization of subcellular structure \cite{peddieVolumeElectronMicroscopy2022, villingerFIBSEMTomography2012, heinrichWholecellOrganelleSegmentation2021a,mccaffertyIntegratingCellularElectron2024} and super-resolution microscopy for the spatiotemporal localization of biochemical species that are important for cellular functions including signaling \cite{stoneSuperResolutionMicroscopyShedding2017,schermellehSuperresolutionMicroscopyDemystified2019}.
Detailed computational models using realistic cellular geometries and reaction-diffusion mechanisms can help us identify the possible biophysical mechanisms underlying such spatial compartmentalization \cite{hakeModellingCardiacCalcium2012a,bellDendriticSpineMorphology2022a,garciaMitochondrialMorphologyProvides2019,holashStochasticSimulationSkeletal2019a}. 
However, representing these details including subcellular geometries such as organelles and the relevant reaction-transport formulations using the appropriate computational description remains an open challenge.

\begin{figure}[htbp!]
    \centering
    \includegraphics[width=6in,keepaspectratio]{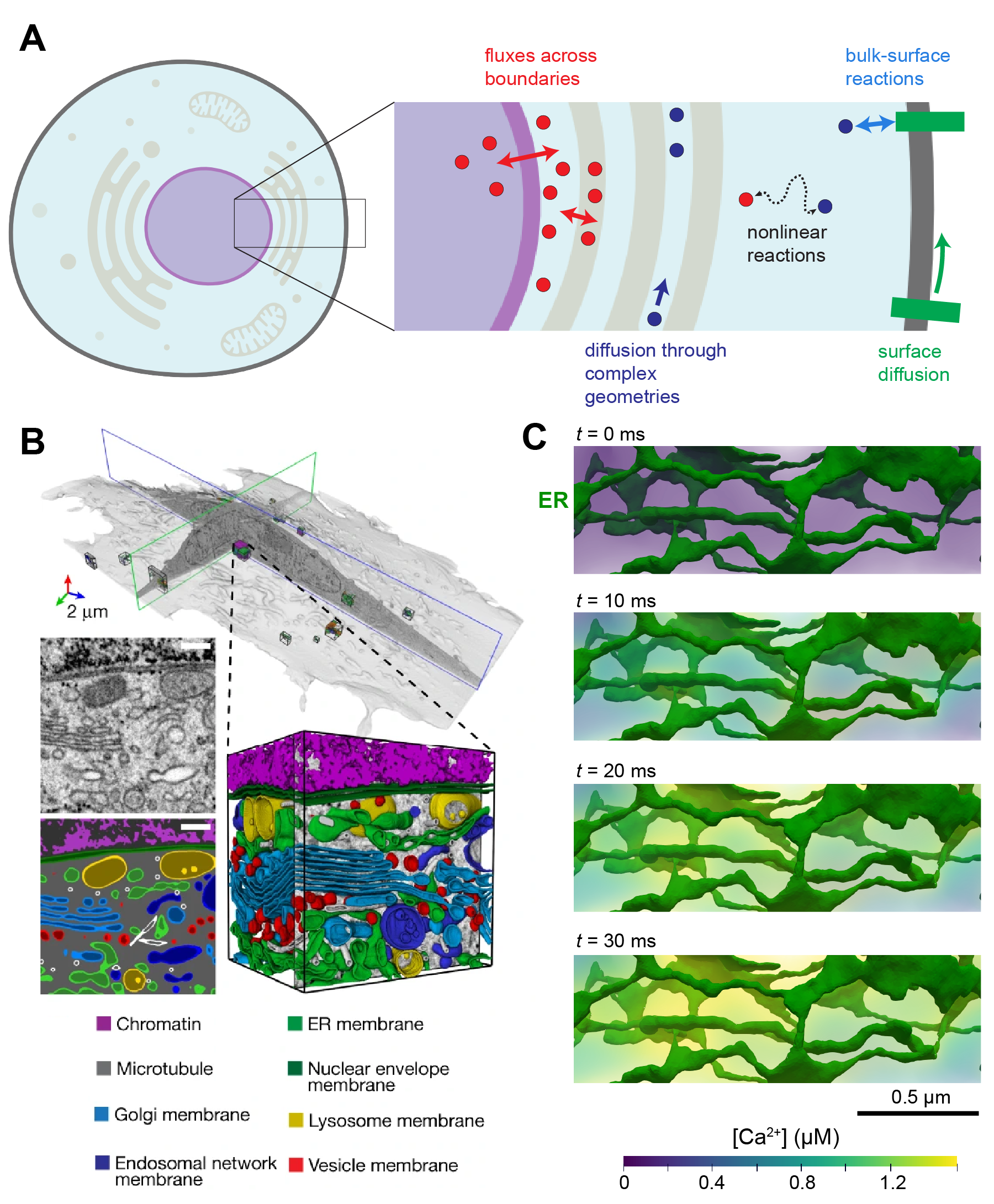}
    \caption{ 
        {\bf Mixed dimensional reaction-transport networks in cells.} 
        A) Schematic of reaction-transport system in SMART.
        Given the topological relationships between different compartments in cells and information on reactions between species, fluxes across boundaries, and diffusion rates of individual species, SMART assembles a finite element system of equations.
        Panel created with Biorender.com.
        A single model in SMART may include both volume species (circles in inset) and surface species (rectangles in inset) that can all diffuse and react with one another.
        B) Whole cell geometry with segmented organelles from volume electron microscopy. 
        Adapted from Figure 1a in Heinrich et al.\@ 2021 \cite{heinrichWholecellOrganelleSegmentation2021a}, permission pending.
        C) Calcium release from the endoplasmic reticulum (ER) in a realistic geometry.
        For illustration, a linear molecular leak flux from the ER ($j_{leak} = j_0 (c_{ER} - c_i)$) was assumed starting at $t = 0$.
        This realistic ER geometry was derived from electron micrographs of dendritic spines \cite{wuContactsEndoplasmicReticulum2017}.
    }
    \label{fig:overview}
\end{figure}

Historically, many mathematical models of cell signaling have neglected spatial effects, treating the cell as a well-mixed volume.
In certain cases, this approximation can be justified, but given the slow diffusion of certain signaling molecules, the crowded intracellular environment, and the complexity of cellular geometries, such approximations can diminish the predictive capability of the models.
Furthermore, due to the variety of membrane-bound organelles present in cells, reaction networks are coupled across subvolumes and involve a complicated set of bulk-surface reactions at organelle membranes (\Cref{fig:overview}A).
As a result, moving from well-mixed models to multicompartment spatiotemporal models of cell signaling presents many technical barriers.
Mathematically, this involves transitioning from systems of ordinary differential equations (ODEs) to systems of mixed-dimensional partial differential equations (PDEs).
Such PDE systems are notoriously difficult to solve numerically due to nonlinearities, stiffness, and instabilities \cite{daversin-cattyAbstractionsAutomatedAlgorithms2021a}.
Furthermore, solving these equations within realistic cellular geometries requires robust discretization of complex geometries \cite{heinrichWholecellOrganelleSegmentation2021a,mccaffertyIntegratingCellularElectron2024} (\Cref{fig:overview}B-C) and is computationally expensive due to the high dimensionality of the systems. 

Recent efforts have been made to accurately represent cellular and subcellular geometries with well-conditioned triangular and tetrahedral meshes \cite{wangRemeshingFlexibleMembranes2022,meansReactionDiffusionModeling2006a,lee3DMeshProcessing2020a,leeOpenSourceMeshGeneration2020a}.
In particular, GAMer2 (Geometry-preserving Adaptive MeshER version 2) allows users to convert microscopy images of cells into suitable meshes for finite element simulations \cite{lee3DMeshProcessing2020a,leeOpenSourceMeshGeneration2020a}.
These meshes can be annotated to mark the location of subcellular structures such as the nucleus, endoplasmic reticulum (ER), or mitochondria, along with their respective membrane boundaries \cite{venkatramanCristaeFormationMechanical2023a,mesaSpineApparatusModulates2023}.
Meshes from GAMer2 can be readily loaded into the open-source finite element software package, FEniCS \cite{alnaesFEniCSProjectVersion2015}, and used to simulate spatial models of cell signaling in realistic geometries.
However, translating systems of PDEs defining cell signaling systems into variational forms defined over subsurfaces and subvolumes of the geometry, and solving these, is nontrivial.

To answer fundamental questions about spatiotemporal compartmentalization of cellular signaling, building on advances in meshing, we introduce Spatial Modeling Algorithms for Reaction and Transport (SMART), a Python-based software package to construct and solve systems of mixed-dimensional reaction-transport equations \cite{laughlinSMARTSpatialModeling2023,laughlinRangamaniLabUCSDSmartV22024a}.
SMART complements other software packages for computational cell biology such as VCell \cite{schaffGeneralComputationalFramework1997, cowanSpatialModelingCell2012} and MCell \cite{kerrFastMonteCarlo2008} by providing a unique toolkit to solve spatial signaling networks over complicated geometries using the finite element method (FEM).
Similar to systems biology frameworks such as the Systems Biology Markup Language (SBML) \cite{huckaSystemsBiologyMarkup2003}, SMART takes user input that specifies species, reactions, compartments, and parameters involved in a given biological network.
Herein, we describe this workflow and then demonstrate the use of SMART in several biological test cases, including YAP/TAZ-mediated mechanotransduction in cells on micropatterned substrates, calcium signaling in neuronal dendritic spines, calcium release within a cardiomyocyte, and synthesis of ATP in a mitochondrion.
Several of these examples use realistic subcellular geometries derived from electron micrographs and conditioned with GAMer2.
We further demonstrate the accuracy of SMART via a series of addition numerical verification tests and summarize its performance and scalability.

\section*{Results}

\phantomsection
\subsection*{Spatial modeling algorithms for reaction and transport to simulate biological networks} \setcurrentname{Spatial modeling algorithms for reaction and transport to simulate biological networks}
\label{sec:smart}

Cell signaling networks rely on non-trivial chains of molecular reactions and transport mechanisms acting within and across compartments: extracellular, intracellular, and subcellular spaces, and membranes (\Cref{fig:overview}A). 
These cellular domains define three-dimensional bulk \emph{volumes} while the membranes can be viewed as two-dimensional manifold \emph{surfaces}. 
The geometry of these compartments can be accurately represented through synthetic or imaging-guided computational meshes in the form of simplicial complexes, with individual volumes and surfaces identified and labeled by tags \cite{lee3DMeshProcessing2020a} (\Cref{fig:overview}B--C, Methods). 
Spatial modeling of cellular pathways and processes describe the distribution and evolution of different \emph{species} present in or on these compartments; \textit{e.g.}, ion concentrations such as \naion in the cytosol or interstitial fluid, \caion in subcellular compartments, or the density of receptors or other channels distributed along the plasma membranes.

\begin{figure}[htbp!]
  \centering
    \includegraphics[width=7in,keepaspectratio]{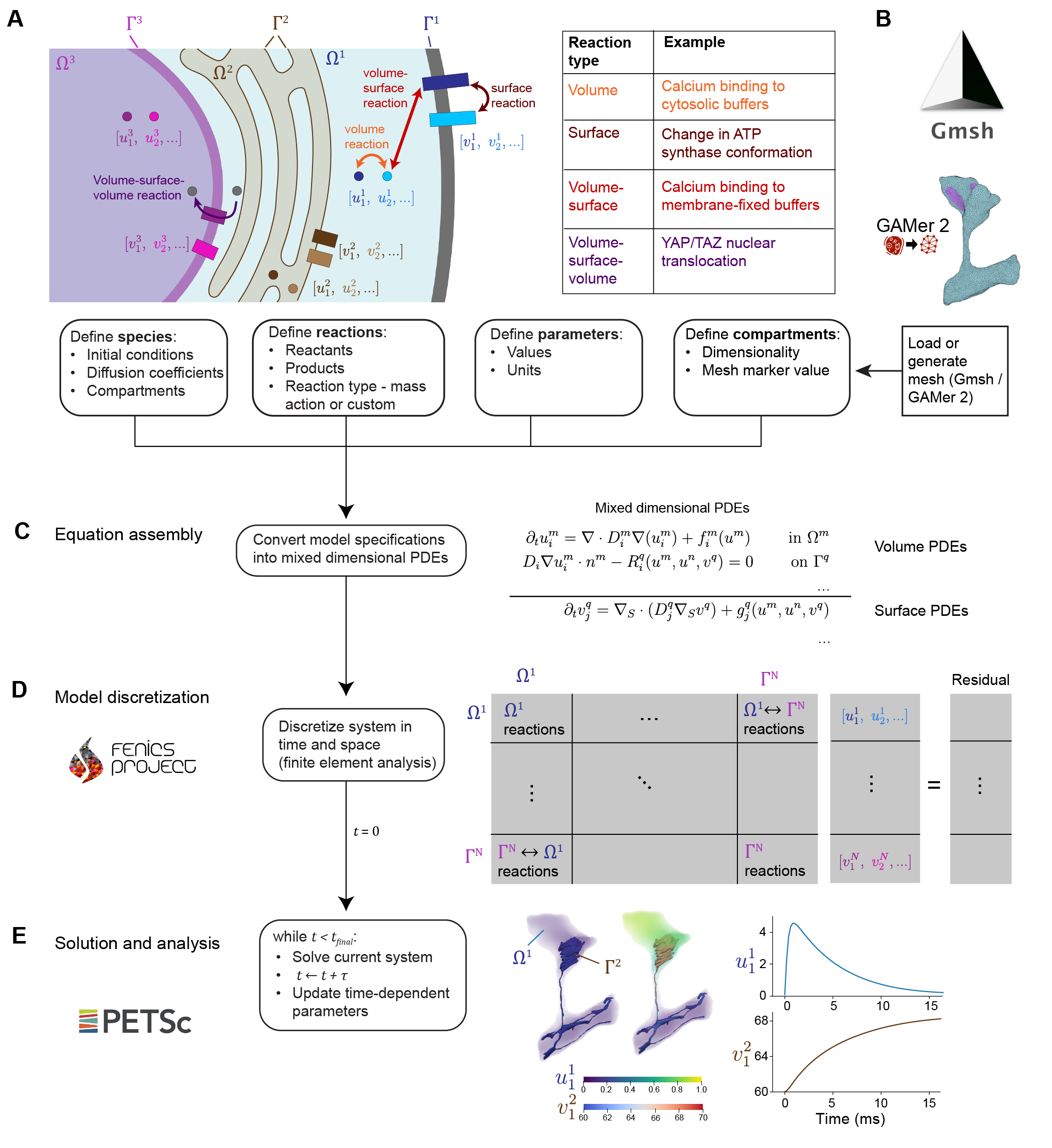}
    \caption{ 
        {\bf SMART workflow.} 
        A) Illustration of the basic components used to define a model in SMART - species, reactions, parameters, and compartments.
        The graphic contains 3 volume compartments ($\Omega^m$) and 3 surface compartments ($\Gamma^q$), with illustrations of the reaction types supported by SMART, including volume, surface, volume-surface, and volume-surface-volume.
        Panel created with Biorender.com.
        B) Model geometry specified by mesh generated in Gmsh or conditioned in GAMer2.
        C) Assembly of equations from reaction specifications in SMART.
        Each volume species and each surface species has an associated PDE with boundary conditions, as shown on the right (equations in \Cref{app:sec:equations}).
        D) SMART model discretization using finite elements.
        The nonlinear system is discretized using linear finite elements in FEniCS and then the block matrix problem is assembled in PETSc.
        The matrix is nested in terms of compartments involved in each interaction, as summarized on the right.
        E) Model solution and postprocessing in SMART.
        The system is solved iteratively at each time step until reaching $t_{final}$.
        Results are post-processed to examine changes in concentration over time and space.
    }
    \label{fig:workflow}
\end{figure}

Mathematically, we describe diffusion of such species and reactions between species, within or across compartments, via coupled systems of time-dependent, nonlinear and mixed-dimensional partial differential equations defined over the computational geometries (\Cref{fig:workflow}, Methods). 
Our modeling assumptions enable species to diffuse within volumes and on surfaces, and to be transported across surfaces to cross between volumes.
Reactions between a single or multiple species occur within compartments (volume or surface reactions) or in adjacent compartments (volume-surface or volume-surface-volume reactions) (\Cref{fig:workflow}A).
The use of a mixed finite element discretization in space allows for high numerical accuracy and geometric flexibility~\cite{daversin-cattyAbstractionsAutomatedAlgorithms2021a}.
Crucially, this approach allows for spatial variation of each species within compartments and transport across compartments, while conserving mass and momentum.
The computational model is allowed to evolve over time, yielding detailed predictions for changes in each species within the specified geometries (\Cref{fig:workflow}E). 
High performance sparse numerical linear algebra~\cite{balayPETScPortableExtensible1998} in combination with scalable finite element algorithms~\cite{alnaesFEniCSProjectVersion2015} allows for computational models with millions of degrees of freedom to be solved efficiently. 

In the following sections, we demonstrate SMART via a series of use cases that model different signaling networks in cellular and subcellular geometries.
Our examples begin at the whole-cell scale and approach the single organelle scale, including both idealized geometries as well as experimentally derived geometries from 3D electron microscopy.

\subsection*{YAP/TAZ mechanotransduction in cells on micropatterned substrates}

To demonstrate use of SMART at the whole-cell length scale, we consider the model of YAP/TAZ mechanotransduction originally developed by Sun et al.\@ \cite{sunComputationalModelYAP2016} and extended to a spatial model by Scott et al.\@ \cite{scottSpatialModelYAP2021}.
This model considers the intracellular signaling cascade induced by a cell adhering to a substrate with a given stiffness, from phosphorylation of focal adhesion kinase (FAK) to downstream activation of the actomyosin cytoskeleton and nuclear translocation of the transcriptional regulatory proteins Yes-associated protein (YAP) and PDZ-binding motif (TAZ) (\Cref{fig:mech}A).
Generally, on stiffer substrates, increased FAK phosphorylation results in increased actin polymerization and myosin contractility, leading to dephosphorylation of cytosolic YAP/TAZ and subsequent translocation of YAP/TAZ into the nucleus.

Here, we test the predicted effects of cell adhesion to micropatterned surfaces on the localization of YAP/TAZ to the nucleus.
We consider three different contact region geometries previously tested experimentally \cite{maduramSubcellularCurvaturePerimeter2008} - circular, rectangular, and star-shaped micropatterns (\Cref{fig:mech}B-C).
In all cases, the cell volume and size of the contact region are the same, but the total plasma membrane surface area is increased for geometries with greater curvature at the contact regions (see \Cref{sec:meshes} for details on generating these meshes).

\begin{figure}[htbp!]
    \centering
    \includegraphics[width=7in, keepaspectratio]{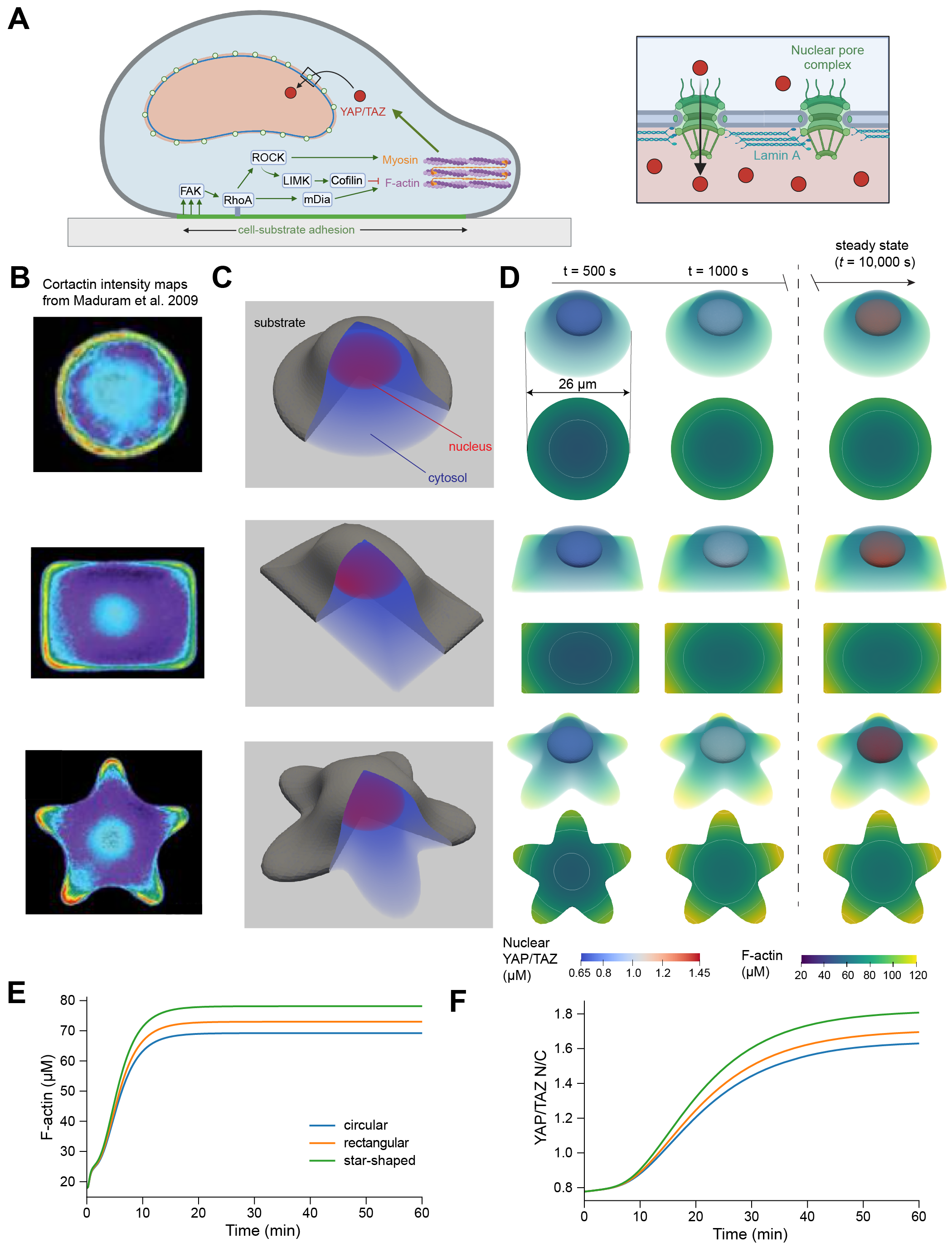}
    \caption{\textbf{Spatial model of YAP/TAZ mechanotransduction for cells on micropatterned surfaces.}
    (cont. on next page)}
\end{figure}

\addtocounter{figure}{-1}
\begin{figure} [t!]
  \caption{(cont. from previous page.) 
    A) Schematic of YAP/TAZ mechanotransduction signaling pathway. 
    Phosphorylation of FAK within the region of cell-substrate adhesion leads to cytoskeletal activation, triggering the dephosphorylation of YAP/TAZ and the opening of nuclear pore complexes (inset), allowing for the transport of YAP/TAZ into the nucleus.
    Panel created with Biorender.com.
    B) Measurements of cytoskeletal activation in cells on micropatterned substrates.
    Modified from Figures 2 and 4 in \cite{maduramSubcellularCurvaturePerimeter2008}, permission pending.
    C) Summary of geometries for cells spread on circular, rectangular, and star-shaped micropatterns.
    D) Simulations of YAP/TAZ mechanotransduction in cells on circular, rectangular, and star-shaped micropatterns.
    For each case, the side view is pictured, including both F-actin in the cytosol and YAP/TAZ in the nucleus, as well as the bottom-up contact region view showing local activation of actin polymerization in regions of higher curvature along the cell contour.
    Bottom-up views also include contours showing lines of constant concentration.
    E-F) F-actin (E) and nuclear YAP/TAZ (F) dynamics plotted over time for all three cases.
    }
    \label{fig:mech}
\end{figure}

We consider the predictions of this model across all three geometries on a very stiff substrate such as glass.
We find that regions of the cell where the plasma membrane surface area to cytosolic volume ratio is the highest have high concentrations of signaling molecules of interest.
In particular, we observe elevated levels of F-actin in these regions, in agreement with the results from Maduram et al.\@ \cite{maduramSubcellularCurvaturePerimeter2008} (\Cref{fig:mech}B, D, \Cref{video:circ,video:rect,video:star}).
Micropatterns with highly curved regions along the perimeter (star or rectangular patterns) show greater increases in overall cytoskeletal activation (\Cref{fig:mech}E) and, consequently, higher nuclear abundance of YAP/TAZ (\Cref{fig:mech}F).
Importantly, regardless of the shape of the contact area, all increases in YAP/TAZ are much lower than those predicted by a well-mixed model of YAP/TAZ mechanotransduction (compare to \Cref{fig:accuracy}D) due to persistent gradients in cytoskeletal activation over the cellular geometry.
That is, the F-actin concentration at the nuclear membrane is much lower in this spatial model compared to the well-mixed case, in which the effects of signal attenuation in regions of the cytosol further away from the plasma membrane are neglected.
This observation highlights the importance of accounting for spatial effects in cell signaling networks.

\subsection*{Intracellular calcium dynamics in realistic subcellular geometries}

We next consider spatial models defined over realistic \emph{subcellular} geometries acquired from 3D electron microscopy.
We utilize previously published models of calcium dynamics within dendritic spines in neurons \cite{bellDendriticSpineGeometry2019a} and cardiomyocyte calcium release units (CRUs) \cite{hakeModellingCardiacCalcium2012a}.
Each case includes calcium influx through the plasma membrane and calcium exchange across the sarcoplasmic or endoplasmic reticulum (SR or ER) membrane, as well as calcium binding and unbinding to buffering proteins within the cytosol and SR/ER (\Cref{fig:calcium}A). 
The calcium dynamics in each system are influenced by the geometry and relative spatial arrangement of organelles \cite{leungSystemsModelingPredicts2021b,liOrganizationCa2Signaling2022}.

Using the model previously implemented for idealized dendritic spine geometries by Bell et al.\@ \cite{bellDendriticSpineGeometry2019a}, we simulate calcium changes within a realistic dendritic spine containing a specialized form of endoplasmic reticulum known as the spine apparatus (SA) \cite{wuContactsEndoplasmicReticulum2017}.
Calcium influx occurs through voltage-sensitive calcium channels (VSCCs) located in the head and a section of the neck and through NMDA receptors localized to a dense region of proteins known as the post-synaptic density (PSD) (\Cref{fig:calcium}B).
Each of these fluxes is written as an analytical expression over time, and dependent on a specified change in the membrane potential. 
Calcium is removed from the cytosol via efflux across the plasma membrane through the sodium calcium exchanger (NCX) and plasma membrane calcium ATPase (PMCA) and influx into the spine apparatus through the sarco/endoplasmic calcium ATPase (SERCA).
Over the short time scale considered in this model, we assume that calcium entry into the SA dominates over release. 
Simulations reveal that calcium elevation is highest within the spine head, also leading to a more substantial increase in calcium concentration in the SA located in this region (\Cref{fig:calcium}C-D, \Cref{video:spine}).
The peak calcium approaches 5 \uM, lower than the 8 $\upmu$M peak observed in the original model \cite{bellDendriticSpineGeometry2019a}.
This lower peak calcium concentration is accounted for by a higher surface area of SA present in the spine head for this realistic geometry compared to the smaller SA in the original simulations (here: 10.6 $\upmu$m\textsuperscript{2} SA per $\upmu$m\textsuperscript{3} cytosol; original: 2.31 um\textsuperscript{2} SA per um\textsuperscript{3} cytosol).

Given the importance of calcium in signaling pathways, to ensure robustness across different model formulations and geometries, we also use SMART to model calcium release from the SR in a CRU.
The CRU geometry\cite{hoshijimaCCDB3603MUS2004} includes a continuous section of SR, one T-tubule volume, and two mitochondria.
We do not consider any species within the mitochondria or T-tubule, but rather assume that the T-tubule calcium concentration remains constant due to its continuity with the extracellular space and treat the mitochondria as passive diffusion barriers, in line with previous modelling by Hake et al.\cite{hakeModellingCardiacCalcium2012a}.
Similarly, the NCX, PMCA, and leak fluxes through the T-tubule membrane, the ryanodine receptor (RyR) and SERCA fluxes through the SR membrane, as well as calcium buffering due to calmodulin, troponin, and ATP in the cytosol and calsequestrin in the SR, all follow established relationships\cite{hakeModellingCardiacCalcium2012a}.
However, we here utilize a full spatial discretization of the sarcoplasmic reticulum interior, which was previously treated as a series of well-mixed subregions\fixme{\cite{hakeModellingCardiacCalcium2012a}}.
For comparison, we consider two conditions -- one with SERCA present throughout the SR membrane and another with no SERCA activity.
In agreement with the original model, calcium reaches a concentration of several $\upmu$M near the T-tubule / SR junction and about 1 $\upmu$M further away from the sites of RyR release (\Cref{fig:calcium}E-F, \Cref{video:cru1}).
This calcium spike is short-lived, as the model assumes that RyRs close upon sufficient reduction of SR calcium.
Including active SERCA in the SR membrane results in a slightly prolonged calcium elevation and robust refilling of the SR calcium store over time (\Cref{fig:calcium}F, \Cref{video:cru2}), again in line with previous findings\cite{hakeModellingCardiacCalcium2012a}.
In both calcium signaling examples presented here, the spatial behavior captured by SMART is critical to the overall dynamics, with the proximity between calcium sources and sinks determining the overall extent of calcium increase. 

\begin{figure}[htbp!]
    \centering
    \includegraphics[width=7in,keepaspectratio]{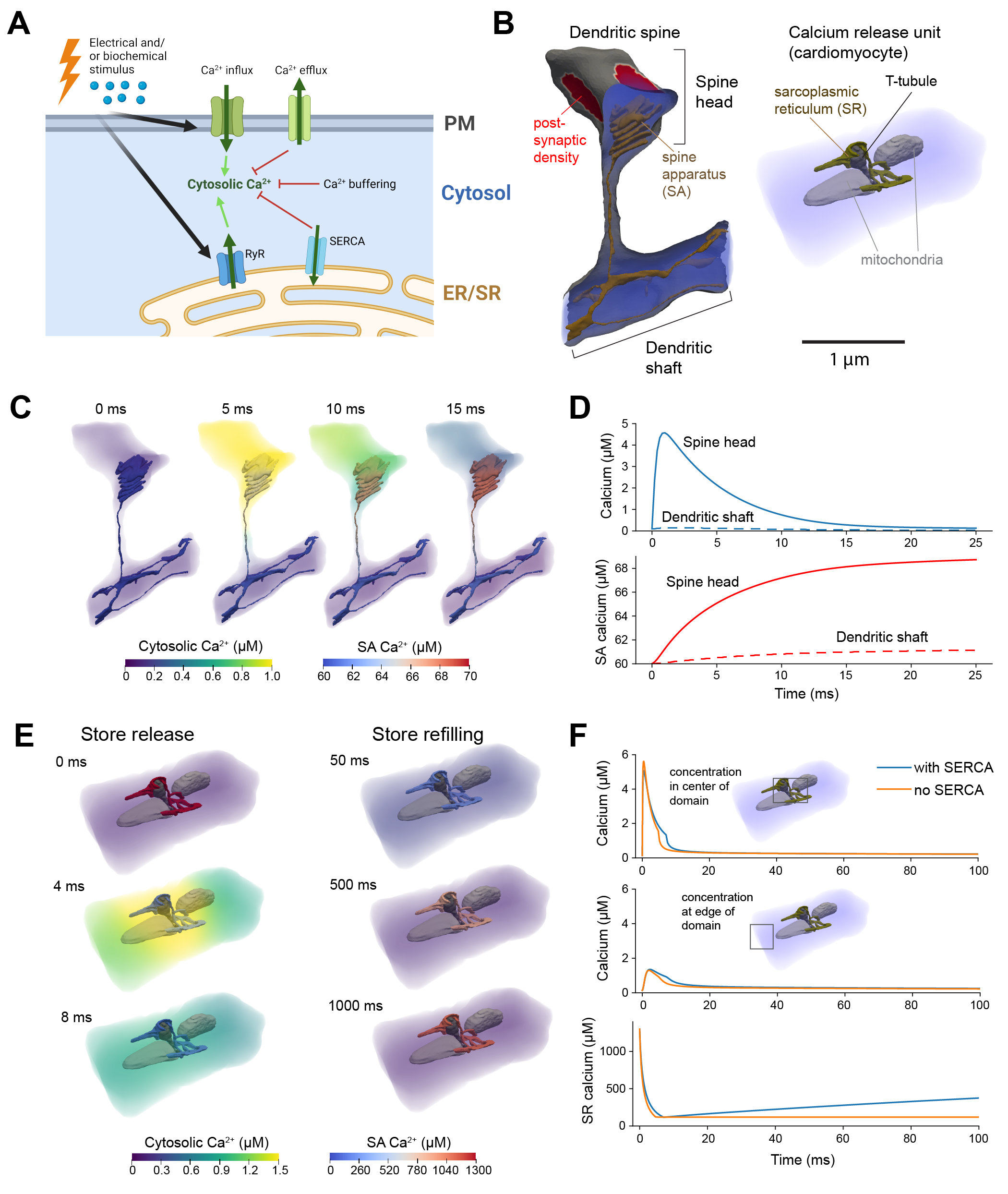}
    \caption{ 
        {\bf Calcium dynamics within a realistic dendritic spine and calcium release unit.}
        (cont. on next page)
        }
\end{figure}

\addtocounter{figure}{-1}
\begin{figure} [t!]
    \caption{(cont. from previous page.) 
        A) General schematic of a calcium signaling cascade in a subcellular region.
        In response to biochemical or electrical stimulus, calcium influx occurs through the PM and calcium is released from the ER/SR store.
        Calcium elevations are counteracted by calcium efflux into the extracellular space and repackaging into the ER/SR through SERCA.
        Finally, the changes in free calcium are dampened due to calcium binding to proteins in the cytosol (buffering).
        Panel created with Biorender.com.
        B) Realistic geometries of a dendritic spine and a cardiomyocyte calcium release unit derived from electron microscopy.
        The dendritic spine branches off from the main dendritic shaft and contains a lamellar ER structure known as the spine apparatus (SA, gold), as well as a denser region of signaling proteins in a subregion of the spine head, the postsynaptic density (PSD, red).
        The calcium release unit consists of a section of SR (gold) closely apposed to T-tubules (dark grey) in a region known as the junctional cleft.
        Mitochondria (light grey) serve as passive diffusion barriers in this case.
        C) Simulation of calcium dynamics in a dendritic spine.
        Calcium elevations are concentrated to the dendritic head; following the initial influx, calcium is pumped into the SA via SERCA.
        D) Plots of cytosolic and SA calcium dynamics in the dendritic spine head vs.\@ the dendritic shaft.
        E) Simulation of calcium dynamics in a CRU.
        Calcium release from the SR results in a calcium spike near the center of the geometry, followed by refilling of the SR with calcium.
        F) Plots comparing the dynamics of cytosolic and SR calcium with vs.\@ without SERCA.
    }
    \label{fig:calcium}
\end{figure}

\subsection*{ATP synthesis in realistic mitochondrial geometries}

We finally consider a biochemical network within a single organelle, modeling the generation and transport of ATP within a realistic mitochondrial geometry.
This mitochondrial geometry was previously reconstructed from serial electron tomograms and the resulting mesh was conditioned in GAMer2 \cite{garciaMitochondrialMorphologyProvides2019, mendelsohnMorphologicalPrinciplesNeuronal2022a}.
It was then used for simulations of ATP generation in a well-mixed context and for particle-based simulations in MCell \cite{garciaMitochondrialMorphologyProvides2019,garciaMitochondrialMorphologyGoverns2023}.
Here, we implement a thermodynamically consistent continuum-based model version\cite{garciaMitochondrialMorphologyGoverns2023} and compare our current results to those from well-mixed ODE simulations and particle-based simulations\cite{garciaMitochondrialMorphologyProvides2019,garciaMitochondrialMorphologyGoverns2023}.

\begin{figure}[htbp!]
    \centering
    \includegraphics[width=7in,keepaspectratio]{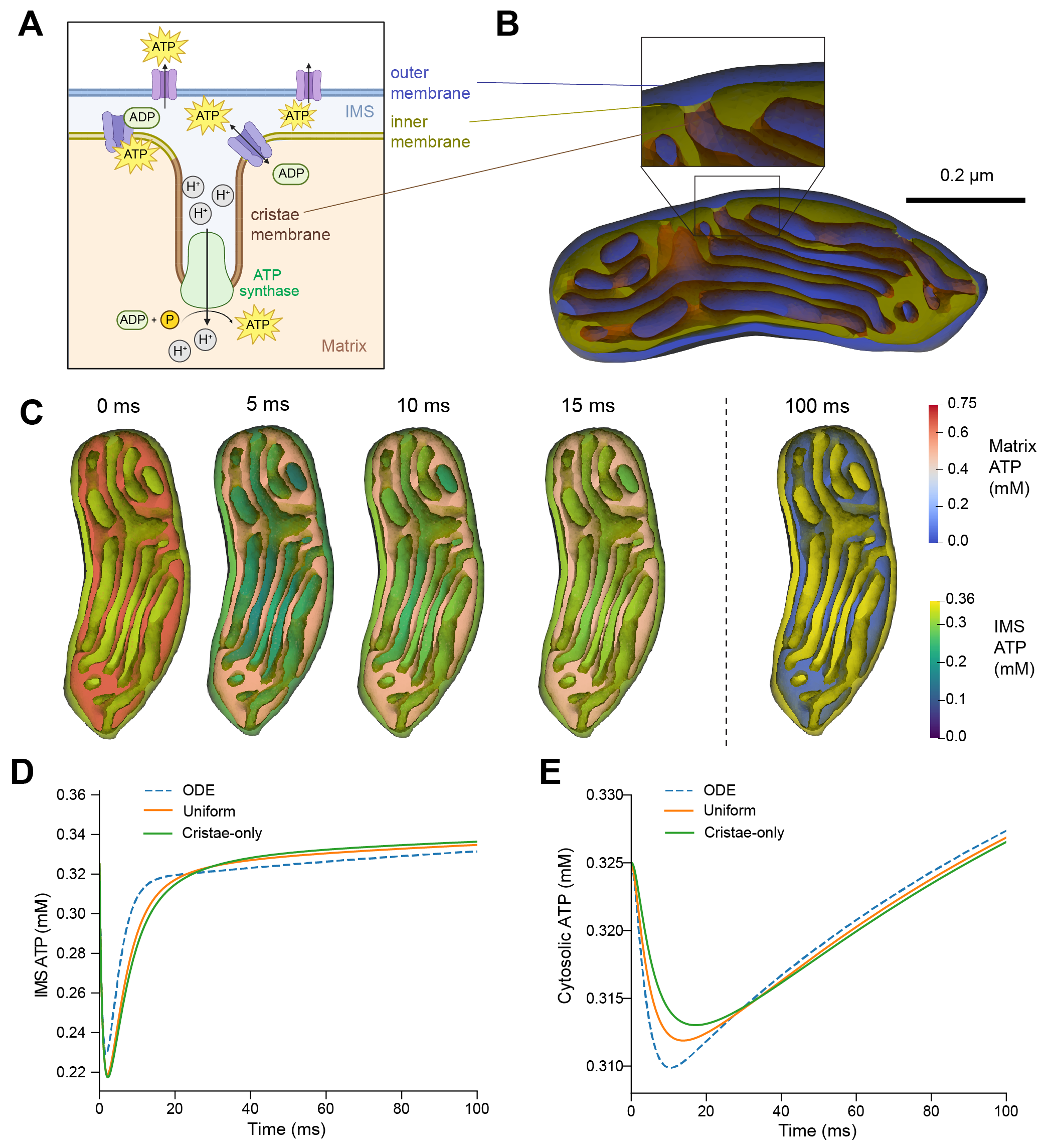}
    \caption{ 
        {\bf ATP synthesis in a realistic mitochondrial geometry.} 
        A) Schematic of the ATP synthesis process in mitochondrion.
        ADP in the mitochondrial matrix is converted into ATP by ATP synthase in the cristae membrane, then transported into the inner membrane space through ANTs and exported into the cytosol through VDACs.
        Panel created with Biorender.com.
        B) Realistic mitchondrial geometry.
        The inner membrane divides two volume compartments - matrix (yellow) and inner membrane space (blue).
        The inner membrane itself is divided into a portion on the periphery (yellow) and invaginations known as cristae (brown).
        C) Dynamics of IMS ATP and matrix ADP in a mitochondrion with uniform distributions of ATP synthase and ANTs in the inner membrane.
        D-E) IMS ATP concentration (D) and cytosolic ATP concentration (E) over time for uniform vs.\@ cristae-localized distributions of IMS proteins.
        ODE model predictions are plotted as well for ease of comparison (dashed lines).
    }
    \label{fig:mito}
\end{figure}

The model considers the generation of ATP in the mitochondrial matrix by ATP synthase, followed by the transport of ATP into the inner membrane space (IMS) by adenine nucleotide transporters (ANTs) and export into the cytosol through voltage-dependent anion channels (VDACs) in the outer membrane (OM) (\Cref{fig:mito}A-B).
Previous experimental data and modeling results suggest that the spatial organization of ATP synthase and ANTs is a key determinant of the magnitude of ATP changes in the cytosol \cite{garciaMitochondrialMorphologyProvides2019}.
Accordingly, we tested two alternative spatial arrangements of ATP synthase and ANTs while keeping the total number of molecules unchanged.
In the first case, we distributed these molecules uniformly throughout the inner membrane (IM), whereas in the second, more physiological case, we colocalized ANTs and ATP synthase in invaginations of the IM known as cristae.
In both cases, we maintained the same total amount of ANTs and ATP synthase molecules.

A uniform distribution of ANTs results in dynamics of IMS and cytosolic ATP similar to those predicted by the ODE model and particle-based simulations(\Cref{fig:mito}D-E).
ATP in the IMS initially decreases rapidly as it binds to ANTs, then gradually increases as ATP synthase produces more ATP.
In comparison, concentrating ANTs and ATP synthase in the cristae results in a less pronounced initial reduction of ATP in the cytosol.
This phenomenon was previously termed ``energy buffering'' \cite{garciaMitochondrialMorphologyProvides2019} and was attributed to the spatial separation between ANTs in the cristae and VDACs in the OM.
Rapid changes to ATP concentration in the inner cristae space do not immediately affect ATP levels in the cytosol as time is required for diffusion between the outer membrane and cristae.
This delay and dampening effect could make the cell more resilient to a noisy environment featuring rapid changes in the availability of ATP and ADP.
This spatial phenomenon is captured via SMART simulations but is not accessible via well-mixed ODE models.

\subsection*{Verification and validation of the SMART computational algorithms.}
\begin{figure}
    \includegraphics[width=6.5in,keepaspectratio]{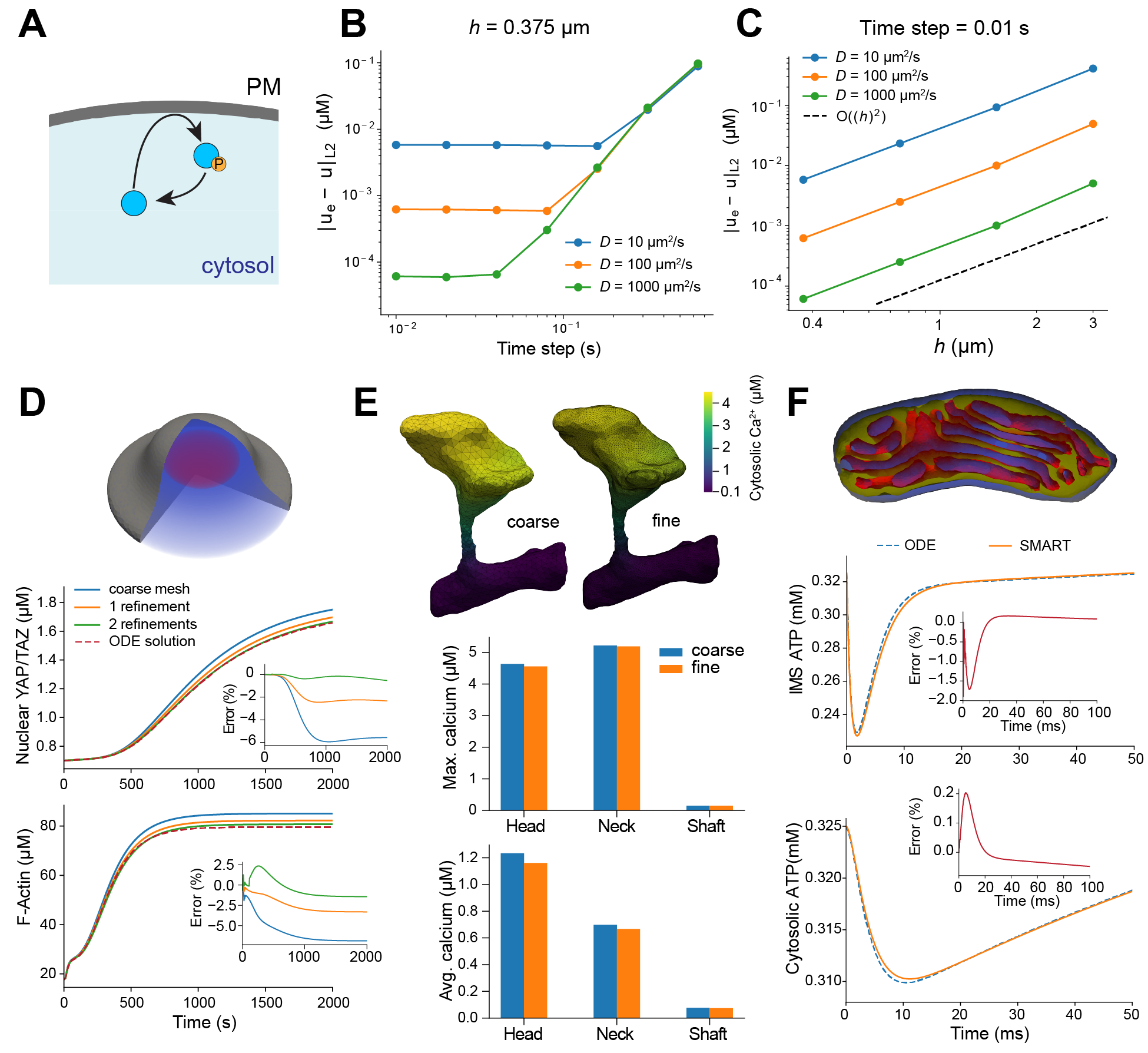}
    \centering
      \caption{{\bf Numerical verification and validation of SMART.}
      A) Model of protein phosphorylation at the plasma membrane. 
      B-C) Convergence of system describing phosphorylation of proteins at the plasma membrane (as described in Ref.\@ \citenum{meyersPotentialControlSignaling2006}) upon time step refinement (B) and mesh refinement (C). The $\mathcal{L}_2$ error between the computed and analytical solutions for all combinations of temporal and spatial refinement are shown in \Cref{tab:rates_D10.0}, \Cref{tab:rates_D100.0} and \Cref{tab:rates_D1000.0} for diffusion coefficients $D$ 10, 100, and \qty{1000}{\micro\meter\squared\per\second} respectively.  
      D) Comparison of ODE solution to full SMART simulation for mechanotransduction example with fast diffusion at different mesh refinements.
      E) Spatial differences in solution for dendritic spine calcium for course vs. refined mesh. Maximum and average calcium concentrations in different regions are reported for the coarse vs. fine mesh. 
      F) Comparison of ODE solution to full SMART simulation of mitochondrial ATP production for fast-diffusing nucleotide species.}
    \label{fig:accuracy}
\end{figure}

To verify our numerical and computational approach, we examine the simulation results for an example with an analytical solution at steady state (details in \Cref{app:sec:phos}).
In this example, originally examined by Meyers et al.\@ \cite{meyersPotentialControlSignaling2006}, a protein is phosphorylated at the cell membrane and dephosphorylated throughout the cytosol (\Cref{fig:accuracy}A).
We considered the case of a thin slab whose spatial solution is well-described by the 1D solution along its thickness.
This geometry mimics the case of an adherent cell spread out on a flat substrate. 
To demonstrate spatial convergence, we uniformly refined the reference mesh four times, resulting in maximum cell sizes of 3 \um, 1.5 \um, 0.75 \um, and 0.375 \um.
Similarly, to show temporal convergence we halved the time step 7 times from 0.64 s to 0.01 s. 
\Cref{fig:accuracy}B-C show the temporal and spatial convergence respectively, for three different diffusion coefficients, using the finest mesh (B) and finest time step (C). 
We observe that the numerical error is consistently lower for higher diffusion coefficients.
Moreover, the mesh resolution error dominates up to some critical value of the time step for each $D$, as indicated by the plateau in each curve (\Cref{fig:accuracy}B).
This critical time-step is smaller for high values of $D$, reflecting the need for smaller time-steps to achieve optimal convergence in cases of rapid diffusion.
At the smallest time step, mesh refinement error dominates in all cases, as demonstrated by the theoretically-expected and optimal second-order convergence in the $\mathcal{L}_2$ norm of the error with respect to element size $h$ (\Cref{fig:accuracy}C).



Next, we focus on the accuracy and convergence of the numerical results for our three main biological scenarios.
In the case of YAP/TAZ mechanotransduction, we can compare our numerical solutions to solutions of the well-mixed ODE approximation.
Considering the case of a cell with a circular contact region on a glass substrate, we set diffusion coefficients for all volume species to \qty{1000}{\micro\meter\squared\per\second} and treat all surface species as uniform and non-diffusing.
We start with a time step of $\Delta t = 0.01$ s and adaptively adjust the step according to the number of Newton iterations required for convergence at the previous time (see \Cref{table:time-step} for details of our adaptive time-stepping scheme).
Upon mesh refinement, the average F-actin concentration and nuclear YAP/TAZ concentration converge to a solution close to the trajectories predicted by the ODEs (\Cref{fig:accuracy}D).
For the main downstream quantity of interest, YAP/TAZ nuclear concentration, the refined mesh differs a maximum of 0.6\% from the well-mixed solution.
Note that the most refined mesh here is coarser than the mesh used in simulations shown in \Cref{fig:mech} (see all mesh statistics in \Cref{tab:mech-meshes} and \Cref{tab:mech-compartments}).

Dendritic spine simulations were initially conducted on a mesh with $h_{min} = 0.009$ \um  and $h_{max} = 0.199$ \um and a time step of $\Delta t = 0.00025$ s.
After two refinements of the mesh, we observe that the average calcium concentrations are similar across refinements, but that the solutions exhibit local spatial differences (\Cref{fig:accuracy}E).
In general, lower peak values are observed in the more refined simulation.
However, across refinements, the peak calcium in the spine head deviated by a maximum of 0.52\%, demonstrating the accuracy of the baseline results (\Cref{fig:accuracy}E).
Similarly, testing time steps from 0.00025 s to 0.002 s reveals convergence to a single solution consistent with that in \Cref{fig:calcium} (\Cref{fig:suppl-spine-testing}).

Finally, we assess the accuracy of our solutions at the single organelle level by comparing our mitochondrial ATP production simulations to a well-mixed approximation.
The mitochondrion mesh resolution is $h_{min} = 0.00661$ \um and $h_{max} = 0.0383$ \um, and our time step is adaptively adjusted from an initial value of $\Delta t = 0.02$ ms.
When setting the diffusion coefficients for all volumetric nucleotide species to \qty{150}{\micro\meter\squared\per\second}, we obtain a close match to the ODE solution (\Cref{fig:accuracy}F).
In particular, the predicted value of cytosolic ATP deviates by a maximum of 0.2\% from the well-mixed solution (\Cref{fig:accuracy}F).
 

\subsection*{Computational performance and scalability}

\begin{figure}
    \includegraphics[width=7in,keepaspectratio]{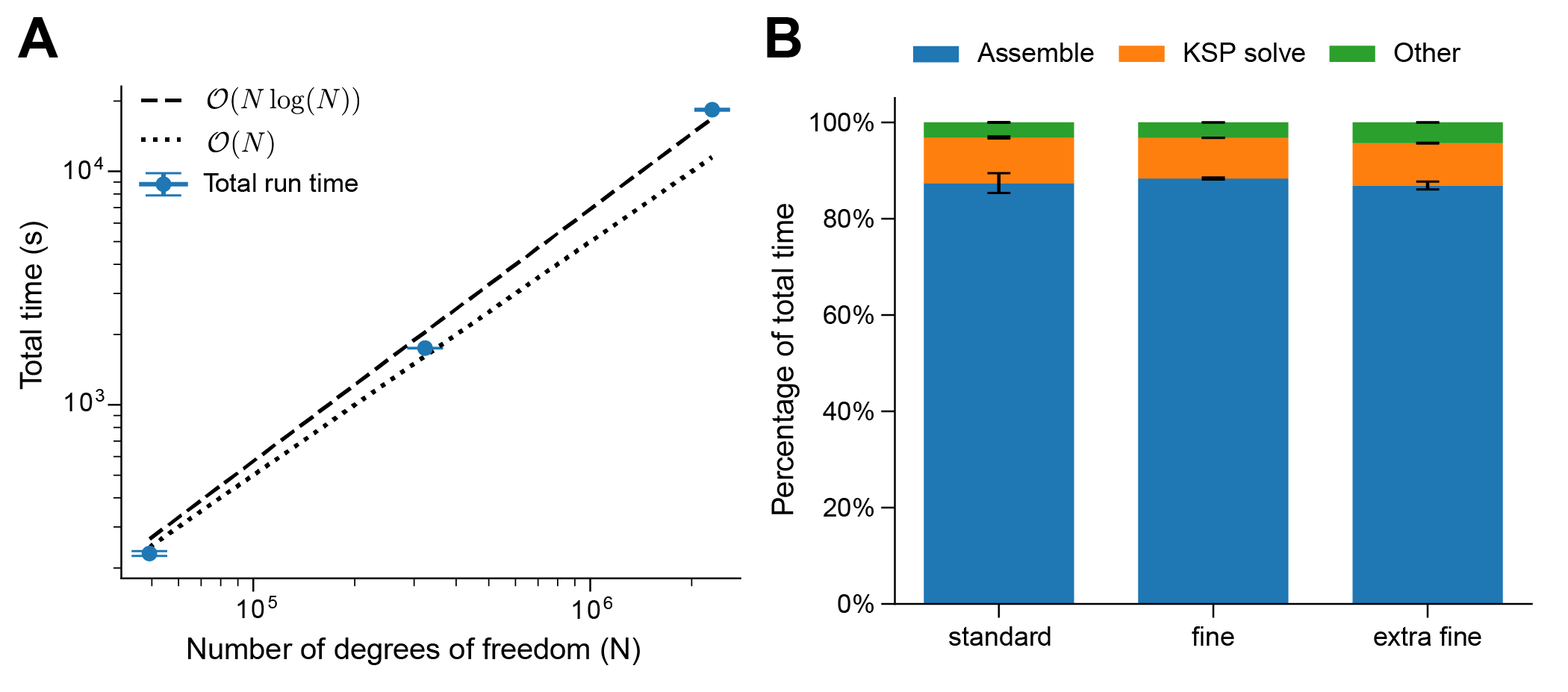}
    \centering
   \caption{{\bf Performance and scalability of SMART for dendritic spine calcium dynamics}
   A) Total run times for 25 simulated time steps of intracellular calcium dynamics for increasing mesh refinement levels (standard, fine and extra fine), averaged over 5 runs for standard and fine meshes and 3 runs for the extra fine mesh.
   B) Relative contributions to the total run time for each level, averaged over 5 runs for standard and fine meshes and 3 runs for the extra fine mesh.
   The computational cost is dominated by the finite element assembly and function evaluation due to the complexity of the coupled nonlinear equations.}
   \label{fig:performance}
\end{figure}

To study the scalability of our computational framework, we consider the simulation of intracellular calcium dynamics in an electron micrograph-based representation of a dendritic spine (Example 2). 
Starting with the baseline computational mesh, we consider two additional levels of uniform mesh refinement resulting in three discrete representations (`standard', `fine', and `extra fine'). 
These meshes are composed of \num{106979} (\num{18649}), \num{855832} (\num{146024}), and \num{6846656} (\num{1154871}) tetrahedral cells (vertices), respectively, thus corresponding to an $8\times$ increase in the number of cells for each refinement and a similar increase ($7.8-7.9\times$) in the number of vertices. 
The corresponding numbers of degrees of freedom ($N$, dimensions of the discrete solutions) are: \num{49194}, \num{323328}, \num{2299090}; thus corresponding to a $6.6\times$ and $7.1\times$ increase between refinements. 
For each simulation, we use a 0.001 s timestep through $t = 0.025$ s. 
We first note that the number of nonlinear iterations per time step is constant ($4$) across time steps and refinement levels. 
Next, inspecting the total run time of each simulation as a function of the computational cost (measured in terms of the number of degrees of freedom $N$), we observe that the simulation time scales close to log-linearly (between $\mathcal{O}(N)$ and $\mathcal{O}(N \log N)$) (\Cref{fig:performance}A).
The computational cost associated with the initialization of the SMART symbolic problem representation is small ($2.0\%$, $2.4\%$, $3.6\%$). 
A break-down of the total run times shows that the simulation time is persistently dominated by the finite element assembly ($87.4\%$, $88.4\%$, $87.0\%$), followed by the iterative solution of the linear systems ($9.5\%$, $8.4\%$, $8.8\%$) (\Cref{fig:performance}B). 


\section*{Discussion}

Cell function is tightly linked to cell shape, as evidenced by the diverse shapes exhibited by different cell types, from neurons with branch-like extensions to cardiomyocytes that assemble in block-like morphologies \cite{rangamaniDecodingInformationCell2013a,nevesCellShapeNegative2008,calizoCellShapeRegulates2020}. 
The importance of geometry in function extends to the single organelle-level; for instance, the inner membrane of mitochondria forms tortuous invaginations (cristae) that facilitate efficient production of ATP.
The importance of cell and organelle geometry are generally well-appreciated, but detailed geometries have proven difficult to include in models of cell signaling.
The models showcased within this work exhibit the use of our simulation technology and software, SMART, to simulate such signaling networks with spatial resolution, with a particular focus on realistic cell and organelle geometries.

SMART offers many possibilities for exploring the impacts of cell shape in spatiotemporal signaling models in cell biology.
This opportunity is driven by a wealth of new imaging data from modalities such as volume electron microscopy and super-resolution fluorescence microscopy \cite{heinrichWholecellOrganelleSegmentation2021a,mccaffertyIntegratingCellularElectron2024,stoneSuperResolutionMicroscopyShedding2017,schermellehSuperresolutionMicroscopyDemystified2019}.
While the experimentally-derived cell geometries here were all extracted from electron micrographs, other types such as e.g. fluorescence data could be utilized for future applications of SMART.
Fluorescence data could offer the additional opportunity to inform not just cell and organelle geometry, but also the spatial localization and dynamics of different molecules such as membrane receptors.
There are a variety of high-quality public datasets available from super-resolution microscopy and volume electron microscopy alike due to projects such as the Allen Institute's Cell Explorer \cite{johnsonBuildingNextGeneration2023,vianaIntegratedIntracellularOrganization2023} and Janelia's OpenOrganelle \cite{heinrichWholecellOrganelleSegmentation2021a,xuOpenaccessVolumeElectron2021}.
Biophysical modelling leveraging such imaging datasets takes full advantage of the recent efforts to improve mesh conditioning for biological samples such as GAMer2 \cite{lee3DMeshProcessing2020a}, VolRover \cite{edwardsVolRoverNEnhancingSurface2014} or fTetWild \cite{huFastTetrahedralMeshing2020}.

Indeed, the examples in this paper demonstrate the fundamental importance of geometry in models of cell signaling.
For instance, in our model of YAP/TAZ mechanotransduction, while results agree qualitatively between a well-mixed model and a full spatial description, quantitative predictions of the model differ considerably between these two modeling regimes.
Furthermore, certain aspects of the signaling network cannot be captured by a well-mixed model, such as increased levels of actin polymerization near the plasma membrane compared to elsewhere in the cell.
Similarly, despite the fast diffusion of small molecular species such as calcium and ATP, we find that spatial factors such as the adjacency between sources and sinks (dendritic spine and CRU examples) or the distribution of surface species (ATP example) are key determinants of system behavior.
Indeed, in the case of ATP generation, changing the distribution of ATP synthase and ANTs in the inner membrane has a non-negligible effect on ATP dynamics in the cytosol.
Considering that model parameters are commonly estimated using well-mixed models, parameterizing spatial models remains an important challenge.
SMART is well-positioned to address these issues via auxiliary tools such as dolfin-adjoint for sensitivity analysis and parameter estimation in FEniCS-based models \cite{mituschDolfinadjoint2018Automated2019}.

In addition to SMART, there exist several complementary software options available to computational biologists and biophysicists for examining the spatiotemporal behavior of cell signaling networks.
Among the most popular are Virtual Cell (VCell) \cite{schaffGeneralComputationalFramework1997, cowanSpatialModelingCell2012} and MCell \cite{kerrFastMonteCarlo2008}, both open source projects focused on continuum and particle-based models of biochemical networks in cells, respectively.
Within this broader context, SMART offers unique capabilities and features for those users wishing to test complex cell geometries in a continuum framework.
For instance, VCell provides a robust option for solving mixed dimensional reaction-diffusion equations using the finite volume method, but currently only pixelized or voxelized grid meshes are supported \cite{novakDiffusionCurvedSurface2007}.
MCell, in contrast, does support surfaces defined by unstructured meshes; however, MCell uses particle-based simulations to describe stochastic reaction-diffusion networks.
Such simulations are much more accurate when considering a small number of molecules but cannot yet be scaled to whole-cell simulations where many billions of molecules may be present.
Alternative stochastic simulation packages such as the STochastic Engine for Pathway Simulation (STEPS) \cite{hepburnSTEPSEfficientSimulation2012,hepburnVesicleReactiondiffusionHybrid2024} or Lattice Microbes \cite{robertsLatticeMicrobesHighperformance2013} use other algorithms to model stochastic reaction-diffusion networks within cells.
Many other options allow users to readily assemble signaling networks and/or perform parameter estimation and sensitivity analysis in the well-mixed case \cite{hoopsCOPASICOmplexPAthway2006, clarkLogicbasedModelingBiological2024}.
These complementary approaches work well alongside SMART to help develop robust models of cell signaling and select parameters for spatial and well-mixed models alike.

In terms of limitations, SMART currently only supports describing a set of compartments of codimension 1, \textit{i.e.}, only 3D/2D or 2D/1D, whereas in principle, 1D or even 0D features could contribute in the full 3D case.
We note that 0D/well-mixed features can currently be included by explicitly updating parameters at each time step, as in our ATP generation example (see \Cref{eq:ode1,eq:ode2}).
SMART also currently only supports diffusion as a transport mechanism, whereas our framework can easily be expanded in the future to support e.g.~reaction-advection-diffusion problems.
This can also naturally be extended to the case in which the cell boundary moves over time according to some prescribed velocity field. SMART does not currently consider the effects of other physical processes, such as electrodiffusion or mechanical forces, that are often tightly coupled to signaling.
In fact, this remains an issue across most of computational cell biophysics.
FEM-based approaches, such as that of SMART, offer one promising approach to efficiently solve such problems, due to their general applicability to solve a range of PDEs.
In summary, SMART represents an advancement towards realistically simulating cellular and subcellular systems.
The ability to do so will not only drive forward the understanding of cell biology, but could act as a tool to predict the response of cells to diverse treatments at an unprecedented level of detail.

\section*{Methods}

The mathematical models, computational algorithms and simulation technology underlying SMART are summarized here. An extended description is provided in \Cref{app:sec:math}.

\subsection*{Domains, geometries and interfaces}
\label{sec:domains}

We consider model geometries embedded in 2D or 3D represented by tessellated curves, surfaces and volumes.
Each geometry is described by a collection of three-dimensional volumes (or two-dimensional surfaces in the 2D case) with boundaries and interfaces between volumes represented by two-dimensional surfaces (or one-dimensional curves in the 2D case).
We refer to the highest dimensional compartments as the \emph{bulk domains} ($\Omega$) and the lower dimensional boundaries or interfaces as the \emph{surfaces} ($\Gamma$).
Moreover, each of these domains are represented by a simplicial tessellation formed by the mesh cells (intervals, triangles or tetrahedra), mesh facets (points, intervals or triangles), and mesh vertices (points).
Importantly, all (sub)regions, boundaries and interfaces are defined relative to a single overarching \emph{parent} mesh \cite{daversin-cattyAbstractionsAutomatedAlgorithms2021a}.
Subdomains $\Omega^m \subseteq \Omega$ are defined as a set of mesh cells with a common label or tag $m \in \mathcal{M}$, and similarly exterior boundaries and interior interfaces $\Gamma^q$ are defined as a set of mesh facets again with a common tag $q \in \mathcal{Q}$.
In the examples presented here, we generated parent meshes and labeled their subdomains using either GAMer2 \cite{lee3DMeshProcessing2020a} to construct meshes from electron microscopy data or Gmsh \cite{geuzaineGmsh3DFinite2009} to generate idealized cell geometries. 

\subsection*{Coupled multi-domain reaction-transport equations}
\label{sec:eqns-overview}

SMART represents the spatially and temporally-varying concentrations of multiple species coexisting within domains via a general abstract modelling framework based on first principles.
Each concentration is defined over a subdomain $\Omega^m$ and/or surface $\Gamma^q$.
The evolution and distribution of these concentrations are described by a coupled system of time-dependent and nonlinear partial differential equations representing conservation of mass, the diffusion of each species, reactions between species within the bulk or surface domains, and fluxes between the bulk and surface domains (see \Cref{app:sec:equations} for a full description of the general framework).
Reactions are assumed to be local in the sense that different species may interact within each subdomain $\Omega^m$, and on or across the surfaces $\Gamma^q$ via the surface itself and its neighboring subdomains.
Specific computational models specify the effective diffusivity of each species and symbolic expressions for model reactions and fluxes, the initial concentration of each species, and any bulk or surface source or sinks.
Input parameters may be constant or spatially or temporally varying.
For further details, we refer to \Cref{sec:appendix:reactions}.

\subsection*{Numerical approximation and solution strategies} 

Each system of mixed-dimensional reaction-transport equations is discretized in time using a first-order accurate implicit Euler scheme with a uniform, variable or adaptive timestep size, yielding a coupled system of nonlinear partial differential equations at each time step. 
For the spatial discretization, we employ a monolithic finite element method via the FEniCS finite element software suite \cite{alnaesFEniCSProjectVersion2015}. 
As unknown discrete fields, we consider the concentration of each species $u_i^m$ defined in each bulk subdomain $\Omega^m$ and the concentration of each species $v_j^q$ defined over each surface $\Gamma^q$. 
All discrete fields are represented by continuous piecewise linear finite elements defined relative to the mesh used to represent the geometry. 
Since the equations are coupled across fields and domains, the resulting nonlinear systems of discrete equations take a block structure \cite{daversin-cattyAbstractionsAutomatedAlgorithms2021a} (\Cref{fig:workflow}D).


All nonlinear systems of equations are by default solved by Newton-Raphson iterations with an exact, automatically- and symbolically-derived Jacobian.
The time step for each system is either set as uniform throughout the course of the simulation or can be set adaptively based on the number of Newton-Raphson iterations required for convergence at the previous time step, as summarized in \Cref{table:time-step}.
In case of nonlinear solver divergence or negative solutions, repeated restarts with associated reductions in time step are invoked as an optional mediation strategy. 
The linear systems are solved iteratively using Krylov solvers in PETSc \cite{balayPETScPortableExtensible1998}. 
For default solver tolerances and settings, we refer to the open-source SMART code \cite{laughlinRangamaniLabUCSDSmartV22024a}.

\subsection*{SMART model specifications} 

As outlined in the first results section, the user must specify species, reactions, parameters, and compartments, and link these to a parent mesh prior to initializing a simulation.
Several minimal use cases are given in the SMART documentation \cite{laughlinRangamaniLabUCSDSmartV22024a} and the code for all examples in this paper is freely available \cite{francisBiologicalTestCases2024}.
In short, each species, reaction, parameter, or compartment is defined as a Python object, each with associated properties detailed in \Cref{sec:software}.
All instances of a given object type are then stored in an ordered Python dictionary, resulting in a single ``container'' for each type.
The parent mesh is either generated using Gmsh or read directly from an hdf5 or xml file.
The containers and mesh are then used to initialize the SMART model.


\subsection*{Model setup for biological test cases}

Here, we briefly outline some of the relevant details for each test case; for full model specifications, refer to \Cref{sec:tables} and our code \cite{francisBiologicalTestCases2024}.

\subsubsection*{Example 1}

This example is described in a previous publication that conducted similar simulations in VCell \cite{scottSpatialModelYAP2021}.
There are 24 species in the original model, 11 of which are eliminated after accounting for mass conservation.
We note that this simplification is only possible when different forms of a given molecule reside in the same compartment and share the same diffusion coefficient.
This simplification, for instance, cannot be applied to actin species (F-actin and G-actin), as they have drastically different diffusion coefficients.
All parameters were used unaltered from Scott et al.\@\cite{scottSpatialModelYAP2021}, with the exception of the altered diffusion coefficients for well-mixed simulations.

We generate meshes for these cell geometries as described in \Cref{sec:meshes}.
We minimized computational expense by exploiting the symmetries of each geometry; in the case of a circular contact region, the geometry is axially symmetric, allowing us to simulate the model over a 2-dimensional mesh while ensuring the correct $r$-dependence in all integrals (\Cref{sec:axisymm}).
The rectangular contact region has two symmetry axes, allowing us to only simulate one-quarter of the full geometry, treating the faces on the symmetry axes as no flux boundaries.
Similarly, the star contact region has five symmetry axes, allowing us to simulate one-tenth of the full mesh.
Simulations were run to $t = 10,000$ s, by which time the system has plateaued to an apparent steady-state.

\subsubsection*{Example 2}

The model of calcium dynamics within a dendritic spine was derived from Bell et al.\cite{bellDendriticSpineGeometry2019a}.
This model involves only four dynamical species - calcium, fixed calcium buffers bound to the plasma membrane, mobile calcium buffers in the cytosol, and calcium in the spine apparatus, which was previously assumed to be constant.
Calcium bound to buffers was not treated as a separate species, but was implicitly included assuming mass conservation and the same diffusion coefficient for buffering protein and buffering protein bound to calcium.
All other quantities that change over time were treated as time-dependent parameters, explicitly defined by functions of time.
As in the original model, NMDA receptors were restricted to the postsynaptic density.
VSCCs were restricted to the spine head and a portion of the neck to match the region of stimulus in Bell et al.\@ \cite{bellDendriticSpineGeometry2019a}.
All other membrane fluxes were uniform throughout the spine head, neck, and dendritic shaft.

Our analysis of the calcium release unit used the model presented by Hake et al.\@ \cite{hakeModellingCardiacCalcium2012a} with minor modifications.
In their simulations, the SR was treated as a composite of several well-mixed compartments and the wider space surrounding the CRU was simulated as connected well-mixed compartments.
Instead, we included the entire SR volume explictly in our spatial simulations, and boundaries of the cytosolic mesh were treated as no-flux surfaces.
As above, complexes between calcium and buffering species (ATP, CMDN, TRPN, and CSQN) were implicitly considered under the assumptions of buffer mass conservation and unchanged diffusion coefficients. 
The main mechanism for calcium elevations in this model is release from the SR through ryanodine receptors, which is assumed to terminate when calcium concentration falls below a certain threshold.
Rather than explicitly encoding this discontinuity in the equations, we enforce this condition manually--we assign zero conductance to the RyRs after the average calcium concentration in the entire SR reaches the threshold.
SERCA channels were either included uniformly through the SR membrane or were excluded entirely.

\subsubsection*{Example 3}

The model of ATP generation was directly adapted from the model by Garcia et al.\@ \cite{garciaMitochondrialMorphologyGoverns2023}.
The model considers 6 states of ATP synthase and 9 states of ANTs, as well as ATP concentration in the matrix and inner membrane space (IMS) and ADP concentration in the matrix.
Assuming mass conservation, only 5 states and 8 states need to be modeled explicitly for ATP synthase and ANTs.
ADP concentration in the IMS is assumed to be constant and ATP concentration in the cytosol was solved for using the following coupling scheme.
Prior to solving the PDEs at each time step ($t = t_n$), the ATP concentration $T_{cyto}$ at the next time point $t_n + \tau_n$ was estimated as:
\begin{equation}
T_{cyto,est}(t_n + \tau_n) = T_{cyto} (t_n) + \tau_n \frac{10^{18} \text{mM} \upmu\text{m}^3 / \text{mol}}{vol_{cyto} N_A} \iint\limits_{\Gamma_{OM}} k_{vdac} [VDAC] (T_{IMS} (t_n) - T_{cyto} (t_n))d\Gamma,
\label{eq:ode1}
\end{equation}
where $T_{IMS}$ is the ATP concentration in the IMS, $vol_{cyto}$ is the volume of cytosol immediately surrounding the mitochondrion, $N_A$ is Avogadro's number, and other parameters are summarized in \Cref{sec:tables}.
This updated value of $T_{cyto}$ was then used in solving the PDE system, after which the estimated value was updated for consistency with the implicit Euler time discretization:
\begin{equation}
T_{cyto}(t_n + \tau_n) = T_{cyto} (t_n) + \tau_n \frac{10^{18} \text{mM} \upmu\text{m}^3 / \text{mol}}{vol_{cyto} N_A} \iint\limits_{\Gamma_{OM}} k_{vdac} [VDAC] (T_{IMS} (t_n + \tau_n) - T_{cyto,est}(t_n + \tau_n))d\Gamma.
\label{eq:ode2}
\end{equation}






\section*{Acknowledgements}

We would like to acknowledge the members of the Rangamani Lab for their valuable insights throughout the development of SMART.
We also thank Dr.\@ C\'ecile Daversin-Catty for her guidance on using mixed dimensional features in FEniCS.

E.A.F. was supported by the National Science Foundation under Grant \# EEC-2127509 to the American Society for Engineering Education and funding from the Wu Tsai Human Performance Alliance at UCSD to P.R. 
P.R. was supported by the Air Force Office of Scientific Research (AFOSR, https://www.wpafb.af.mil/afrl/afosr/) Multidisciplinary University Research Initiative (MURI) FA9550-18-1-0051.
J.G.L. was supported by a fellowship from the UCSD Center for Transscale Structural Biology and Biophysics/Virtual Molecular Cell Consortium and C.T.L. was supported by a Hartwell Foundation Postdoctoral Fellowship and a Kavli Institute of Brain and Mind Postdoctoral Fellowship.

M.E.R was supported by the European Research Council (ERC) under the European Union's Horizon 2020 research and innovation programme under
grant agreement 714892 (Waterscales), by the Research Council of Norway under grant \#324239 (EMIx), and by the Foundation Kristian Gerhard Jebsen via the K.~G.~Jebsen Center for Brain Fluid Research, and by the Fulbright Foundation.

Simulation results presented in this paper has benefited from the Experimental Infrastructure for Exploration of Exascale Computing (eX3), which is financially supported by the Research Council of Norway under contract 270053.

\section*{Author contributions statement}

J.G.L. wrote the original code and conducted preliminary simulations.
E.A.F. conducted the simulations shown in this work, wrote the tests, analyzed data, and prepared figures.
J.S.D. and H.F. contributed to software development and conducted numerical and performance testing.
C.T.L. assisted in the original development of SMART and worked on mesh processing.
M.E.R. and P.R. designed and supervised the project and wrote sections of the manuscript.
All authors reviewed the manuscript. 

\section*{Competing interests}

The authors declare no competing interests.

\newpage
\beginsupplement
\appendix

\section{A general framework for multi-domain and multi-species reaction and transport}

\label{app:sec:math}

SMART is designed to represent and solve coupled systems of nonlinear ordinary and partial differential equations, describing reactions and/or transport due to diffusion, convection or drift in multi-domain geometries including subdomains of co-dimension zero or one (volume-surface problems). This section defines the range and scope of SMART by describing its foundational mathematical modelling framework.
SMART uses a model specification framework inspired by the Systems Biology Markup Language\cite{huckaSystemsBiologyMarkup2003}, consisting of compartments, species, reactions, and parameters.
Throughout this mathematical overview, we refer to objects within SMART that correspond to different terms in the equations.

\subsection{Notation}

\begin{itemize}
\item $\mathcal{M}$: index/label set for domains $\Omega^m$ ($m \in \mathcal{M}$). 
\item $\mathcal{Q}$: index/label set for the surfaces $\Gamma^q$ ($q \in \mathcal{Q}$).
\item $\mathcal{I}$, $\mathcal{I}^m$, $\mathcal{I}^q$: index/label sets for the species, the species in $\Omega^m$ and species on $\Gamma^q$, respectively.
\item $\mathcal{K}$, $\mathcal{K}^m$, $\mathcal{K}^q, \mathcal{K}^{mq}, \mathcal{K}^{mqn}$: index/label sets for the reactions, reactions in $\Omega^m$, reactions on $\Gamma^q$, volume-surface reactions between species in $\Omega^m$ and $\Gamma^q$, and volume-surface-volume reactions between species in $\Omega^m$, $\Gamma^q$, and $\Omega^n$.
\end{itemize}

\subsection{Multi-domain and multi-surface geometry representation}

We consider an open topologically $D$-dimensional manifold $\Omega \subset \R^d$ for $D \leqslant d = 1, 2, 3$, and assume that $\Omega$ is partitioned into $|\mathcal{M}|$ open and disjoint \emph{domains} $\Omega^m \subset \R^d$:
\begin{equation}
  \Omega = \bigcup_{m \in \mathcal{M}} \Omega^m ,    
\end{equation}
each with (internal or external) boundary $\partial \Omega^m$, and boundary $\partial \Omega$. 
In the SMART framework, each domain $\Omega^m$ is referred to as a volume compartment.
The \emph{boundary normal} $\mathbf{n}^m$ on $\partial \Omega^m$ is the outward pointing normal vector field to the boundary. The interface $\Gamma^{mn}$ between domains $\Omega^m$ and $\Omega^n$ is defined as the intersection of the closure of the domains:
\begin{equation}
  \Gamma^{mn} = \overline{\Omega^m} \cap \overline{\Omega^n},
\end{equation}
for $m, n \in \mathcal{M}$, and may or may not be empty. 
Each interface $\Gamma^{mn}$ may be further partitioned into one or more \emph{surfaces}:
\begin{equation}
    \Gamma^{mn} = \bigcup_{q \in \mathcal{Q}^{mn}} \Gamma^{q} ,
\end{equation}
where each $\Gamma^{q}$ is referred to as a surface compartment within SMART.
We denote the total set of surfaces by $\mathcal{Q} = \cup_{m, n \in \mathcal{M}} \cup \mathcal{Q}_0$ where $\mathcal{Q}_0$ also includes (sub)surfaces on the boundary of $\Omega$. 
For each surface, we define the surface-to-neighbors map from the surface index $q$ to its neighboring domain indices $m, n$: $q \mapsto m, n$.
If the surface is a real boundary surface, the map returns the single neighbor $q \mapsto m, $.

Moreover, let time $t \in [0, T]$ for $T > 0$. 

\subsection{Coupled multi-domain reaction-transport equations}
\label{app:sec:equations}

Different physical processes may occur on domains or on surfaces. 
For instance, in the context of computational neuroscience at the cellular level, different species and processes dominate in the dendrites and axons, and on the plasma membrane versus at post-synaptic densities. 
In general, SMART is designed to model different species coexisting, interacting and moving within a (sub)domain or (sub)surface, between (sub)domains and across (sub)surfaces.
Within the SMART framework, information about each species is stored within an object, and the equations describing their interactions are stored within reaction objects.
Any other quantities involved in the reaction equations (\textit{e.g.}, reaction rates and binding affinities) are defined separately and stored within parameter objects.

\subsubsection*{Domain species}
We define and denote the set of species coexisting in $\Omega^m$ by $\mathcal{I}^m$ for each (sub)domain $\Omega^m$.
For each $m$, the species $i \in \mathcal{I}^m$ with concentrations $u_i^m = u_i^m(x, t)$ for $x \in \Omega^m$ and $t \in (0, T]$ satisfy reaction-transport equations of the form
\begin{equation}
  \partial_t u_i^m + \mathcal{T}_i^m(u_i^m) - f_i^m(u^m) = 0 \qquad \text{ in } \Omega^m, 
  \label{eq:domain}
\end{equation}
where $\partial_t$ is the time-derivative, $\mathcal{T}_i^m$ defines transport terms, and $f_i^m$ are volume reactions i.e.~reactions within the domain, typically given by non-linear and non-trivial relations involving one or more of the species $u^m = \{u_i^m\}_{i \in \mathcal{I}^m}$. 
In the case of transport by diffusion alone,
\begin{equation}
    \mathcal{T}_i^m (u) = - \nabla \cdot ( D_i^m \nabla (u) ) ,
\end{equation}
where $\nabla \cdot$ is the spatial divergence operator, $\nabla$ is the spatial gradient, and $D_i^m$ is the diffusion coefficient of species $i$ in domain $\Omega^m$ which may be heterogeneous i.e., spatially varying and/or anisotropic i.e., tensor-valued. 
Remaining relative to $\Omega^m$, on (all) surfaces $\Gamma^{q} \subseteq \Gamma^{mn}$, we assume that the flux of species $i$ is governed by a relation of the form:
\begin{equation}
    D_i \nabla u_i^m \cdot n^m - R_i^{q} (u^m, u^n, v^q) = 0 \qquad \text{ on } \Gamma^{q} ,
    \label{eq:robin}
\end{equation}
where $R_i^q$ defines surface fluxes, and the surface concentrations $v^q$ are defined below. 

\subsubsection*{Surface species}
The surface concentrations $v^q = \{ v_j^q \}_{j \in \mathcal{I}_q}$, entering in \eqref{eq:robin} above, are defined on $\Gamma^q \subseteq \Gamma^{mn}$, either as prescribed fields or via surface equations as follows: find $v^q = v^q(x, t)$ for $x \in \Gamma^q$, $t \in (0, T]$ such that
\begin{equation}
  \partial_t v_j^q + \mathcal{T}_i^q(v_i^q ) - g_j^q ( u^m, u^n, v^q ) = 0 \qquad \text{ on } \Gamma^{q} ,
  \label{eq:surface}
\end{equation}
where $g_j^q$ are surface reactions for each species $j \in \mathcal{I}^q$. In the case of transport by surface diffusion, with surface diffusion coefficient $D_j^q$ for species $j$, surface gradient $\nabla_S$ and surface divergence $\nabla_S \cdot$, we have:
\begin{equation}
    \mathcal{T}_j^q( v ) = - \nabla_S \cdot (D_j^q \nabla_S v ) .
\end{equation}

\subsubsection*{Boundary conditions}

For any (sub)boundary $\Gamma^q \subset \partial \Omega^{m}$ such that $\Gamma^q \subseteq \partial \Omega$, the following boundary conditions are prescribed:
\begin{equation} \label{eq:bc}
    D_i^m \nabla u_i^m \cdot n^m - R_i^{q} (u^m, v^q) = 0 ,
\end{equation}
which can be seen as a special case of \Cref{eq:robin} in which there is no adjacent volume compartment.
For any species $v^q$ on any surface $\Gamma^q$ with non-empty boundary, zero flux boundary conditions are prescribed:
\begin{equation}
    D_j^q \nabla_S v_j^q \cdot n^q = 0 ,
\end{equation}
where $n^q$ is the normal to the boundary line.

\subsubsection*{Initial conditions}
Initial conditions are required for any $u_i^m$ for $m \in \mathcal{M}$, $i \in \mathcal{I}^m$, and $v_i^q$ for $q \in \mathcal{Q}$, $i \in \mathcal{I}^q$.

\subsection{SMART model generation from reaction specifications}
\label{sec:appendix:reactions}

Here, we outline the conventions used internally by SMART to convert reaction specifications into associated terms in PDEs and boundary conditions.
In SMART, each reaction has several ``flux" objects associated with it - one for every reactant and every product.
The manner in which these fluxes are computed depends on the type of reaction, generally classified as ``volume", ``surface", or ``volume-surface" (\Cref{fig:workflow}).

Considering the set of all expressions associated with volume reactions within compartment $\Omega^m$, $\mathcal{F}^m = \{\mathcal{F}^m_k\}_{k\in\mathcal{K}^m}$, the flux associated with a reactant or product species $i$ for reaction $k$ is given by:
\begin{equation}
    f_{i,k}^m = \mathcal{S}_{i,k}^m \mathcal{F}^m_k ,
\end{equation}
where $\mathcal{S}_{i,k}^m$ is the stoichiometric coefficient, which is positive for products and negative for reactants and whose magnitude matches the number of molecules generated or consumed.

Surface reactions are handled analogously.
The set of all surface reactions within compartment $\Gamma^q$ is $\mathcal{G}^q = \{\mathcal{G}^q_k\}_{k\in\mathcal{K}^q}$, and the flux associated with a reactant or product species $j$ for reaction $k$ is given by:
\begin{equation}
    g_{j,k}^q = \mathcal{S}_{j,k}^q \mathcal{G}^q_k ,
\end{equation}
where $\mathcal{S}_{i,k}^m$ is the stoichiometric coefficient as before.

Volume-surface reactions generally contribute to both PDEs and boundary conditions as follows.
Considering the set of all expressions associated with volume-surface reaction $k$ between compartments $\Omega^m$ and $\Gamma^q$, $\mathcal{R}^{mq} = \{\mathcal{R}^{mq}_k\}_{k\in\mathcal{K}^{mq}}$, the flux associated with a reactant or product volume species $i$ for reaction $k$ is given by:
\begin{equation}
    R_{i,k}^{mq} = \alpha_{i,k}^{mq} \mathcal{S}_{i,k}^{mq} \mathcal{R}^{mq}_k \label{eq:vol-surf1},
\end{equation}
where $\mathcal{S}_{i,k}^m$ is the stoichiometric coefficient and $\alpha_{i,k}^{mq}$ is a scaling factor converting molecular flux (molecules per unit surface per unit time) into the dimensionally equivalent units of volume concentration multiplied by length per unit time.
For any surface species $j$ that acts as a product or reactant, such reactions have an associated flux:
\begin{equation}
    g_{j,k}^{mq} = \mathcal{S}_{j,k}^{mq} \mathcal{R}^{mq}_k \label{eq:vol-surf2},
\end{equation}
where $\mathcal{S}_{j,k}^{mq}$ is the stoichiometric coefficient.

The remaining case of a surface-mediated reaction between two adjacent volume compartments (``volume-surface-volume" reaction) is a natural extension of the above ``volume-surface" case.
The set of expressions for all such reactions between species in $\Omega^m$, $\Gamma^q$, and $\Omega^n$ is $\mathcal{R}^{mqn} = \{\mathcal{R}^{mqn}_k\}_{k\in\mathcal{K}^{mqn}}$ and the contribution to reactants and products are given by replacing all cases of $mq$ with $mqn$ in \Cref{eq:vol-surf1,eq:vol-surf2}.

All of these fluxes are included in \Cref{eq:domain,eq:robin,eq:surface,eq:bc} by summing the associated contributions for a given species.
In particular, the total reaction term $f^m_i$ in \Cref{eq:domain} is given by summing the contribution from all volume reactions involving species $i$, $\mathcal{K}_i^m$:
\begin{equation}
f^m_i = \sum_{k\in\mathcal{K}_i^m} {f_{i,k}^m}.
\end{equation}

The total reaction term $R^q_i$ in \Cref{eq:robin} or \Cref{eq:bc} is given by summing the contribution from all volume-surface or volume-surface-volume reactions involving species $i$ ($\mathcal{K}_i^{mq}$, $\mathcal{K}_i^{mqn}$):
\begin{equation}
R^q_i = \sum_{k\in\mathcal{K}_i^{mq}} {R_{i,k}^{mq}} + \sum_{k\in\mathcal{K}_i^{mqn}} {R_{i,k}^{mqn}},
\end{equation}
where the second term is equal to zero when applied to \Cref{eq:bc}, where there is no adjacent volume compartment.

Finally, the total reaction term $g^q_j$ in \Cref{eq:surface} is given by summing the contribution from all surface reactions involving species $j$, $\mathcal{K}_j^q$ as well as any volume-surface or volume-surface-volume reactions involving species $j$ ($\mathcal{K}_j^{mq}$, $\mathcal{K}_j^{mqn}$):
\begin{equation}
g^q_j = \sum_{k\in\mathcal{K}_j^q} {g_{j,k}^q} + \sum_{k\in\mathcal{K}_j^{mq}} {g_{j,k}^{mq}} + \sum_{k\in\mathcal{K}_j^{mqn}} {g_{j,k}^{mqn}}.
\end{equation}

\subsubsection{Example of assembly from reactions}

\begin{figure} [htbp!]
\includegraphics[width=6.5in,keepaspectratio]{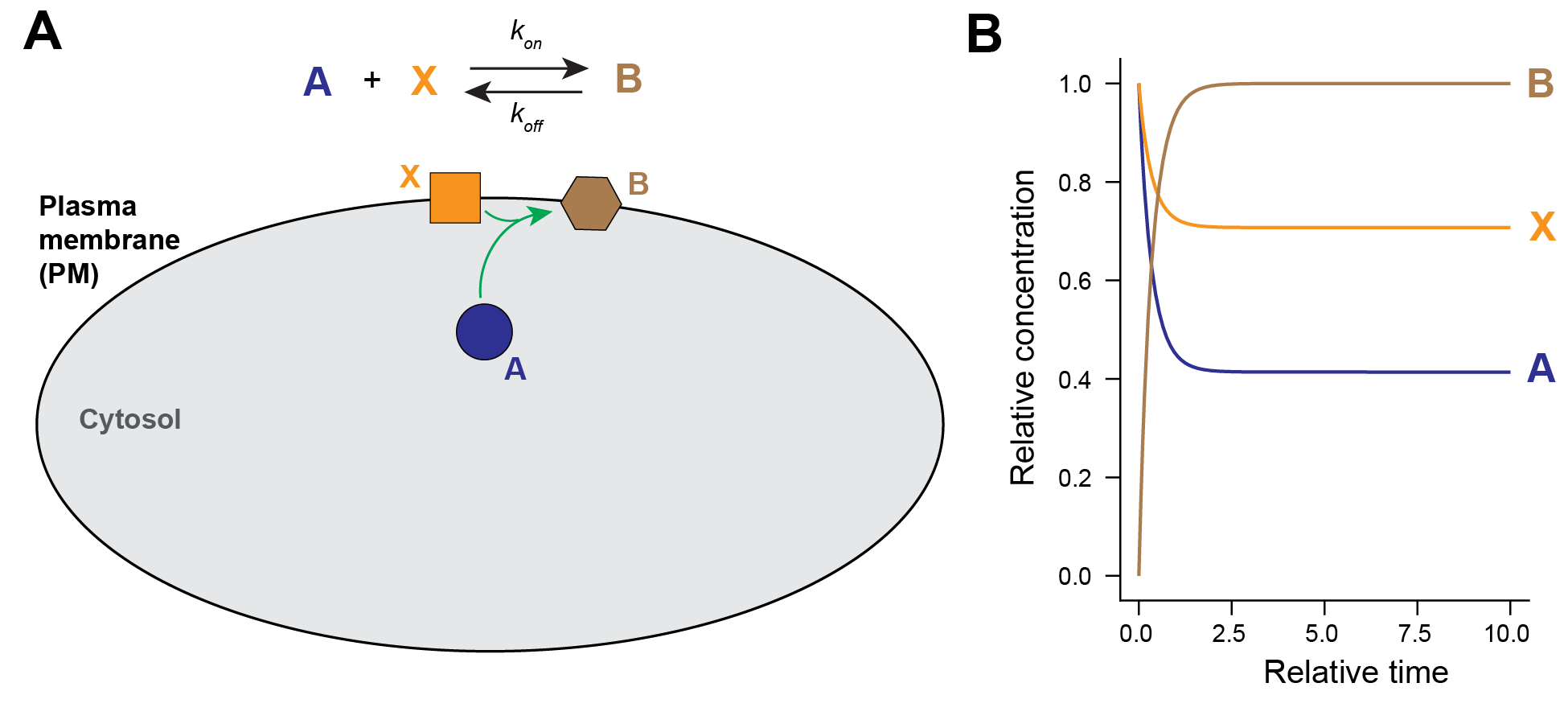}
    \centering
      \caption{{\bf Simple example of a bulk-surface reaction in a cell.}
      A) Schematic summarizing the reaction of volume species A with surface species X to form surface species B.
      B) Normalized concentration over time for each of the three species.}
\label{fig:axb}
\end{figure}

As a simple case of model generation in SMART, we consider the bulk-surface reaction from Rangamani et al.\@ \cite{rangamaniDecodingInformationCell2013a}, which is also included as Example 2 in the SMART documentation.
In brief, this model involves a volume species $A$ in the cytosolic domain $\Omega_{cyto}$ that can bind to the surface species $X$ on the plasma membrane domain $\Gamma_{PM}$ to form surface species $B$ (\Cref{fig:axb}A).
Treating $A$ and $X$ as reactants and $B$ as a product, the rate of this volume-surface reaction is defined according to mass conservation:
\begin{equation}
    \mathcal{R}_1^{cyto-PM} = k_{on} u_A v_X - k_{off} v_B
\end{equation}
From \Cref{eq:vol-surf1}, the boundary flux for species $A$ is then:
\begin{equation}
    R_{A,1}^{cyto-PM} = -\alpha_{A,1}^{cyto-PM} \mathcal{R}_1^{cyto-PM},
\end{equation} 
where $\alpha_{A,1}^{cyto-PM}$ depends on the chosen units.
For instance, in a 3D/2D system with volume concentration units of \uM~ and surface concentration units of molecules/\um$^2$, $\alpha_{A,1}^{cyto-PM} = \frac{1\times10^{21}}{N_A}$ \unit{\micro\molar\per\micro\meter\cubed}, where $N_A$ is Avogadro's number.
From \Cref{eq:vol-surf2}, the surface reaction rate for $X$ is:
\begin{equation}
    g_{X,1}^{cyto-PM} = -\mathcal{R}_1^{cyto-PM},
\end{equation}
and the surface reaction rate for $B$ is:
\begin{equation}
    g_{B,1}^{cyto-PM} = \mathcal{R}_1^{cyto-PM}.
\end{equation}

In terms of these expressions, the system of PDEs is:
\begin{align}
\frac{\partial{u_A}}{\partial{t}} &= D_A \nabla ^2 u_A \quad \text{in} \; \Omega_{cyto}\\
D_A (\textbf{n} \cdot \nabla u_A)  &= R_{A,1}^{cyto-PM} \quad \text{on} \; \Gamma_{PM} \\
\frac{\partial{v_X}}{\partial{t}} &= D_X \nabla ^2 v_X + g_{X,1}^{cyto-PM} \quad \text{on} \; \Gamma_{PM}\\
\frac{\partial{v_B}}{\partial{t}} &= D_B \nabla ^2 v_B + g_{B,1}^{cyto-PM} \quad \text{on} \; \Gamma_{PM}.
\end{align}

\section{Numerical approximation and computational strategy}
\label{sec:numerics}

SMART solves the multi-domain reaction-transport equations outlined in Section \ref{app:sec:math} via finite difference discretizations in time and a finite element discretization in space, using a \emph{monolithic} approach to address the system couplings. 
We describe this monolithic approach here, considering linear diffusion-only transport relations for concreteness and to illustrate linear-nonlinear strategies. 

\subsection{Monolithic solution of the multi-domain reaction-diffusion equations}

Formally, we consider the Sobolev spaces $H^1(\Omega^m)$, $m \in \mathcal{M}$ of square-integrable functions on $\Omega^m$ with square-integrable weak derivatives, and analogously for $H^1(\Gamma^q)$, $q \in Q$, as well as the vector function spaces $H^1(\Omega^m, \R^d) = H^1(\Omega)^d$. 
For solving the multi-domain reaction-diffusion equations, we introduce two product spaces $U$ and $V$ consisting of bulk fields and surface fields, respectively:
\begin{equation}
    U = \bigotimes_{m \in \mathcal{M}} H^1(\Omega^m; \R^{|\mathcal{I}^m|}), \quad 
    V = \bigotimes_{q \in \mathcal{Q}} H^1(\Gamma^q; \R^{|\mathcal{I}^q|}) .
\end{equation}
To represent the solution fields $u, v$ comprised of the separate bulk and surface components for the different species, we label
\begin{equation}
    u 
    = \{ u^m \}_{m \in \mathcal{M}} 
    = \{ \{u_i^m \}_{i \in \mathcal{I}^m} \}_{m \in \mathcal{M}}, 
    \quad 
    v 
    = \{ v^q \}_{q \in \mathcal{Q}} 
    = \{ \{ v_i^q \}_{i \in \mathcal{I}^q} \}_{q \in \mathcal{Q}} .
\end{equation}
By standard techniques (integrating by test functions and integrating by parts), we rephrase the multi-domain reaction-diffusion equations in variational form, resulting in the following coupled variational problem: find $u \in U$ and $v \in V$ such that for all $\phi \in U$ and $\psi \in V$:
\begin{equation}
    F(u, v; \phi) + G(u, v; \psi) = 0,
\end{equation}
where both forms $F$ and $G$ are composed of sums over domains or surfaces and species:
\begin{equation}
    F(u, v; \phi) = \sum_{m \in \mathcal{M}} \sum_{i \in \mathcal{I}^m} F_i^m(u, v; \phi_i^m), \qquad
    G(u, v; \psi) = \sum_{q \in \mathcal{Q}} \sum_{i \in \mathcal{I}^q} G_i^q(u, v; \psi_i^q) .
\end{equation}
Furthermore, with the $L^2(O)$-inner product over any given domain $O \subset \Omega$ defined as
\begin{equation*}
    \inner{a}{b}_{O} = \int_{O} a \cdot b \, \mathrm{d}x ,
\end{equation*}
we have defined
\begin{equation}
    F_i^m(u, v, \phi_i^m) 
    = \inner{\partial_t u_i^m}{\phi_i^m}_{\Omega^m} 
    + \inner{D_i^m \nabla u_i^m}{\nabla \phi_i^m}_{\Omega^m}
    - \inner{f_i^m(u^m)}{\phi_i^m}_{\Omega^m}
    - \sum_{q \in \mathcal{Q}^{mn}} \inner{R_i^q(u^m, u^n, v^q)}{\phi_i^m}_{\Gamma^q} 
\end{equation}
and, for any $q \in \mathcal{Q}^{mn}$ for given $\Omega^m$ and $\Omega^n$ interfacing $\Omega^m$ via $\Gamma^q$, finally:
\begin{equation}
    G_i^q(u, v, \psi_i^q) 
    = \inner{\partial_t v_i^q}{\psi_i^q}_{\Gamma^q}
     + \inner{D_i^q \nabla_S v_i^q}{\nabla_S \psi_i^q}_{\Gamma^q}
     - \inner{g_i^q(u^m, u^n, v^q)}{\psi_i^q}_{\Gamma^q} .
\end{equation}

\subsection{Discretization in time and space}
SMART discretizes in time via an implicit (first-order) Euler scheme with time steps $0 = t_0 < t_1 < \ldots < t_{n-1} < t_n = T$ with timestep $\tau_n = t_n - t_{n-1}$. 
To discretize in space, we employ a conforming finite element mesh $\mathcal{T}$ defined relative to the parcellation of $\Omega$ into domains and surfaces, such that $\mathcal{T}^m \subseteq \mathcal{T}$ defines a submesh of $\Omega^m$ for $m \in \mathcal{M}$, and $\mathcal{G}^q$ defines a (co-dimension  one) mesh of $\Gamma^q$ for $q \in \mathcal{Q}$. 
We define the finite element spaces of continuous piecewise linears $\mathcal{P}_1$ defined relative to each (sub)mesh, and define the product spaces
\begin{equation}
    U_h = \bigotimes_{m \in \mathcal{M}} \mathcal{P}_1^{|\mathcal{I}_m|}(\mathcal{T}^m), \qquad
    V_h = \bigotimes_{q \in \mathcal{Q}} \mathcal{P}_1^{|\mathcal{I}_q|}(\mathcal{G}^q) .
\end{equation}
For each time step $t_i$, given discrete solutions $u_h^{-}$ and $v_h^{-}$ at the previous time $t_{i-1}$, we then solve the nonlinear, coupled time-discrete problem: find $u_h \in U_h$ and $v_h \in V_h$
\begin{equation}
    \label{eq:nonlinear}
    H(u_h, v_h) = F_{\tau_n}(u_h, v_h, \phi) + G_{\tau_n}(u_h, v_h, \psi) = 0
\end{equation}
for all $\phi \in U_h$, $\psi \in V_h$, where $F_{\tau_n}$ and $G_{\tau_n}$ are defined by the forms $F$ and $G$ after implicit Euler time-discretization and depend on the given $u_h^{-}$ and $v_h^{-}$. 
\subsection{Nonlinear solution algorithms}

By default, the monolithic nonlinear discrete system \eqref{eq:nonlinear} is solved by Newton-Raphson iteration with a symbolically derived discrete Jacobian. 

\section{Software implementation}
\label{sec:software}

The SMART abstractions and algorithms are implemented via the open source and generally available FEniCS finite element software package (2022-version) \cite{alnaesFEniCSProjectVersion2015}. FEniCS supports high-level specification of variational forms via the Unified Form Language (UFL) \cite{alnaesUnifiedFormLanguage2014}, symbolic differentiation of variational forms e.g.\@ for derivation for Jacobians, automated assembly of general and nonlinear forms over finite element meshes, and high-performance linear algebra and nonlinear solvers via e.g.\@ PETSc \cite{balayPETScPortableExtensible1998}.
Below, we describe the abstractions used by SMART to translate user-level specifications into the variational form of the reaction-transport equations summarized above.

\subsection{SMART model components}
A model in SMART consists of a collection of four classes of objects - compartments, species, reactions, and parameters.

A compartment is initialized by a name, dimensionality, associated length units, and marker value in the parent mesh.
Each compartment is linked to a parent mesh by marker values stored in discrete functions defined over mesh entities, $M_{facet}$ defined over all mesh facets (faces for 2D parent mesh, edges for 3D parent mesh) and $M_{cell}$ defined over all mesh cells.
For instance, in a 3D/2D model, $M_{cell}$ stores a marker value associated with each tetrahedron in the geometry; if the cytosol is linked to marker value ``1", then the cytosol is comprised of all tetrahedra with value ``1" in $M_{cell}$.
All information about the compartment topology is implicit in the mesh functions, but evaluation can be sped up by users explicitly specifying which compartments are nonadjacent.

A species is initialized by a name, initial condition, units, diffusion coefficient, diffusion coefficient units, and compartment.
Importantly, each species is associated with only one compartment; a single molecule that moves between compartments will be specified by two species (\textit{e.g.}, cytosolic calcium and ER calcium).
The initial condition can be specified as a constant value or as a string giving a spatially dependent expression.
The unknown symbols in an expression can include spatial coordinates $x, y, z$ or, in the case of surface species, the curvature $curv$.

A reaction is initialized by a name, a list of reactants, a list of products, equations for the forward and reverse reaction rates, and Python dictionaries mapping strings from the equations to parameters and species initialized separately.
The reactant and product lists also specify the stoichiometry of each reaction, \textit{e.g.}~, if a product is listed twice, it has a stoichiometric coefficient of 2.
Reactions are integrated into the variational forms of each PDE as detailed above, and SMART carries out internal checks to ensure consistency in units using the Python package pint \cite{greccoPintPhysicalQuantities}.
If no forward or reverse reaction strings are specified, SMART defaults to assuming mass action kinetics, with an on-rate parameter \textit{on} and an off-rate parameter \textit{off}.
For non-mass-action kinetics, SMART supports a range of non-algebraic functions, including trigonometric functions, exponential and logarithmic functions, and absolute value and sign functions.
Furthermore, reactions can be defined as curvature sensitive by including $curv$ in the reaction expression.

Finally, a parameter is initialized by a name, value, and units.
The value of a parameter can be either a scalar or a string specifying a time and/or space-dependent expression, where the free symbols should only include $x, y, z, t$.
A time-dependent parameter can also be loaded from a text file as shown in Example 5 of the SMART documentation.
If a time-dependent parameter with no dependence on species concentrations is used to define a flux, the user may optionally set \verb|pre_integration| to \verb|True|.
In this case, SMART uses a pre-integrated expression or a numerical approximation to update parameter values while ensuring consistent mass transfer; that is, given a time-dependent flux $f(t)$, flux values are updated as follows:
\begin{equation}
    f (t+\tau_n) = \tau_n^{-1} \int_t^{t+\tau_n}\, f(T) dT
\end{equation}
When using this updating scheme together with implicit Euler time integration, the amount of mass transferred is independent of the time step.

\subsection{Solver specifications and adaptive time-stepping}

SMART uses PETSc4py for all linear and nonlinear solves \cite{dalcinParallelDistributedComputing2011a}.
Linear systems are solved iteratively using Krylov solvers with field-split biconjugate gradient preconditioning.
The nonlinear system is solved using the Newton line search SNES solver within PETSc.
Default relative tolerances in each case are set to $1\times10^{-5}$, but these and other solver parameters can be readily adjusted as illustrated in Example 6 of the SMART documentation.

By default, SMART takes uniform time-steps matching the initial specification.
However, there is an optional function for adaptive time-stepping using the number of Newton iterations required to converge at the previous step.
The rules were set \textit{ad hoc} and are summarized in \Cref{table:time-step}.
The reported ``time-step adjustment factor" $\zeta$ is the multiplicative factor used to alter the time-step at the end of a given time step, \textit{i.e.}, the new time-step $\tau_{new}$ is given by a in terms of the previous time-step $\tau_{old}$ as $\tau_{new} = \zeta \tau_{old}$. 
\\

\begin{table}[!ht]
\caption{Adaptive time stepping rules in SMART.}
\centering
\begin{tabular}{ll}
\toprule
Number of  & Time-step \\
Nonlinear Iterations & Adjustment Factor \\
\midrule
0-1 & 1.1 \\
2-4 & 1.05 \\
5-10 & 1.0 \\
11-20 & 0.8 \\
>20 & 0.5 \\
\end{tabular}
\label{table:time-step}
\end{table}

\subsection{Axisymmetric models}
\label{sec:axisymm}

SMART offers the ability to assume axisymmetry given a 2D mesh.
Given a 2D mesh in the $r-z$ plane, the symmetry axis is assumed to be $r = 0$.
When this feature is turned on in SMART, the variational forms are adjusted accordingly, with the only necessary change being the multiplication by $r$ in each integral (factors of $2\pi$ cancel throughout):

\begin{align}
    F_i^m(u, v, \phi_i^m) 
    &= \inner{r\partial_t u_i^m}{\phi_i^m}_{\Omega^m} 
    + \inner{rD_i^m \nabla u_i^m}{\nabla \phi_i^m}_{\Omega^m}
    - \inner{rf_i^m(u^m)}{\phi_i^m}_{\Omega^m}
    - \sum_{q \in \mathcal{Q}^{mn}} \inner{rR_i^q(u^m, u^n, v^q)}{\phi_i^m}_{\Gamma^q} \\
    G_i^q(u, v, \psi_i^q) 
    &= \inner{r\partial_t v^q}{\psi_i^q}_{\Gamma^q}
     + \inner{rD_i^q \nabla_S v_i^q}{\nabla_S \psi_i^q}_{\Gamma^q}
     - \inner{rg_i^q(u^m, u^n, v^q)}{\psi_i^q}_{\Gamma^q} .
\end{align}

In this case, the compartments must be initialized as a 2D/1D system, as shown in the mechanotransduction code and Example 3 with axisymmetry in the SMART repository.

\section{Numerical testing}

\subsection{Numerical testing of protein phosphorylation model}
\label{app:sec:phos}

As an idealized test case for SMART, we considered the simple system described by Meyers et al.\@, in which a single protein is phosphorylated at the plasma membrane with rate $k_{kin}$ and dephosphorylated through the cytosol with rate $k_{p}$ \cite{meyersPotentialControlSignaling2006}.
The SMART specifications are summarized in \Cref{tab:phos-compartments,tab:phos-species,tab:phos-reactions,tab:phos-params}.

\begin{longtable}{lllllll}
\caption{Compartments in phosphorylation model. Mesh statistics are given for the coarsest mesh ($h=3.0$)} \\
\toprule
 & Dimensionality & Species & Vertices & Cells & Marker value & Size \\
\midrule
\endfirsthead
\toprule
 & Dimensionality & Species & Vertices & Cells & Marker value & Size \\
\midrule
\endhead
\midrule
\multicolumn{7}{r}{Continued on next page} \\
\midrule
\endfoot
\bottomrule
\endlastfoot
PM & 2 & 0 & 242 & 400 & 10 & $\SI[]{8.000e+02}{\micro\meter\squared}$ \\
Cyto & 3 & 1 & 363 & 1200 & 1 & $\SI[]{8.000e+02}{\micro\meter\cubed}$
\label{tab:phos-compartments}
\end{longtable}

\begin{longtable}{llll}
\caption{Species in phosphorylation model.}\\
\toprule
 & Compartment & $D (\unit{\micro\meter\squared\per\second})$ & Initial condition \\
\midrule
\endfirsthead
\toprule
 & Compartment & $D (\unit{\micro\meter\squared\per\second})$ & Initial condition \\
\midrule
\endhead
\midrule
\multicolumn{4}{r}{Continued on next page} \\
\midrule
\endfoot
\bottomrule
\endlastfoot
Aphos & Cyto & 10 & $\SI[]{1.000e-01}{\micro\molar}$ 
\label{tab:phos-species}
\end{longtable}

\begin{longtable}{lllll}
\caption{Reactions in phosphorylation model.}\\
\toprule
 & Reactants & Products & Equation & Type \\
\midrule
\endfirsthead
\toprule
 & Reactants & Products & Equation & Type \\
\midrule
\endhead
\midrule
\multicolumn{5}{r}{Continued on next page} \\
\midrule
\endfoot
\bottomrule
\endlastfoot
r1 & [] & [`Aphos'] & $(VolSA) k_{kin} (A_{tot} - A_{phos})$ & volume\_surface \\
r2 & [`Aphos'] & [] & $k_{dephos} A_{phos}$ & volume 
\label{tab:phos-reactions}
\end{longtable}

\begin{longtable}{lll}
\caption{Parameters in phosphorylation model.}\\
\toprule
 & Value/Equation & Description \\
\midrule
\endfirsthead
\toprule
 & Value/Equation & Description \\
\midrule
\endhead
\midrule
\multicolumn{3}{r}{Continued on next page} \\
\midrule
\endfoot
\bottomrule
\endlastfoot
A\_tot & $\SI[]{1.000e+00}{\micro\molar}$ & Total cytosolic A \\
k\_kin & $\SI[]{5.000e+01}{\per\second}$ &  Rate constant for A phosphorylation \\
VolSA & $\SI[]{5.000e-02}{\micro\meter}$ &  Cytosolic volume to PM surface area ratio \\
k\_p & $\SI[]{1.000e+01}{\per\second}$ &  Rate constant for A dephosphorylation
\label{tab:phos-params}
\end{longtable}

Given the two reactions summarized above, we can define the following flux terms in accordance with \Cref{sec:appendix:reactions}:

\begin{align}
    R_{A_{phos},r1}^{cyto-PM} &= \alpha_{A_{phos},r1}^{cyto-PM} [VolSA] k_{kin} (A_{tot} - u_{A_{phos}}),\\
    f_{A_{phos},r2}^{cyto} &= - k_{p} A_{phos}.
\end{align}

Accordingly the governing equation and boundary conditions are:

\begin{align}
    \frac{\partial{u_{A_{phos}}}}{\partial{t}} &= D_{A_{phos}} \nabla ^2 u_{A_{phos}} + f_{A_{phos},r2}^{cyto} \quad \text{in} \; \Omega_{Cyto},\\
    \quad D_{A_{phos}}  (\textbf{n} \cdot \nabla u_{A_{phos}}) &= R_{A_{phos},r1}^{cyto-PM} \quad \text{on} \; \Gamma_{PM}.
\end{align}

Due to the simplicity of this problem, a closed form solution exists for the steady state concentration in 1D.
This 1D solution approximates the 3D case in which a cell is modeled as a very thin sheet, where the solution is used along the thin cell dimension in the $z$ direction.
A cell of thickness $\Delta z$ with one region of membrane at $z = 0$ and the other at $z = \Delta z$ is then predicted to have the following concentration profile at steady state:

\begin{align}
    u_{A_{phos},SS} &= A_{tot} C_1 \left(\exp\left(\frac{\Delta z - z}{z_0}\right) + \exp\left(\frac{z}{z_0}\right) \right), \\
    \text{where} \qquad C_1 &= \frac{k_{phos} z_0}{D_{A_{phos}} \left(\exp\left(\frac{\Delta z}{z_0}\right) - 1\right) + k_{phos} z_0 \left(1 + \exp\left(\frac{\Delta z}{z_0}\right)\right)} \\
    \text{and} \qquad z_0 &= \sqrt{\frac{D_{A_{phos}}}{k_{dephos}}}.
\end{align}

We ran these simulations for a thin slab where the thickness is ten times smaller than the length in the $x$ and $y$ directions, testing the effects of time-step refinement, mesh refinement, and changes in the diffusion coefficient.
As an error metric, we computed the $\mathcal{L}_2$ norm of the difference between the analytical steady state solution and our numerical solution at $t = 1$ s ($u_{A_{phos},num})$:

\begin{equation}
    \left\lvert\left\lvert u_{A_{phos},SS} - u_{A_{phos},num} \right\rvert\right\rvert_{\mathcal{L}2} = \sqrt{~\iiint\limits_{\Omega_{Cyto}} {(u_{A_{phos},SS} - u_{A_{phos},num})^2 d\Omega}}
\end{equation}

\begin{table}[h!]
\centering
\caption{$\mathcal{L}_2$ error (in \uM) between computed and analytical solution for different time steps and mesh sizes for $D=10.0~\unit{\micro\meter\squared\per\second}$.}
\label{tab:rates_D10.0}
\begin{tabular}{l|cccc}
\toprule
  & $h=0.375$ \um & $h=0.75$ \um & $h=1.5$ \um & $h=3$ \um \\
\hline
$\tau=0.64$ s & \SI{8.87e-02}{} & \SI{8.58e-02}{} & \SI{1.06e-01}{} & \SI{3.68e-01}{} \\
$\tau=0.32$ s & \SI{1.97e-02}{} & \SI{2.56e-02}{} & \SI{8.77e-02}{} & \SI{3.96e-01}{} \\
$\tau=0.16$ s & \SI{5.58e-03}{} & \SI{2.24e-02}{} & \SI{9.14e-02}{} & \SI{4.04e-01}{} \\
$\tau=0.08$ s & \SI{5.69e-03}{} & \SI{2.31e-02}{} & \SI{9.21e-02}{} & \SI{4.06e-01}{} \\
$\tau=0.04$ s & \SI{5.78e-03}{} & \SI{2.32e-02}{} & \SI{9.22e-02}{} & \SI{4.06e-01}{} \\
$\tau=0.02$ s & \SI{5.80e-03}{} & \SI{2.32e-02}{} & \SI{9.23e-02}{} & \SI{4.06e-01}{} \\
$\tau=0.01$ s & \SI{5.80e-03}{} & \SI{2.32e-02}{} & \SI{9.23e-02}{} & \SI{4.06e-01}{} \\
\bottomrule
\end{tabular}
\end{table}

\begin{table}[h!]
\centering
\caption{$\mathcal{L}_2$ error (in \uM) between computed and analytical solution for different time steps and mesh sizes for $D=100.0~\unit{\micro\meter\squared\per\second}$.}
\label{tab:rates_D100.0}
\begin{tabular}{l|cccc}
\toprule
  & $h=0.375$ \um & $h=0.75$ \um & $h=1.5$ \um & $h=3$ \um \\
\hline
$\tau=0.64$ s & \SI{9.64e-02}{} & \SI{9.58e-02}{} & \SI{9.35e-02}{} & \SI{8.41e-02}{} \\
$\tau=0.32$ s & \SI{2.09e-02}{} & \SI{2.04e-02}{} & \SI{1.98e-02}{} & \SI{4.27e-02}{} \\
$\tau=0.16$ s & \SI{2.53e-03}{} & \SI{2.93e-03}{} & \SI{9.33e-03}{} & \SI{4.78e-02}{} \\
$\tau=0.08$ s & \SI{5.89e-04}{} & \SI{2.40e-03}{} & \SI{9.86e-03}{} & \SI{4.89e-02}{} \\
$\tau=0.04$ s & \SI{6.06e-04}{} & \SI{2.48e-03}{} & \SI{9.96e-03}{} & \SI{4.90e-02}{} \\
$\tau=0.02$ s & \SI{6.19e-04}{} & \SI{2.49e-03}{} & \SI{9.97e-03}{} & \SI{4.90e-02}{} \\
$\tau=0.01$ s & \SI{6.22e-04}{} & \SI{2.49e-03}{} & \SI{9.98e-03}{} & \SI{4.90e-02}{} \\
\bottomrule
\end{tabular}
\end{table}

\begin{table}[h!]
\centering
\caption{$\mathcal{L}_2$ error (in \uM) between computed and analytical solution for different time steps and mesh sizes for $D=1000.0~\unit{\micro\meter\squared\per\second}$.}
\label{tab:rates_D1000.0}
\begin{tabular}{l|cccc}
\toprule
  & $h=0.375$ \um & $h=0.75$ \um & $h=1.5$ \um & $h=3$ \um \\
\hline
$\tau=0.64$ s & \SI{9.72e-02}{} & \SI{9.71e-02}{} & \SI{9.69e-02}{} & \SI{9.49e-02}{} \\
$\tau=0.32$ s & \SI{2.11e-02}{} & \SI{2.10e-02}{} & \SI{2.08e-02}{} & \SI{1.92e-02}{} \\
$\tau=0.16$ s & \SI{2.65e-03}{} & \SI{2.59e-03}{} & \SI{2.48e-03}{} & \SI{4.40e-03}{} \\
$\tau=0.08$ s & \SI{3.04e-04}{} & \SI{3.27e-04}{} & \SI{9.39e-04}{} & \SI{4.90e-03}{} \\
$\tau=0.04$ s & \SI{6.54e-05}{} & \SI{2.38e-04}{} & \SI{9.91e-04}{} & \SI{5.02e-03}{} \\
$\tau=0.02$ s & \SI{5.95e-05}{} & \SI{2.48e-04}{} & \SI{1.00e-03}{} & \SI{5.04e-03}{} \\
$\tau=0.01$ s & \SI{6.12e-05}{} & \SI{2.50e-04}{} & \SI{1.01e-03}{} & \SI{5.04e-03}{} \\
\bottomrule
\end{tabular}
\end{table}

\FloatBarrier

\subsection{Extra details on the numerical testing of biological test cases}

In our numerical testing of the mechanotransduction and dendritic spine examples, we considered refined versions of the reference mesh.
For the model of mechanotransduction, we tested refined versions of the 2D mesh used to model an axisymmetric spread cell on a circular contact region.
The mesh statistics and total degrees of freedom (DOFs) are given in \Cref{tab:mech-meshes}.

\FloatBarrier

\begin{table} [htbp!]
\centering
\caption{Refined meshes for numerical testing of the mechanotransduction model.}
\begin{tabular}{lrrr}
\toprule
Refinements & Vertices & Triangles & DOFs \\
\midrule
0 & 308 & 548 & 3078 \\
1 & 1163 & 2192 & 11058 \\
2 & 4517 & 8768 & 41772 \\
3 & 17801 & 35072 & 162216 \\
\bottomrule
\end{tabular}
\label{tab:mech-meshes}
\end{table}

\FloatBarrier

\begin{figure} [h!]
\includegraphics[width=6.5in,keepaspectratio]{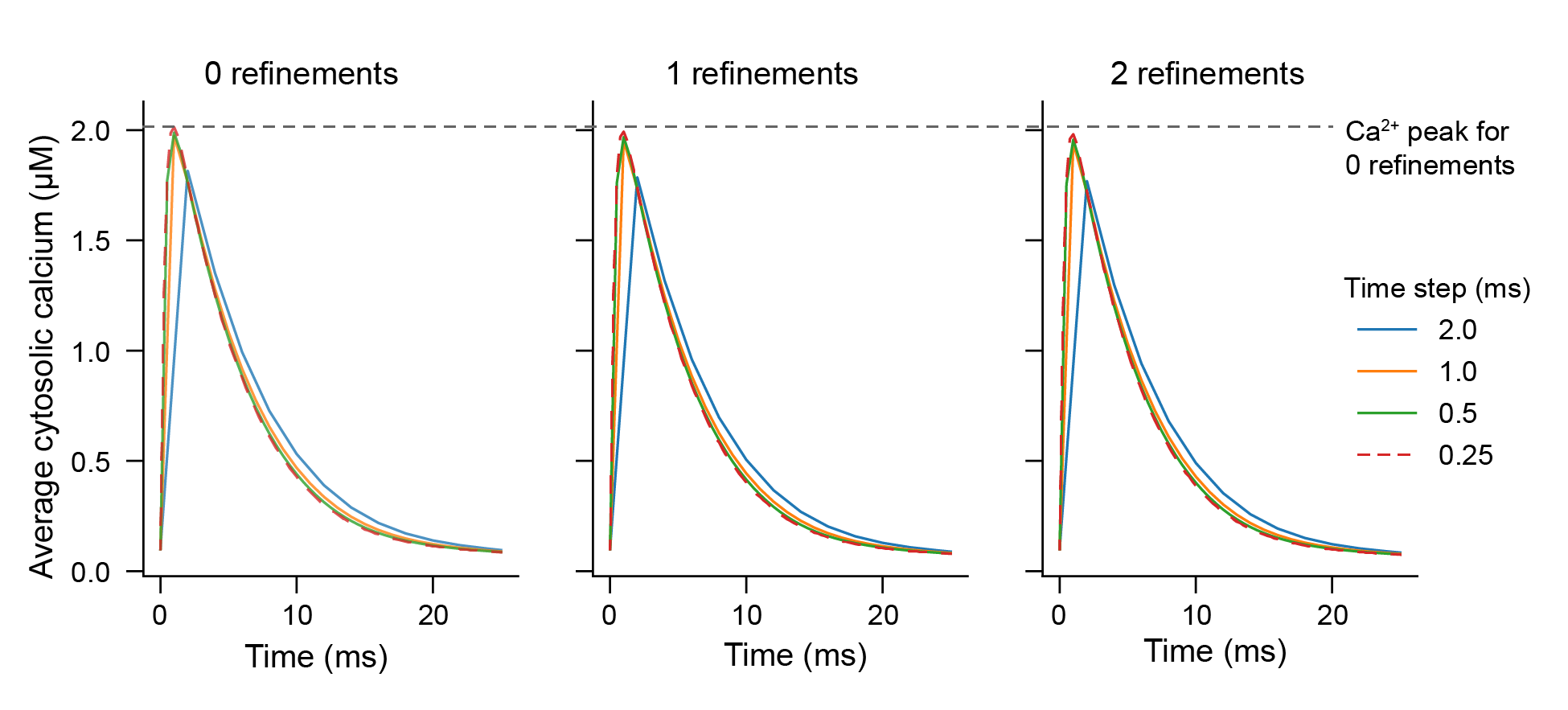}
    \centering
      \caption{{\bf Summary of calcium dynamics in dendritic spine upon mesh refinement and time-step refinement.}
      Curves are plotted for three different mesh refinements and 5 different time steps, as indicated.
      The calcium peak for the coarsest mesh is indicated by the horizontal dashed line throughout, showing the reduction in calcium peak upon mesh refinement.}
\label{fig:suppl-spine-testing}
\end{figure}

In the case of the dendritic spine example, we tested a range of different mesh refinements and time steps, as summarized in \Cref{fig:suppl-spine-testing}.
Overall, we found the effects of time-step refinement were almost identical across three different mesh refinements.
We observe almost identical solutions for time steps 0.001 s or less, justifying our choice of 0.001 s time steps for the results shown in \Cref{fig:calcium}.
At the smallest time step tested, the error due to mesh refinement is very small, as indicated by the minimal differences across mesh refinements.
As shown in \Cref{fig:accuracy}E in the main text, mesh refinement does influence the spatial solution, but this effect appears to be quite small at the starting resolution of the dendritic spine mesh.

The mesh statistics for each refinement are provided in \Cref{tab:spine-refinement}.

\FloatBarrier

\begin{table} [htbp!]
\centering
\caption{Refined meshes for numerical testing of the dendritic spine model.}
\begin{tabular}{lrrr}
\toprule
Refinements & Vertices & Tetrahedra & DOFs \\
\midrule
0 & 18649 & 106979 & 49194 \\
1 & 146024 & 855832 & 323328\\
2 & 1154871 & 6846656 & 2299090 \\
\bottomrule
\end{tabular}
\label{tab:spine-refinement}
\end{table}

\FloatBarrier

\section{Mesh generation for mechanotransduction model}
\label{sec:meshes}

Cell geometries on micropatterned substrates were generated in Gmsh \cite{geuzaineGmsh3DFinite2009}.
As a reference geometry, we used the axisymmetric cell shape previously defined in VCell \cite{scottSpatialModelYAP2021}, where the plasma membrane surface satisfies
\begin{equation}
\frac{\left(1 - \frac{z^4}{2000+z^4}\right) (r^2 + z^2) + 0.4(r^2 + (z+9.72)^2) z^4}{15 + z^4} = 169
\end{equation}
and the nuclear membrane surface satisfies
\begin{equation}
\left(\frac{r}{5.3}\right)^2 + \left(\frac{z-4.8}{2.4}\right)^2 = 1,
\end{equation}
where $r$ and $z$ are cylindrical coordinates.
These implicit boundary representations were constructed by solving for $r$ at each $z$ value using \texttt{solveset} in Sympy \cite{meurerSymPySymbolicComputing2017}.

Shapes with rectangular or star-shaped contact regions were constructed by introducing an angular dependence to the reference shape.
In particular, we scale the $r$ value at the surface as a function of the cylindrical coordinate $\theta$.
Considering an implicitly (as above) or explicitly defined contour as a function of arc length $r(s), z(s)$, the scaled contour at a given value of $\theta$ is $r(s) T(\theta), z(s)$.
We additionally require that this shape conserves cell volume, resulting in the constraint:
\begin{equation}
    \int_0^{2\pi} {T^2(\theta) d\theta} = 2\pi.
\end{equation}
For the circular contact region, $T(\theta) = 1$ and the above constraint is automatically satisfied.

For a rectangular contact region with sides of length $a$ and $b$:
\begin{equation}
    T_{rect}(\theta) = 
    \begin{cases}
    \frac{a}{2\cos{\theta}} & -\arctan(\frac{b}{a}) < \theta \leq \arctan(\frac{b}{a}) \\
    \frac{b}{2\sin{\theta}} & \arctan(\frac{b}{a}) < \theta \leq \pi-\arctan(\frac{b}{a}) \\
    -\frac{a}{2\cos{\theta}} & \pi-\arctan(\frac{b}{a}) < \theta \leq \pi+\arctan(\frac{b}{a}) \\
    -\frac{b}{2\sin{\theta}} & \pi+\arctan(\frac{b}{a}) < \theta \leq 2\pi-\arctan(\frac{b}{a}) \\
    \end{cases},
\end{equation}
resulting in the constraint $ab = \pi$.
For our example, we chose $a = \sqrt{0.6 \pi}$ and $b = \sqrt{\frac{\pi}{0.6}}$ for an aspect ratio of 0.6.

For a star-shaped contact region:
\begin{equation}
    T_{star}(\theta) = T_0 + T_1 \cos (5\theta),
\end{equation}
resulting in the constraint $2 T_0 ^2 + T_1 ^2 = 2$.
For our example, we chose $T_0 = 0.98$ and $T_1 = 0.2814$.

Using these shapes directly results in sharp corners and unrealistic artifacts, so we applied additional smoothing steps without altering the cell volume.
We first constructed an additional shape with an elliptical contact region; $T(\theta)$ for an ellipse with axes $c, d$ is:
\begin{equation}
T_{ellipse}(\theta) = \frac{cd}{\sqrt{\left(d\cos(\theta)\right)^2 + \left(c\sin(\theta)\right)^2}},
\end{equation}
and the associated constraint is $cd = 1$.
We choose the aspect ratio of the ellipse to be $\frac{\text{range}( x_{contact} )}{\text{range}( y_{contact} )}$, where $x_{contact}$ and $y_{contact}$ are the set of coordinates within the star or rectangular contact region.
Accordingly, this shape is a smoother option with a contact region that very roughly approximates the desired contour.
To smooth out the cell shape away from the substrate, we can define new $T$ functions that also depend on $z$:
\begin{align}
T_{rect,smooth}^2 (\theta,z) &= \frac{z-z_{max}}{z_{max}} T_{rect}^2 + \frac{z}{z_{max}} T_{ellipse}^2,\\ 
T_{star,smooth}^2 (\theta,z) &= \frac{z-z_{max}}{z_{max}} T_{star}^2 + \frac{z}{z_{max}} T_{ellipse}^2,
\end{align}
where $z_{max}$ is the height of the cell contour.
Using these definitions, the desired contour shape at the surface is perfectly preserved and cell volume is conserved.
In the case of the rectangular mesh, we also rounded off the sharp corners of the contact region, replacing them with arcs of radius 0.2 $\upmu$m.
The nucleus shape remained unaltered across all simulations.

\section{Model specifications for all examples and biological test cases}
\label{sec:tables}


Here, we summarize each model in terms of its compartments, species, reactions and parameters.
We note that these tables were directly output from SMART using the \texttt{print\_to\_latex} functions associated with each class of container, with some minor modifications for readability.

\subsection{Mechanotransduction model}

\begin{longtable}{llllllll}
\caption{Compartments in mechanotransduction model.} \\
\toprule
 & Dimensionality & Species & Vertices & Cells & Marker value & Size \\
\midrule
\endfirsthead
\toprule
 & Dimensionality & Species & Vertices & Cells & Marker value & Size \\
\midrule
\endhead
\midrule
\multicolumn{8}{r}{Continued on next page} \\
\midrule
\endfoot
\bottomrule
\endlastfoot
Circular contact region & & & & & & ($\int{2\pi rdr}$)\\
\midrule
Cyto & 2 & 11 & 7920 & 15360 & 1 & $\SI[]{1.925e+03}{\micro\meter\squared}$ \\
PM & 1 & 2 & 305 & 304 & 10 & $\SI[]{1.295e+03}{\micro\meter}$ \\
Nuc & 2 & 1 & 649 & 1159 & 2 & $\SI[]{2.823e+02}{\micro\meter\squared}$ \\
NM & 1 & 2 & 125 & 124 & 12 & $\SI[]{2.345e+02}{\micro\meter}$ \\
\midrule
Rectangular contact region & & & & & & (1/4 symmetry)\\
\midrule
Cyto & 3 & 11 & 17479 & 84954 & 1 & $\SI[]{4.794e+02}{\micro\meter\cubed}$ \\
PM & 2 & 2 & 5113 & 10030 & 10 & $\SI[]{3.314e+02}{\micro\meter\squared}$ \\
Nuc & 3 & 1 & 1668 & 6406 & 2 & $\SI[]{7.044e+01}{\micro\meter\cubed}$ \\
NM & 2 & 2 & 901 & 1716 & 12 & $\SI[]{5.856e+01}{\micro\meter\squared}$ \\
\midrule
Star-shaped contact region & & & & & & (1/10 symmetry) \\
\midrule
Cyto & 3 & 11 & 7687 & 34466 & 1 & $\SI[]{1.899e+02}{\micro\meter\cubed}$ \\
PM & 2 & 2 & 2158 & 4110 & 10 & $\SI[]{1.360e+02}{\micro\meter\squared}$ \\
Nuc & 3 & 1 & 799 & 2759 & 2 & $\SI[]{2.818e+01}{\micro\meter\cubed}$ \\
NM & 2 & 2 & 380 & 674 & 12 & $\SI[]{2.342e+01}{\micro\meter\squared}$
\label{tab:mech-compartments}
\end{longtable}

\begin{longtable}{llll}
\caption{Species in mechanotransduction model.} \\
\toprule
 & Compartment & $D$ ($\unit{\micro\meter\squared\per\second}$)  & Initial Condition \\
\midrule
\endfirsthead
\toprule
 & Compartment & $D$ ($\unit{\micro\meter\squared\per\second}$)  & Initial Condition \\
\midrule
\endhead
\midrule
\multicolumn{4}{r}{Continued on next page} \\
\midrule
\endfoot
\bottomrule
\endlastfoot
Emod$^*$ & PM & 0 & $70 \left(\dfrac{1-\text{sign}(z-0.0001)}{2}\right)$ GPa \\
\\
pFAK & Cyto & 10 & $\SI[]{3.000e-01}{\micro\molar}$ \\
RhoA\_GDP & Cyto & 1 & $\SI[]{1.000e+00}{\micro\molar}$ \\
RhoA\_GTP & PM & 0.3 & $\SI[]{3.360e+01}{\molecule\per\micro\meter\squared}$ \\
ROCK\_A & Cyto & 75 & $\SI[]{0.000e+00}{\micro\molar}$ \\
mDia\_A & Cyto & 1 & $\SI[]{0.000e+00}{\micro\molar}$ \\
Myo\_A & Cyto & 0.8 & $\SI[]{1.500e+00}{\micro\molar}$ \\
LIMK\_A & Cyto & 10 & $\SI[]{1.000e-01}{\micro\molar}$ \\
Cofilin\_NP & Cyto & 10 & $\SI[]{1.800e+00}{\micro\molar}$ \\
FActin & Cyto & 0.6 & $\SI[]{1.790e+01}{\micro\molar}$ \\
GActin & Cyto & 13.37 & $\SI[]{4.824e+02}{\micro\molar}$ \\
LaminA & NM & 0.001 & $\SI[]{0.000e+00}{\molecule\per\micro\meter\squared}$ \\
NPC\_A & NM & 0.001 & $\SI[]{0.000e+00}{\molecule\per\micro\meter\squared}$ \\
YAPTAZ & Cyto & 19 & $\SI[]{7.000e-01}{\micro\molar}$ \\
YAPTAZ\_phos & Cyto & 19 & $\SI[]{2.000e-01}{\micro\molar}$ \\
YAPTAZ\_nuc & Nuc & 19 & $\SI[]{7.000e-01}{\micro\molar}$ \\
\caption*{$^*$Emod represents the substrate stiffness and also specifies the location of the substrate ($z=0$).}
\label{tab:mech-species}
\end{longtable}

\renewcommand{\arraystretch}{1.2}
\begin{longtable}{lllll}
\caption{Reactions in mechanotransduction model.} \\
\toprule
 & Reactants & Products & Equation & Type \\
\midrule
\endfirsthead
\toprule
 & Reactants & Products & Equation & Type \\
\midrule
\endhead
\midrule
\multicolumn{5}{r}{Continued on next page} \\
\midrule
\endfoot
\bottomrule
\endlastfoot
a1 & [] & [`pFAK'] & $(cyto_{Convert}) ([FAK_{tot}] - [pFAK]) (\dfrac{E_{mod} k_{sf}}{C + Emod} + k_{f})$ & vol-surf \\
\\
a2 & [`pFAK'] & [] & $k_{df} [pFAK]$ & vol \\
\\
a3 & [`RhoA\_GDP'] & [`RhoA\_GTP'] & $[RhoA_{GDP}] cyto_{Convert} k_{fkrho} (\gamma_{const} pFAK^5 + 1)$ & vol-surf \\
& & & $-[RhoA_{GTP}] k_{drho}$ & \\
\\
a4 & [] & [`ROCK\_A'] & $[RhoA_{GTP}] k_{rrho} ([ROCK_{tot}] - [ROCK_{A}])$ & vol-surf \\
\\
a5 & [`ROCK\_A'] & [] & $[ROCK_{A}] k_{drock}$ & vol \\
\\
b1 & [] & [`mDia\_A'] & $[RhoA_{GTP}] k_{mrho} ([mDia_{tot}] - [mDia_{A}])$ & vol-surf \\
\\
b2 & [`mDia\_A'] & [] & $k_{dmdia} [mDia_{A}]$ & vol \\
\\
b3 & [] & [`Myo\_A'] & $k_{mr} ([Myo_{tot}] - [Myo_{A}]) \cdot$ & vol \\
& & & $\left([ROCK_{A}] \epsilon \dfrac{1 + \tanh\left(sc1 ([ROCK_{A}] - [ROCK_{B}])\right)}{2} + 1\right)$ & \\
& & & $ - [Myo_{A}] k_{dmy}$ & \\
\\
b4 & [] & [`LIMK\_A'] & $k_{lr} ([LIMK_{tot}] - [LIMK_{A}]) \cdot$ & vol \\
& & & $\left([ROCK_{A}] \tau \dfrac{1 + \tanh\left(sc1 ([ROCK_{A}] - [ROCK_{B}])\right)}{2} + 1\right)$ & \\
& & & $- [LIMK_{A}] k_{dl}$ & \\
\\
b5 & [] & [`Cofilin\_NP'] & $k_{turnover} ([Cofilin_{tot}]-[Cofilin_{NP}])$ & vol \\
& & & $- \dfrac{[Cofilin_{NP}] [LIMK_{A}] k_{catCof}}{Cofilin_{NP} + k_{mCof}}$ & \\
\\
b6 & [`GActin'] & [`FActin'] & $[GActin] k_{ra} \cdot$ & vol \\
& & & $\left(\alpha [mDia_{A}]\dfrac{1 + \tanh\left(sc1 ([mDia_{A}] - [mDia_{B}])\right)}{2} + 1\right)$ & \\
& & & $ - [FActin] ([Cofilin_{NP}] k_{fc1} + k_{dep})$ & \\
\\
c1 & [`YAPTAZ\_phos'] & [`YAPTAZ'] & $[YAPTAZ_{phos}] ([FActin] [Myo_{A}] k_{CY} + k_{CN})$ & vol \\
& & & $ -[YAPTAZ] k_{NC}$ & \\
\\
c3 & [] & [`LaminA'] & $\dfrac{[FActin]^{2.6} k_{fl} p ([LaminA_{tot}]-[LaminA])}{C_{LaminA} + [FActin]^{2.6} p} - [LaminA] k_{rl}$ & vol-surf \\
\\
c4 & [] & [`NPC\_A'] & $[FActin] [LaminA] [Myo_{A}] k_{fNPC} ([NPC_{tot}]-[NPC_{A}])$ & vol-surf \\
& & & $ - [NPC_{A}] k_{rNPC}$ & \\
\\
c5 & [`YAPTAZ'] & [`YAPTAZ\_nuc'] & $[YAPTAZ] ([NPC_{A}] k_{in2} + k_{insolo}) - [YAPTAZ_{nuc}] k_{out}$ & vol-surf-vol
\label{tab:mech-reactions}
\end{longtable}

\renewcommand{\arraystretch}{1}
\begin{longtable}{lllll}
\caption{Parameters in mechanotransduction model. Values were adopted directly from those reported in \cite{scottSpatialModelYAP2021}} \\
\toprule
 & Value & Description \\
\midrule
\endfirsthead
\toprule
 & Value & Description \\
\midrule
\endhead
\midrule
\multicolumn{5}{r}{Continued on next page} \\
\midrule
\endfoot
\bottomrule
\endlastfoot
FAK\_tot & $\SI[]{1.000e+00}{\micro\molar}$ &  Total cytosolic FAK \\
k\_f & $\SI[]{1.500e-02}{\per\second}$ & rate constant for baseline FAK phosphorylation\\
k\_sf & $\SI[]{3.790e-01}{\per\second}$ & substrate-stiffness-dependent FAK phosphorylation rate constant\\
C & $\SI[]{3.250e+00}{\kilo\pascal}$ &  critical substrate stiffness\\
cytoConvert & $\SI[]{1.825e+00}{\micro\meter}$ & vol/SA ratio\\
k\_df & $\SI[]{3.500e-02}{\per\second}$ & FAK dephosphorylation rate constant\\
k\_fkrho & $\SI[]{1.680e-02}{\per\second}$ &  RhoA activation rate constant\\
gammaConst & $\SI[]{7.756e+01}{\per\micro\molar\tothe{5}}$ &  Scaling factor \\
k\_drho & $\SI[]{6.250e-01}{\per\second}$ &  RhoA deactivation rate constant\\
k\_rrho & $\SI[]{6.480e-01}{\per\micro\molar\per\second}$ &  ROCK activation rate constant\\
ROCK\_tot & $\SI[]{1.000e+00}{\micro\molar}$ &  Total cytosolic ROCK\\
k\_drock & $\SI[]{8.000e-01}{\per\second}$ &  ROCK deactivation rate constant\\
k\_mrho & $\SI[]{2.000e-03}{\per\micro\molar\per\second}$ &  mDia activation rate constant\\
mDia\_tot & $\SI[]{8.000e-01}{\micro\molar}$ &  Total cytosolic mDia\\
k\_dmdia & $\SI[]{5.000e-03}{\per\second}$ &  mDia deactivation rate constant\\
Myo\_tot & $\SI[]{5.000e+00}{\micro\molar}$ &  Total cytosolic myosin\\
k\_mr & $\SI[]{3.000e-02}{\per\second}$ &  Myosin activation rate constant\\
ROCK\_B & $\SI[]{3.000e-01}{\micro\molar}$ &  Critical activated ROCK concentration\\
epsilon & $\SI[]{3.600e+01}{\per\micro\molar}$ &  Scaling factor\\
sc1 & $\SI[]{2.000e+01}{\per\micro\molar}$ &  Scaling factor\\
k\_dmy & $\SI[]{6.700e-02}{\per\second}$ &  Myosin deactivation rate constant \\
LIMK\_tot & $\SI[]{2.000e+00}{\micro\molar}$ &  Total cytosolic LIMK\\
k\_lr & $\SI[]{7.000e-02}{\per\second}$ &  LIMK activation rate constant\\
tau & $\SI[]{5.549e+01}{\per\micro\molar}$ &  Scaling factor\\
k\_dl & $\SI[]{2.000e+00}{\per\second}$ &  LIMK deactivation rate constant\\
Cofilin\_tot & $\SI[]{2.000e+00}{\micro\molar}$ &  Total cytosolic cofilin\\
k\_turnover & $\SI[]{4.000e-02}{\per\second}$ &  Cofilin dephosphorylation rate constant \\
k\_catCof & $\SI[]{3.400e-01}{\per\second}$ & Cofilin phosphorylation rate constant\\
k\_mCof & $\SI[]{4.000e+00}{\micro\molar}$ & Critical cofilin concentration\\
k\_ra & $\SI[]{4.000e-01}{\per\second}$ &  Actin polymerization rate constant\\
alpha & $\SI[]{5.000e+01}{\per\micro\molar}$ &  Scaling factor\\
mDia\_B & $\SI[]{1.650e-01}{\micro\molar}$ &  Critical activated mDia concentration \\
k\_dep & $\SI[]{3.500e+00}{\per\second}$ &  Actin depolymerization rate constant\\
k\_fc1 & $\SI[]{4.000e+00}{\per\micro\molar\per\second}$ &  Cofilin-mediated actin polymerization rate constant\\
k\_CN & $\SI[]{5.600e-01}{\per\second}$ &  Baseline YAP/TAZ dephosphorylation rate constant\\
k\_CY & $\SI[]{7.600e-04}{\per\micro\molar\squared\per\second}$ & stress-fiber-mediated YAP/TAZ dephosphorylation rate constant\\
k\_NC & $\SI[]{1.400e-01}{\per\second}$ &  YAP/TAZ phosphorylation rate constant\\
LaminA\_tot & $\SI[]{3.500e+03}{\molecule\per\micro\meter\squared}$ &  Total lamin A in NM\\
k\_fl & $\SI[]{4.600e-01}{\per\second}$ & Lamin A dephosphorylation rate constant \\
p & $\SI[]{9.000e-06}{\kilo\pascal\per\micro\molar\tothe{2.600}}$ & Scaling factor for cell stiffness \\
C\_LaminA & $\SI[]{1.000e+02}{\kilo\pascal}$ &  Critical cell stiffness for lamin A dephosphorylation \\
k\_rl & $\SI[]{1.000e-03}{\per\second}$ &  Lamin A phosphorylation rate constant \\
NPC\_tot & $\SI[]{6.500e+00}{\molecule\per\micro\meter\squared}$ &  Total NPC density in NM\\
k\_fNPC & $\SI[]{2.800e-07}{\micro\meter\squared\per\micro\molar\squared\per\second}$ &  NPC opening rate constant \\
k\_rNPC & $\SI[]{8.700e+00}{\per\second}$ & NPC closing rate constant\\
k\_insolo & $\SI[]{1.000e+00}{\molecule\per\micro\meter\squared\per\micro\molar\per\second}$ &  NPC-independent YAP/TAZ nuclear import rate \\
k\_in2 & $\SI[]{1.000e+01}{\per\micro\molar\per\second}$ &  NPC-dependent YAP/TAZ nuclear import rate \\
k\_out & $\SI[]{1.000e+00}{\molecule\per\micro\meter\squared\per\micro\molar\per\second}$ &  YAP/TAZ nuclear export rate
\label{tab:mech-params}
\end{longtable}

\subsection{Dendritic spine calcium model}

\begin{longtable}{lllllll}
\caption{Compartments in dendritic spine model.} \\
\toprule
 & Dimensionality & Species & Vertices & Cells & Marker value & Size \\
\midrule
\endfirsthead
\toprule
 & Dimensionality & Species & Vertices & Cells & Marker value & Size \\
\midrule
\endhead
\midrule
\multicolumn{7}{r}{Continued on next page} \\
\midrule
\endfoot
\bottomrule
\endlastfoot
Cyto & 3 & 2 & 17742 & 81206 & 1 & $\SI[]{6.463e-01}{\micro\meter\cubed}$ \\
PM & 2 & 3 & 1661 & 3243 & 10 & $\SI[]{6.236e+00}{\micro\meter\squared}$ \\
SA & 3 & 1 & 8727 & 25773 & 2 & $\SI[]{2.692e-02}{\micro\meter\cubed}$ \\
SAm & 2 & 0 & 7820 & 15672 & 12 & $\SI[]{2.387e+00}{\micro\meter\squared}$
\label{tab:spine-compartments}
\end{longtable}

\begin{longtable}{llll}
\caption{Species in dendritic spine model.} \\
\toprule
 & Compartment & $D$ ($\unit{\micro\meter\squared\per\second}$)  & Initial Condition \\
\midrule
\endfirsthead
\toprule
 & Compartment & $D$ ($\unit{\micro\meter\squared\per\second}$)  & Initial Condition \\
\midrule
\endhead
\midrule
\multicolumn{4}{r}{Continued on next page} \\
\midrule
\endfoot
\bottomrule
\endlastfoot
Ca & Cyto & 220 & $\SI[]{1.000e-01}{\micro\molar}$ \\
NMDAR & PM & 0 & $\SI[]{1.000e+00}{}^*$ \\
VSCC & PM & 0 & $\dfrac{1 + \text{sign}(z + 0.25)}{2}$ \\
Bf & PM & 0 & $\SI[]{7.957e+00}{\micro\meter\micro\molar}$ \\
Bm & Cyto & 20 & $\SI[]{2.000e+01}{\micro\molar}$ \\
CaSA & SA & 45 & $\SI[]{6.000e+01}{\micro\molar}$ \\
\caption*{$^*$NMDAR restricted to postsynaptic density.}
\label{tab:spine-species}
\end{longtable}

\begin{longtable}{lllll}
\caption{Reactions in dendritic spine model.} \\
\toprule
 & Reactants & Products & Equation & Type \\
\midrule
\endfirsthead
\toprule
 & Reactants & Products & Equation & Type \\
\midrule
\endhead
\midrule
\multicolumn{5}{r}{Continued on next page} \\
\midrule
\endfoot
\bottomrule
\endlastfoot
a1 & [] & [`\caion'] & $G_{NMDAR}(J0_{NMDAR}) [NMDAR] (V_m - V_{rev})$ & volume\_surface \\
\\
a2 & [] & [`\caion'] & $J_{VSCC} [VSCC]$ & volume\_surface \\
\\
a3 & [`\caion'] & [] & $\dfrac{100 n_{PMr} \left(\dfrac{[Ca^{2+}]^5 (Vmax_{hr23})}{[Ca^{2+}]^5 + (Km_{hr23})^5} + \dfrac{[Ca^{2+}]^2 (Vmax_{lr23})}{[Ca^{2+}]^2 + (Km_{lr23})^2}\right)}{1 + \dfrac{(Kme)(Prtote)}{([Ca^{2+}] + (Kme))^2} + \dfrac{(Kmx) (Prtotex)}{([Ca^{2+}] + (Kmx))^2}}$ & volume\_surface \\
\\
a4 & [`\caion'] & [] & $\dfrac{1000 n_{PMr} \dfrac{(Vmax_{r22}) [Ca^{2+}]}{[Ca^{2+}] + (Km_{r22})}}{1 + \dfrac{(Kme) (Prtote)}{([Ca^{2+}] + (Kme))^2} + \dfrac{(Kmx) (Prtotex)}{([Ca^{2+}] + (Kmx))^2}}$ & volume\_surface \\
\\
a5 & [`\caion', `Bf'] & [] & $(kBf_{on}) [Bf] [Ca^{2+}] - (kBf_{off}) ((Bf_{tot})-[Bf])$ & volume\_surface \\
\\
b1 & [`\caion', `Bm'] & [] & $(kBm_{on}) [Bm] [Ca^{2+}] - (kBm_{off}) ((Bm_{tot})-[Bm])$ & volume \\
\\
c1 & [`\caion'] & [`CaSA'] & $\dfrac{1000 n_{SAr} \dfrac{[Ca^{2+}]^2 (Vmax_{r19})}{[Ca^{2+}]^2 + (KP_{r19})^2}}{1 + \dfrac{(Kme) (Prtote)}{([Ca^{2+}] + (Kme))^2} + \dfrac{(Kmx) (Prtotex)}{([Ca^{2+}] + (Kmx))^2}}$ & volume\_surface\_volume \\
\\
c2 & [`CaSA'] & [`\caion'] & $k_{leak} n_{SAr} ([CaSA] - [Ca^{2+}])$ & volume\_surface\_volume
\label{tab:spine-reactions}
\end{longtable}

\begin{longtable}{lll}
\caption{Parameters in dendritic spine model.} \\
\toprule
 & Value/Equation & Description \\
\midrule
\endfirsthead
\toprule
 & Value/Equation & Description \\
\midrule
\endhead
\midrule
\multicolumn{3}{r}{Continued on next page} \\
\midrule
\endfoot
\bottomrule
\endlastfoot
n\_PMr & $\SI[]{1.011e-01}{\micro\meter}$ &  Experimental vol to surface area ratio\\
Vm & $\big[ \left(55.51 \exp\left(-\dfrac{t}{0.003}\right) + 
10.29 \exp(-40.0 t)\right) \dfrac{\text{sign}(t) + 1}{2}$ & PM voltage\\
& $+ 25 \left(\exp\left(-\dfrac{t}{0.05}\right) - \exp\left(-\dfrac{t}{0.005}\right)\right) 
\dfrac{\text{sign}(t) + 1}{2} - 65 \big]$ mV & \\
\\
Vrev & $\SI[]{9.000e+01}{\milli\volt}$ &  Reversal voltage for NMDAR\\
G\_NMDAR & $\SI[]{1.377e+05}{\molecule\per\milli\volt\per\second}$ &  NMDAR conductance\\
J0\_NMDAR & $-\dfrac{0.00456 \left(\exp\left(-\dfrac{t}{0.05}\right) + \exp\left(-\dfrac{t}{0.05}\right)\right) (\text{sign}(t) + 1)}{110.8 \exp(-0.092 V_m) + 1}$ \um$^{-2}$ &  \caion influx through NMDAR \\
\\
J\_VSCC & $(0.393 - 2.245\exp(-0.01044 V_m) (\exp(-34700 t) - \exp(-3680 t))\cdot$ &  \caion influx through VSCC \\
& $(\text{sign}(t) + 1)\dfrac{-1120 (2 V_m + 1456)}{192970 - 85942 \exp(0.01044 Vm)} \text{s}^{-1}$ \um$^{-2}$ & \\
\\
Prtote & $\SI[]{1.910e+02}{\micro\molar}$ &  \caion buffer 1 concentration \\
Kme & $\SI[]{2.430e+00}{\micro\molar}$ &  \caion buffer 1 affinity \\
Prtotex & $\SI[]{8.770e+00}{\micro\molar}$ &  \caion buffer 2 concentration \\
Kmx & $\SI[]{1.390e-01}{\micro\molar}$ &  \caion buffer 2 affinity \\
Vmax\_lr23 & $\SI[]{1.130e-01}{\micro\molar\per\second}$ &  PMCA conductance 1 \\
Km\_lr23 & $\SI[]{4.420e-01}{\micro\molar}$ &  PMCA activation constant 1\\
Vmax\_hr23 & $\SI[]{5.900e-01}{\micro\molar\per\second}$ &  PMCA conductance 2 \\
Km\_hr23 & $\SI[]{4.420e-01}{\micro\molar}$ &  PMCA activation constant 2\\
Vmax\_r22 & $\SI[]{1.000e-01}{\micro\molar\per\second}$ & NCX conductance \\
Km\_r22 & $\SI[]{1.000e+00}{\micro\molar}$ &  NCX activation constant\\
kBf\_on & $\SI[]{1.000e+00}{\per\micro\molar\per\second}$ &  On-rate for fixed \caion buffer \\
kBf\_off & $\SI[]{2.000e+00}{\per\second}$ &  Off-rate for fixed \caion buffer \\
Bf\_tot & $\SI[]{7.957e+00}{\micro\meter\micro\molar}$ &  Total amount of fixed \caion buffer\\
kBm\_on & $\SI[]{1.000e+00}{\per\micro\molar\per\second}$ & On-rate for mobile \caion buffer\\
kBm\_off & $\SI[]{1.000e+00}{\per\second}$ & Off-rate for mobile \caion buffer\\
Bm\_tot & $\SI[]{2.000e+01}{\micro\molar}$ &  Total conc. of mobile \caion buffer\\
n\_SAr & $\SI[]{1.130e-02}{\micro\meter}$ &  Cytosolic volume to SA surf. area ratio\\
Vmax\_r19 & $\SI[]{1.140e+02}{\micro\molar\per\second}$ &  SERCA conductance \\
KP\_r19 & $\SI[]{2.000e-01}{\micro\molar}$ &  SERCA activation constant\\
k\_leak & $\SI[]{1.608e-01}{\per\second}$ & SR \caion leak rate constant
\label{tab:spine-params}
\end{longtable}

\subsection{Calcium release unit model}

\begin{longtable}{llllllll}
\caption{Compartments in CRU model.} \\
\toprule
 & Dimensionality & Species & Vertices & Cells & Marker value & Size \\
\midrule
\endfirsthead
\toprule
 & Dimensionality & Species & Vertices & Cells & Marker value & Size \\
\midrule
\endhead
\midrule
\multicolumn{8}{r}{Continued on next page} \\
\midrule
\endfoot
\bottomrule
\endlastfoot
Cyto & 3 & 4 & 45557 & 204894 & 1 & $\SI[]{8.029e+08}{\nano\meter\cubed}$ \\
SR & 3 & 2 & 12178 & 42030 & 2 & $\SI[]{3.438e+06}{\nano\meter\cubed}$ \\
TTM & 2 & 0 & 13282 & 26552 & 10 & $\SI[]{7.190e+05}{\nano\meter\squared}$ \\
SRM & 2 & 1 & 9749 & 19518 & 12 & $\SI[]{4.129e+05}{\nano\meter\squared}$
\label{tab:cru-compartments}
\end{longtable}

\begin{longtable}{llll}
\caption{Species in CRU model.} \\
\toprule
 & Compartment & $D$ ($\unit{\micro\meter\squared\per\second}$)  & Initial Condition \\
\midrule
\endfirsthead
\toprule
 & Compartment & $D$ ($\unit{\micro\meter\squared\per\second}$)  & Initial Condition \\
\midrule
\endhead
\midrule
\multicolumn{4}{r}{Continued on next page} \\
\midrule
\endfoot
\bottomrule
\endlastfoot
Ca & Cyto & 220 & $\SI[]{1.400e-01}{\micro\molar}$ \\
ATP & Cyto & 140 & $\SI[]{4.547e+02}{\micro\molar}$ \\
CMDN & Cyto & 25 & $\SI[]{2.353e+01}{\micro\molar}$ \\
TRPN & Cyto & 0 & $\dfrac{56.8}{1+\exp\left(-\dfrac{(x+125)^2+(z+75)^2}{100^2}\right)}$ \uM $^*$ \\
\\
CaSR & SR & 220 & $\SI[]{1.300e+03}{\micro\molar}$ \\
CSQN & SR & 25 & $\SI[]{2.110e+03}{\micro\molar}$ \\
RyR & SRM & 0 & $\SI[]{1.000e+00}{}^{**}$ \\
\caption*{$^*$ TRPN distribution chosen to approximate that in Ref. \citenum{hakeModellingCardiacCalcium2012a}\@, using the distance from a central point in the T-tubule geometry rather than the distance to the junctional T-tubule boundary. \\
$^{**}$ As in Ref. \citenum{hakeModellingCardiacCalcium2012a}\@, RyR was localized to the SR-T-tubule junction; here defined as any region of the SR membrane less than 20 nm from the T-tubule membrane.}
\label{tab:cru-species}
\end{longtable}

\renewcommand{\arraystretch}{1.3}
\begin{longtable}{lllll}
\caption{Reactions in CRU model.} \\
\toprule
 & Reactants & Products & Equation & Type \\
\midrule
\endfirsthead
\toprule
 & Reactants & Products & Equation & Type \\
\midrule
\endhead
\midrule
\multicolumn{5}{r}{Continued on next page} \\
\midrule
\endfoot
\bottomrule
\endlastfoot
s1 & [`CaSR'] & [`\caion'] & $(NO_{RyR}) [RyR] g_{RyR} ([CaSR] - [Ca^{2+}])$ & vol-surf-vol \\
\\
s2$^*$ & [`\caion'] & [`CaSR'] & $2 (SR_{VtoA}) S_{SERCA} \rho_{SERCA} \cdot$ & vol-surf-vol \\
 & & & $\dfrac{[Ca^{2+}]^2 \alpha_{1,plus} \alpha_{2,plus} - [CaSR]^2 \alpha_{1,minus} \alpha_{2,minus}}{[Ca^{2+}]^2 \alpha_{1,plus} + [CaSR]^2 \alpha_{1,minus} + \alpha_{2,minus} + \alpha_{2,plus}}$ & \\
\\
t1 & [] & [`\caion'] & $(Jmax) \left(1 + (VFactor2)(ksat)\right) \cdot$ & vol-surf \\
 & & &  $\dfrac{(CaTT) (Nai)^3 (VFactor1) - [Ca^{2+}] (Nao)^3 (VFactor2)}{[Ca^{2+}] (denom2) + (denom1)}$ & \\
\\
t2 & [`\caion'] & [] & $(Jp_{max}) \dfrac{(Km_{pCa})^2}{[Ca^{2+}]^2 + (Km_{pCa})^2}$ & vol-surf  \\
\\
t3 & [`\caion'] & [] & $(JCab) \left((Voltage) - 0.5 \dfrac{(R)(T)}{F} \log \left(\dfrac{(CaTT)}{[Ca^{2+}]}\right)\right)$ & vol-surf  \\
\\
b1 & [`\caion', `ATP'] & [] & $(kon_{ATP}) [ATP] [Ca^{2+}]$ & vol \\
& & &  $- (koff_{ATP}) ((ATP_{tot})-[ATP])$ & \\
\\
b2 & [`\caion', `CMDN'] & [] & $(kon_{CMDN}) [CMDN] [Ca^{2+}]$ & vol \\
& & & $- (koff_{CMDN}) ((CMDN_{tot})-[CMDN])$ & \\
\\
b3 & [`\caion', `TRPN'] & [] & $(kon_{TRPN}) [Ca^{2+}] [TRPN] $ & vol \\
& & & $- (koff_{TRPN}) ((TRPN_{tot})-[TRPN])$ & \\
\\
b4 & [`CaSR', `CSQN'] & [] & $(kon_{CSQN}) [CSQN] [CaSR]$ & vol \\
& & & $ - (koff_{CSQN}) ((CSQN_{tot})-[CSQN])$ & \\
\caption*{$^*$In simulations without SERCA, this reaction was excluded from the model.}
\label{tab:cru-reactions}
\end{longtable}
\renewcommand{\arraystretch}{1}

\begin{longtable}{lllll}
\caption{Parameters in CRU model.} \\
\toprule
 & Value & Description \\
\midrule
\endfirsthead
\toprule
 & Value & Description \\
\midrule
\endhead
\midrule
\multicolumn{5}{r}{Continued on next page} \\
\midrule
\endfoot
\bottomrule
\endlastfoot
ATP\_tot & $\SI[]{4.550e+02}{\micro\molar}$  &  Total cytosolic ATP\\
CMDN\_tot & $\SI[]{2.400e+01}{\micro\molar}$  &  Total cytosolic CMDN\\
TRPN\_tot & $\SI[]{7.000e+01}{\micro\molar}$  &  Total cytosolic TRPN\\
CSQN\_tot & $\SI[]{6.390e+03}{\micro\molar}$ &  Total cytosolic CSQN\\
CaTT & $\SI[]{1.800e+03}{\micro\molar}$  &  \caion conc. in T-tubules \\
NO\_RyR & $\SI[]{5.000e+00}{}$  &  Number of open RyR in release event\\
g\_RyR & $\SI[]{2.620e+04}{\nano\meter\per\second}$  &  RyR conductance\\
S\_SERCA & $\SI[]{1.500e+00}{}$  &  SERCA scale parameter \\
rho\_SERCA & $\SI[]{7.500e+01}{\micro\molar}$  &  Volume density of SERCA \\
SR\_VtoA & $\SI[]{3.070e+02}{\nano\meter}$  &  cytosolic volume to SA surf. area ratio 0\\
alpha1\_plus & $\SI[]{1.067e+02}{\per\micro\molar\squared\per\second}$  & SERCA rate constant  \\
alpha2\_plus & $\SI[]{5.350e+00}{\per\second}$  &  SERCA rate constant\\
alpha1\_minus & $\SI[]{5.069e-05}{\per\micro\molar\squared\per\second}$  &  SERCA rate constant\\
alpha2\_minus & $\SI[]{3.868e-02}{\per\second}$  &  SERCA rate constant\\
Voltage & $\SI[]{-8.200e-02}{\volt}$  &  TT voltage \\
R & $\SI[]{8.310e+00}{\joule\per\mole\per\kelvin}$  &  Universal gas constant \\
T & $\SI[]{2.9515e+02}{\kelvin}$ & Temperature \\
F & $\SI[]{9.650e+04}{\coulomb\per\mole}$ & Faraday's constant \\
Jmax & $\SI[]{4.083e+02}{\micro\meter\micro\molar\per\second}$  &  Max. NCX flux\\
Nai & $\SI[]{1.420e+04}{\micro\molar}$  & Cytosolic sodium conc.\\
Nao & $\SI[]{1.400e+05}{\micro\molar}$  &  Extracellular sodium conc.\\
VFactor1 & $\SI[]{3.233e-01}{}$  &  NCX lumped parameter\\
VFactor2 & $\SI[]{8.142e+00}{}$  &  NCX lumped parameter\\
ksat & $\SI[]{2.700e-01}{}$  &  NCX saturation parameter\\
denom1 & $\SI[]{3.852e+16}{\micro\molar\tothe{4}}$  &  NCX lumped parameter\\
denom2 & $\SI[]{4.278e+15}{\micro\molar\cubed}$  &  NCX lumped parameter\\
Jp\_max & $\SI[]{4.974e+00}{\micro\meter\micro\molar\per\second}$  &  Max. \caion pump flux\\
Km\_pCa & $\SI[]{2.890e-01}{\micro\molar}$  &  \caion pump \caion affinity \\
JCab & $\SI[]{2.611e+02}{\micro\meter\micro\molar\per\second\per\volt}$  & TT \caion leak flux \\
kon\_ATP & $\SI[]{2.250e+02}{\per\micro\molar\per\second}$  &  \caion-ATP binding on rate \\
koff\_ATP & $\SI[]{4.500e+04}{\per\second}$  & \caion-ATP binding off rate \\
kon\_CMDN & $\SI[]{3.400e+01}{\per\micro\molar\per\second}$  & \caion-CMDN binding on rate \\
koff\_CMDN & $\SI[]{2.380e+02}{\per\second}$  & \caion-CMDN binding off rate \\
kon\_TRPN & $\SI[]{3.270e+01}{\per\micro\molar\per\second}$  & \caion-TRPN binding on rate \\
koff\_TRPN & $\SI[]{1.960e+01}{\per\second}$  & \caion-TRPN binding off rate \\
kon\_CSQN & $\SI[]{1.020e+02}{\per\micro\molar\per\second}$  & \caion-CSQN binding on rate \\
koff\_CSQN & $\SI[]{6.500e+04}{\per\second}$ & \caion-CSQN binding off rate
\label{tab:cru-params}
\end{longtable}

\subsection{ATP production in mitochondria}
\begin{longtable}{llllllll}
\caption{Compartments in ATP generation model.} \\
\toprule
 & Dimensionality & Species & Vertices & Cells & Marker value & Size \\
\midrule
\endfirsthead
\toprule
 & Dimensionality & Species & Vertices & Cells & Marker value & Size \\
\midrule
\endhead
\midrule
\multicolumn{7}{r}{Continued on next page} \\
\midrule
\endfoot
\bottomrule
\endlastfoot
IMS & 3 & 1 & 36667 & 128395 & 1 & $\SI[]{2.031e-02}{\micro\meter\cubed}$ \\
OM & 2 & 0 & 11514 & 23024 & 10 & $\SI[]{6.320e-01}{\micro\meter\squared}$ \\
Mat & 3 & 2 & 22974 & 85773 & 2 & $\SI[]{1.647e-02}{\micro\meter\cubed}$ \\
IM & 2 & 13 & 18708 & 37598 & 12 & $\SI[]{1.545e+00}{\micro\meter\squared}$ \\
\caption*{$^*$In simulations with cristae-localized membrane species, the IM was split into two regions as shown in \Cref{fig:mito}B. The cristae portion had an area of 1.0197 \um$^2$, with 28578 total surface elements.}
\label{tab:mito-compartments}
\end{longtable}

\begin{longtable}{llll}
\caption{Species in ATP generation model.} \\
\toprule
 & Compartment$^*$ & $D$ ($\unit{\micro\meter\squared\per\second}$)  & Initial Condition $^{**}$ \\
\midrule
\endfirsthead
\toprule
 & Compartment$^*$ & $D$ ($\unit{\micro\meter\squared\per\second}$)  & Initial Condition $^{**}$ \\
\midrule
\endhead
\midrule
\multicolumn{4}{r}{Continued on next page} \\
\midrule
\endfoot
\bottomrule
\endlastfoot
E\_Mat & IM & 0 & $\SI[]{0.000e+00}{\molecule\per\micro\meter\squared}$ \\
E\_IMS & IM & 0 & $\SI[]{1.728e+02}{}/\SI{2.618e+02}{\molecule\per\micro\meter\squared}$ \\
E\_Mat\_H3Star & IM & 0 & $\SI[]{0.000e+00}{\molecule\per\micro\meter\squared}$ \\
E\_Mat\_H3S & IM & 0 & $\SI[]{0.000e+00}{\molecule\per\micro\meter\squared}$ \\
E\_Mat\_H3 & IM & 0 & $\SI[]{0.000e+00}{\molecule\per\micro\meter\squared}$ \\
L & IM & 0 & $\SI[]{1.066e+04}{}/\SI{1.615e+04}{\molecule\per\micro\meter\squared}$ \\
TL & IM & 0 & $\SI[]{0.000e+00}{\molecule\per\micro\meter\squared}$ \\
LT & IM & 0 & $\SI[]{0.000e+00}{\molecule\per\micro\meter\squared}$ \\
DL & IM & 0 & $\SI[]{0.000e+00}{\molecule\per\micro\meter\squared}$ \\
LD & IM & 0 & $\SI[]{0.000e+00}{\molecule\per\micro\meter\squared}$ \\
TLD & IM & 0 & $\SI[]{0.000e+00}{\molecule\per\micro\meter\squared}$ \\
DLT & IM & 0 & $\SI[]{0.000e+00}{\molecule\per\micro\meter\squared}$ \\
DLD $^\dagger$ & IM & 0 & $\SI[]{0.000e+00}{\molecule\per\micro\meter\squared}$ \\
D\_Mat & Mat & 15 & $\SI[]{7.200e-01}{\milli\molar}$ \\
T\_Mat & Mat & 15 & $\SI[]{6.500e-01}{\milli\molar}$ \\
T\_IMS & IMS & 15 & $\SI[]{3.250e-01}{\milli\molar}$ \\
\caption*{$^*$In simulations with cristae-localized membrane species, all species listed as ``IM" here were localized to the Cristae compartment instead. \\
$^{**}$ Two values are given for some species - first value associated with uniform distribution of proteins in the IM, second value for cristae-localized proteins, ensuring consistency in total number of molecules. \\
$^\dagger$ Note that $TLT'$ and $DLD'$ introduced in Ref.\@ \citenum{garciaMitochondrialMorphologyGoverns2023} are not used here, as they are indistinguishable states that can be eliminated by substituting $TLT \gets TLT + TLT'$ and $DLD \gets DLD + DLD'$.}
\label{tab:mito-species}
\end{longtable}

\renewcommand{\arraystretch}{1.3}
\begin{longtable}{lllll}
\caption{Reactions in ATP generation model.} \\
\toprule
 & Reactants & Products & Equation & Type\\
\midrule
\endfirsthead
\toprule
 & Reactants & Products & Equation & Type\\
\midrule
\endhead
\midrule
\multicolumn{4}{r}{Continued on next page} \\
\midrule
\endfoot
\bottomrule
\endlastfoot
E1 & [`E\_Mat'] & [`E\_IMS'] & $k_{16} E_{Mat}-k_{61} E_{IMS}$ & surf\\
E2 & [`E\_IMS'] & [] & $E_{IMS} k_{65} - k_{56} E_{IMS,H3} \qquad ^*$ & surf\\
E3 & [] & [`E\_Mat\_H3Star'] & $k_{54} E_{IMS,H3} - E_{Mat,H3Star} k_{45} \qquad ^*$ & surf\\
E4 & [] & [`E\_Mat\_H3'] & $k_{52} E_{IMS,H3} - E_{Mat,H3} k_{25} \qquad ^*$ & surf\\
E5 & [`E\_Mat\_H3Star', `D\_Mat'] & [`E\_Mat\_H3S'] & $k_{43} E_{Mat,H3Star} D_{Mat}-k_{34} E_{Mat,H3S}$ & vol-surf\\
E6 & [`E\_Mat\_H3S'] & [`E\_Mat\_H3', `T\_Mat'] & $k_{32} E_{Mat_H3S}-k_{23} E_{Mat,H3} T_{Mat}$ & vol-surf\\
E7 & [`E\_Mat\_H3'] & [`E\_Mat'] & $k_{21} E_{Mat,H3}-k_{12} E_{Mat}$ & surf\\
L1 & [`L', `T\_Mat'] & [`LT'] & $k_{on,Tm} [L] T_{Mat}-k_{off,Tm} [LT]$ & vol-surf\\
L2 & [`L', `D\_Mat'] & [`LD'] & $k_{on,Dm} [L] D_{Mat}-k_{off,Dm} [LD]$ & vol-surf\\
L3 & [`L', `T\_IMS'] & [`TL'] & $k_{on,Ti} [L] T_{IMS}-k_{off,Ti} [TL]$ & vol-surf\\
L4 & [`L'] & [`DL'] & $ D_{IMS} [L] k_{on,Di} - [DL] k_{off,Di}$ & surf\\
L5 & [`TL', `T\_Mat'] & [] & $[TL] T_{Mat} k_{on,Tm} - k_{off,Tm} [TLT] \qquad ^{**}$ & vol-surf\\
L6 & [`TL', `D\_Mat'] & [`TLD'] & $k_{on,Dm} [TL] D_{Mat}-k_{off,Dm} [TLD]$ & vol-surf\\
L7 & [`LT', `T\_IMS'] & [] & $[LT] T_{IMS} k_{on,Ti} - k_{off,Ti} [TLT] \qquad ^{**}$ & vol-surf\\
L8 & [`LT'] & [`DLT'] & $D_{IMS} [LT] k_{on,Di} - [DLT] k_{off,Di}$ & surf\\
L9 & [`DL', `T\_Mat'] & [`DLT'] & $k_{on,Tm} [DL] T_{Mat}-k_{off,Tm} [DLT]$ & vol-surf\\
L10 & [`DL', `D\_Mat'] & [`DLD'] & $k_{on,Dm} [DL] D_{Mat}-k_{off,Dm} [DLD]$ & vol-surf\\
L11 & [`LD', `T\_IMS'] & [`TLD'] & $k_{on,Ti} [LD] T_{IMS}-k_{off,Ti} [TLD]$ & vol-surf\\
L12 & [`LD'] & [`DLD'] & $D_{IMS} [LD] k_{on,Di} - [DLD] k_{off,Di}$ & surf\\
L13 & [`DLT'] & [`TLD'] & $k_{p} [DLT]-k_{cp} [TLD]$ & surf\\
V1 & [`T\_IMS'] & [`T\_cyto'] & $[VDAC] k_{vdac} (T_{IMS} - T_{cyto})$ & vol-surf \\
\caption*{$^*$ Due to mass conservation, $E_{IMS,H3}$ is not treated as a separate variable, but solved for as $E_{IMS,H3} = E_{tot} - E_{Mat} - E_{IMS} - E_{Mat,H3Star} - E_{Mat,H3} - E_{Mat,H3S}$. \\
$^{**}$ Due to mass conservation, $TLT$ is not treated as a separate variable, but solved for as $TLT = L_{tot} - L - LT - TL - LD - DL - DLD - DLT - TLD$}
\label{tab:mito-reactions}
\end{longtable}

\renewcommand{\arraystretch}{1}
\begin{longtable}{llllll}
\caption{Parameters for mitochondrial ATP generation.} \\
\toprule
 & Value & Description\\
\midrule
\endfirsthead
\toprule
 & Value & Description\\
\midrule
\endhead
\midrule
\multicolumn{6}{r}{Continued on next page} \\
\midrule
\endfoot
\bottomrule
\endlastfoot
E\_tot & $\SI[]{1.73e+02}{\molecule\per\micro\meter\squared}$ & Total ATP synthase \\
L\_tot & $\SI[]{1.07e+04}{\molecule\per\micro\meter\squared}$ & Total ANTs \\
D\_IMS & $\SI[]{4.50e-02}{\milli\molar}$ & IMS ADP concentration \\
\hline
k\_16 & $\SI[]{1.48e+05}{\per\second}$ & ATP-synthase-associated rate constants \\
k\_61 & $\SI[]{3.37e+04}{\per\second}$ &  \\
k\_65 & $\SI[]{3.97e+03}{\per\second}$ &  \\
k\_56 & $\SI[]{1.00e+03}{\per\second}$ &  \\
k\_54 & $\SI[]{1.00e+02}{\per\second}$ &  \\
k\_45 & $\SI[]{1.00e+02}{\per\second}$ & \\
k\_52 & $\SI[]{1.00e-20}{\per\second}$ &  \\
k\_25 & $\SI[]{5.85e-30}{\per\second}$ &  \\
k\_43 & $\SI[]{2.00e+03}{\per\milli\molar\per\second}$ &  \\
k\_34 & $\SI[]{1.00e+02}{\per\second}$ & \\
k\_32 & $\SI[]{5.00e+03}{\per\second}$ & \\
k\_23 & $\SI[]{5.00e+03}{\per\milli\molar\per\second}$ &  \\
k\_21 & $\SI[]{4.00e+01}{\per\second}$ & \\
k\_12 & $\SI[]{1.00e+02}{\per\second}$ &  \\
\hline
koff\_Tm & $\SI[]{4.00e+04}{\per\second}$ & ANT-associated rate constants \\
kon\_Tm & $\SI[]{6.40e+03}{\per\milli\molar\per\second}$ & \\
koff\_Ti & $\SI[]{2.00e+02}{\per\second}$ & \\
kon\_Ti & $\SI[]{4.00e+02}{\per\milli\molar\per\second}$ & \\
koff\_Dm & $\SI[]{4.00e+04}{\per\second}$ & \\
kon\_Dm & $\SI[]{4.00e+03}{\per\milli\molar\per\second}$ & \\
koff\_Di & $\SI[]{1.00e+02}{\per\second}$ &  \\
kon\_Di & $\SI[]{4.00e+03}{\per\milli\molar\per\second}$ & \\
k\_p & $\SI[]{9.20e+01}{\per\second}$ & \\
k\_cp & $\SI[]{3.50e+00}{\per\second}$ & \\
\hline
k\_vdac & $\SI[]{1.00e+03}{\per\milli\molar\per\second}$ & opening rate of VDACs \\
VDAC & $\SI[]{1.00e+04}{\molecule\per\micro\meter\squared}$ & Density of VDACs in OM
\label{tab:mito-params}
\end{longtable}

\pagebreak

\section{Video captions}

\begin{video}[htbp!]
\caption{{\bf F-actin and YAP/TAZ dynamics in a cell on a circular micropattern.}
        F-actin concentration in the cytosol and YAP/TAZ concentration in the nucleus plotted within a cell on a circular contact region with a radius of 13 \um .
        Substrate stiffness matches that of a glass coverslip (70 GPa).
        }
        \label{video:circ}
\end{video}

\begin{video}[htbp!]
\caption{{\bf F-actin and YAP/TAZ dynamics in a cell on a rectangular micropattern.}
        F-actin concentration in the cytosol and YAP/TAZ concentration in the nucleus plotted within a cell on a rectangular contact region with the same contact area and cell volume as the cell in \Cref{video:circ}.
        Substrate stiffness matches that of a glass coverslip (70 GPa).
        }
        \label{video:rect}
\end{video}

\begin{video}[htbp!]
\caption{{\bf F-actin and YAP/TAZ dynamics in a cell on a star-shaped micropattern.}
        F-actin concentration in the cytosol and YAP/TAZ concentration in the nucleus plotted within a cell on a star-shaped contact region with the same contact area and cell volume as the cell in \Cref{video:circ}.
        Substrate stiffness matches that of a glass coverslip (70 GPa).
        }
        \label{video:star}
\end{video}

\begin{video}[htbp!]
\caption{{\bf Calcium dynamics in a dendritic spine.}
        Calcium influx into the dendritic spine over the duration of a single voltage peak.
        Both cytosolic and SR \caion concentrations are shown, with the SR in the spine head accumulating \caion at later time points due to entry via SERCA.
        }
        \label{video:spine}
\end{video}

\begin{video}[htbp!]
\caption{{\bf Calcium dynamics in a cardiomyocyte CRU with SERCA.}
        Cytosolic and SR calcium concentrations during a single calcium release event from the SR.
        \caion is slowly replenished in the SR after release due to entry through SERCA.
        A smaller calcium flux occurs through the T-tubules (dark grey) and the two mitochondria (light grey) act as a diffusional barrier.
        }
        \label{video:cru1}
\end{video}

\begin{video}[htbp!]
\caption{{\bf Calcium dynamics in a cardiomyocyte CRU without SERCA.}
        Cytosolic and SR calcium concentrations during a single calcium release event from the SR.
        In contrast to \Cref{video:cru1}, \caion remains depleted in the SR here as SERCA is not present in the SR membrane.
        }
        \label{video:cru2}
\end{video}

\begin{video}[htbp!]
\caption{{\bf ATP dynamics in a mitochondrion.}
        ATP concentrations in the matrix (innermost volume) and IMS (region between the outer and inner membranes) are displayed over 100 ms of simulation.
        Rapid initial decrease in both compartments is followed by a slower increase of IMS ATP and a slow decrease in matrix ATP.
        }
        \label{video:mito}
\end{video}

\bibliography{references}

\begin{thebibliography}{10}
\urlstyle{rm}
\expandafter\ifx\csname url\endcsname\relax
  \def\url#1{\texttt{#1}}\fi
\expandafter\ifx\csname urlprefix\endcsname\relax\def\urlprefix{URL }\fi
\expandafter\ifx\csname doiprefix\endcsname\relax\def\doiprefix{DOI: }\fi
\providecommand{\bibinfo}[2]{#2}
\providecommand{\eprint}[2][]{\url{#2}}

\bibitem{yueComputationalSystemsBiology2022}
\bibinfo{author}{Yue, R.} \& \bibinfo{author}{Dutta, A.}
\newblock \bibinfo{journal}{\bibinfo{title}{Computational systems biology in
  disease modeling and control, review and perspectives}}.
\newblock {\emph{\JournalTitle{npj Systems Biology and Applications}}}
  \textbf{\bibinfo{volume}{8}}, \bibinfo{pages}{1--16},
  \doiprefix\url{10.1038/s41540-022-00247-4} (\bibinfo{year}{2022}).

\bibitem{mogilnerQuantitativeModelingCell2006}
\bibinfo{author}{Mogilner, A.}, \bibinfo{author}{Wollman, R.} \&
  \bibinfo{author}{Marshall, W.~F.}
\newblock \bibinfo{journal}{\bibinfo{title}{Quantitative {{Modeling}} in {{Cell
  Biology}}: {{What Is It Good}} for?}}
\newblock {\emph{\JournalTitle{Developmental Cell}}}
  \textbf{\bibinfo{volume}{11}}, \bibinfo{pages}{279--287},
  \doiprefix\url{10.1016/j.devcel.2006.08.004} (\bibinfo{year}{2006}).

\bibitem{kitanoComputationalSystemsBiology2002}
\bibinfo{author}{Kitano, H.}
\newblock \bibinfo{journal}{\bibinfo{title}{Computational systems biology}}.
\newblock {\emph{\JournalTitle{Nature}}} \textbf{\bibinfo{volume}{420}},
  \bibinfo{pages}{206--210}, \doiprefix\url{10.1038/nature01254}
  (\bibinfo{year}{2002}).

\bibitem{wainwrightFormFunctionOrganisms1988}
\bibinfo{author}{WAINWRIGHT, S.~A.}
\newblock \bibinfo{journal}{\bibinfo{title}{Form and {{Function}} in
  {{Organisms}}}}.
\newblock {\emph{\JournalTitle{American Zoologist}}}
  \textbf{\bibinfo{volume}{28}}, \bibinfo{pages}{671--680},
  \doiprefix\url{10.1093/icb/28.2.671} (\bibinfo{year}{1988}).

\bibitem{hermanUnifyingFrameworkUnderstanding2022}
\bibinfo{author}{Herman, M.~A.} \emph{et~al.}
\newblock \bibinfo{journal}{\bibinfo{title}{A {{Unifying Framework}} for
  {{Understanding Biological Structures}} and {{Functions Across Levels}} of
  {{Biological Organization}}}}.
\newblock {\emph{\JournalTitle{Integrative and Comparative Biology}}}
  \textbf{\bibinfo{volume}{61}}, \bibinfo{pages}{2038--2047},
  \doiprefix\url{10.1093/icb/icab167} (\bibinfo{year}{2022}).

\bibitem{peddieVolumeElectronMicroscopy2022}
\bibinfo{author}{Peddie, C.~J.} \emph{et~al.}
\newblock \bibinfo{journal}{\bibinfo{title}{Volume electron microscopy}}.
\newblock {\emph{\JournalTitle{Nature Reviews Methods Primers}}}
  \textbf{\bibinfo{volume}{2}}, \bibinfo{pages}{1--23},
  \doiprefix\url{10.1038/s43586-022-00131-9} (\bibinfo{year}{2022}).

\bibitem{villingerFIBSEMTomography2012}
\bibinfo{author}{Villinger, C.} \emph{et~al.}
\newblock \bibinfo{journal}{\bibinfo{title}{{{FIB}}/{{SEM}} tomography with
  {{TEM-like}} resolution for {{3D}} imaging of high-pressure frozen cells}}.
\newblock {\emph{\JournalTitle{Histochemistry and Cell Biology}}}
  \textbf{\bibinfo{volume}{138}}, \bibinfo{pages}{549--556},
  \doiprefix\url{10.1007/s00418-012-1020-6} (\bibinfo{year}{2012}).

\bibitem{heinrichWholecellOrganelleSegmentation2021a}
\bibinfo{author}{Heinrich, L.} \emph{et~al.}
\newblock \bibinfo{journal}{\bibinfo{title}{Whole-cell organelle segmentation
  in volume electron microscopy}}.
\newblock {\emph{\JournalTitle{Nature}}} \textbf{\bibinfo{volume}{599}},
  \bibinfo{pages}{141--146}, \doiprefix\url{10.1038/s41586-021-03977-3}
  (\bibinfo{year}{2021}).

\bibitem{mccaffertyIntegratingCellularElectron2024}
\bibinfo{author}{McCafferty, C.~L.} \emph{et~al.}
\newblock \bibinfo{journal}{\bibinfo{title}{Integrating cellular electron
  microscopy with multimodal data to explore biology across space and time}}.
\newblock {\emph{\JournalTitle{Cell}}} \textbf{\bibinfo{volume}{187}},
  \bibinfo{pages}{563--584}, \doiprefix\url{10.1016/j.cell.2024.01.005}
  (\bibinfo{year}{2024}).

\bibitem{stoneSuperResolutionMicroscopyShedding2017}
\bibinfo{author}{Stone, M.~B.}, \bibinfo{author}{Shelby, S.~A.} \&
  \bibinfo{author}{Veatch, S.~L.}
\newblock \bibinfo{journal}{\bibinfo{title}{Super-{{Resolution Microscopy}}:
  {{Shedding Light}} on the {{Cellular Plasma Membrane}}}}.
\newblock {\emph{\JournalTitle{Chemical Reviews}}}
  \textbf{\bibinfo{volume}{117}}, \bibinfo{pages}{7457--7477},
  \doiprefix\url{10.1021/acs.chemrev.6b00716} (\bibinfo{year}{2017}).

\bibitem{schermellehSuperresolutionMicroscopyDemystified2019}
\bibinfo{author}{Schermelleh, L.} \emph{et~al.}
\newblock \bibinfo{journal}{\bibinfo{title}{Super-resolution microscopy
  demystified}}.
\newblock {\emph{\JournalTitle{Nature Cell Biology}}}
  \textbf{\bibinfo{volume}{21}}, \bibinfo{pages}{72--84},
  \doiprefix\url{10.1038/s41556-018-0251-8} (\bibinfo{year}{2019}).

\bibitem{hakeModellingCardiacCalcium2012a}
\bibinfo{author}{Hake, J.} \emph{et~al.}
\newblock \bibinfo{journal}{\bibinfo{title}{Modelling cardiac calcium sparks in
  a three-dimensional reconstruction of a calcium release unit: {{Calcium}}
  sparks in reconstructed release unit}}.
\newblock {\emph{\JournalTitle{The Journal of Physiology}}}
  \textbf{\bibinfo{volume}{590}}, \bibinfo{pages}{4403--4422},
  \doiprefix\url{10.1113/jphysiol.2012.227926} (\bibinfo{year}{2012}).

\bibitem{bellDendriticSpineMorphology2022a}
\bibinfo{author}{Bell, M.~K.}, \bibinfo{author}{Holst, M.~V.},
  \bibinfo{author}{Lee, C.~T.} \& \bibinfo{author}{Rangamani, P.}
\newblock \bibinfo{journal}{\bibinfo{title}{Dendritic spine morphology
  regulates calcium-dependent synaptic weight change}}.
\newblock {\emph{\JournalTitle{The Journal of General Physiology}}}
  \textbf{\bibinfo{volume}{154}}, \bibinfo{pages}{e202112980},
  \doiprefix\url{10.1085/jgp.202112980} (\bibinfo{year}{2022}).

\bibitem{garciaMitochondrialMorphologyProvides2019}
\bibinfo{author}{Garcia, G.~C.} \emph{et~al.}
\newblock \bibinfo{journal}{\bibinfo{title}{Mitochondrial morphology provides a
  mechanism for energy buffering at synapses}}.
\newblock {\emph{\JournalTitle{Scientific Reports}}}
  \textbf{\bibinfo{volume}{9}}, \bibinfo{pages}{18306},
  \doiprefix\url{10.1038/s41598-019-54159-1} (\bibinfo{year}{2019}).

\bibitem{holashStochasticSimulationSkeletal2019a}
\bibinfo{author}{Holash, R.~J.} \& \bibinfo{author}{MacIntosh, B.~R.}
\newblock \bibinfo{journal}{\bibinfo{title}{A stochastic simulation of skeletal
  muscle calcium transients in a structurally realistic sarcomere model using
  {{MCell}}}}.
\newblock {\emph{\JournalTitle{PLoS Computational Biology}}}
  \textbf{\bibinfo{volume}{15}}, \bibinfo{pages}{e1006712},
  \doiprefix\url{10.1371/journal.pcbi.1006712} (\bibinfo{year}{2019}).

\bibitem{daversin-cattyAbstractionsAutomatedAlgorithms2021a}
\bibinfo{author}{{Daversin-Catty}, C.}, \bibinfo{author}{Richardson, C.~N.},
  \bibinfo{author}{Ellingsrud, A.~J.} \& \bibinfo{author}{Rognes, M.~E.}
\newblock \bibinfo{journal}{\bibinfo{title}{Abstractions and {{Automated
  Algorithms}} for {{Mixed Domain Finite Element Methods}}}}.
\newblock {\emph{\JournalTitle{ACM Transactions on Mathematical Software}}}
  \textbf{\bibinfo{volume}{47}}, \bibinfo{pages}{31:1--31:36},
  \doiprefix\url{10.1145/3471138} (\bibinfo{year}{2021}).

\bibitem{wangRemeshingFlexibleMembranes2022}
\bibinfo{author}{Wang, X.} \& \bibinfo{author}{Danuser, G.}
\newblock \bibinfo{journal}{\bibinfo{title}{Remeshing flexible membranes under
  the control of free energy}}.
\newblock {\emph{\JournalTitle{PLOS Computational Biology}}}
  \textbf{\bibinfo{volume}{18}}, \bibinfo{pages}{e1010766},
  \doiprefix\url{10.1371/journal.pcbi.1010766} (\bibinfo{year}{2022}).

\bibitem{meansReactionDiffusionModeling2006a}
\bibinfo{author}{Means, S.} \emph{et~al.}
\newblock \bibinfo{journal}{\bibinfo{title}{Reaction {{Diffusion Modeling}} of
  {{Calcium Dynamics}} with {{Realistic ER Geometry}}}}.
\newblock {\emph{\JournalTitle{Biophysical Journal}}}
  \textbf{\bibinfo{volume}{91}}, \bibinfo{pages}{537--557},
  \doiprefix\url{10.1529/biophysj.105.075036} (\bibinfo{year}{2006}).

\bibitem{lee3DMeshProcessing2020a}
\bibinfo{author}{Lee, C.~T.} \emph{et~al.}
\newblock \bibinfo{journal}{\bibinfo{title}{{{3D}} mesh processing using
  {{GAMer}} 2 to enable reaction-diffusion simulations in realistic cellular
  geometries}}.
\newblock {\emph{\JournalTitle{PLOS Computational Biology}}}
  \textbf{\bibinfo{volume}{16}}, \bibinfo{pages}{e1007756},
  \doiprefix\url{10.1371/journal.pcbi.1007756} (\bibinfo{year}{2020}).

\bibitem{leeOpenSourceMeshGeneration2020a}
\bibinfo{author}{Lee, C.~T.} \emph{et~al.}
\newblock \bibinfo{journal}{\bibinfo{title}{An {{Open-Source Mesh Generation
  Platform}} for {{Biophysical Modeling Using Realistic Cellular
  Geometries}}}}.
\newblock {\emph{\JournalTitle{Biophysical Journal}}}
  \textbf{\bibinfo{volume}{118}}, \bibinfo{pages}{1003--1008},
  \doiprefix\url{10.1016/j.bpj.2019.11.3400} (\bibinfo{year}{2020}).

\bibitem{venkatramanCristaeFormationMechanical2023a}
\bibinfo{author}{Venkatraman, K.} \emph{et~al.}
\newblock \bibinfo{journal}{\bibinfo{title}{Cristae formation is a mechanical
  buckling event~controlled by the inner mitochondrial membrane lipidome}}.
\newblock {\emph{\JournalTitle{The EMBO Journal}}}
  \textbf{\bibinfo{volume}{42}}, \bibinfo{pages}{e114054},
  \doiprefix\url{10.15252/embj.2023114054} (\bibinfo{year}{2023}).

\bibitem{mesaSpineApparatusModulates2023}
\bibinfo{author}{Mesa, M.~H.}, \bibinfo{author}{Garcia, G.~C.},
  \bibinfo{author}{Hoerndli, F.~J.}, \bibinfo{author}{McCabe, K.~J.} \&
  \bibinfo{author}{Rangamani, P.}
\newblock \bibinfo{title}{Spine apparatus modulates {{Ca2}}+ in spines through
  spatial localization of sources and sinks},
  \doiprefix\url{10.1101/2023.09.22.558941} (\bibinfo{year}{2023}).

\bibitem{alnaesFEniCSProjectVersion2015}
\bibinfo{author}{Aln{\ae}s, M.} \emph{et~al.}
\newblock \bibinfo{journal}{\bibinfo{title}{The {{FEniCS Project Version}}
  1.5}}.
\newblock {\emph{\JournalTitle{{$<$}p{$>$}Archive of Numerical Software}}}
  \textbf{\bibinfo{volume}{Vol 3}}, \bibinfo{pages}{{$<$}strong{$>$}Starting
  Point and Frequency: {$<$}/strong{$>$}Year: 2013{$<$}/p{$>$}},
  \doiprefix\url{10.11588/ANS.2015.100.20553} (\bibinfo{year}{2015}).

\bibitem{laughlinSMARTSpatialModeling2023}
\bibinfo{author}{Laughlin, J.~G.} \emph{et~al.}
\newblock \bibinfo{journal}{\bibinfo{title}{{{SMART}}: {{Spatial Modeling
  Algorithms}} for {{Reactions}} and {{Transport}}}}.
\newblock {\emph{\JournalTitle{Journal of Open Source Software}}}
  \textbf{\bibinfo{volume}{8}}, \bibinfo{pages}{5580},
  \doiprefix\url{10.21105/joss.05580} (\bibinfo{year}{2023}).

\bibitem{laughlinRangamaniLabUCSDSmartV22024a}
\bibinfo{author}{Laughlin, J.} \emph{et~al.}
\newblock \bibinfo{title}{{{RangamaniLabUCSD}}/smart: V2.2.1}.
\newblock \bibinfo{howpublished}{Zenodo},
  \doiprefix\url{10.5281/zenodo.11268687} (\bibinfo{year}{2024}).

\bibitem{schaffGeneralComputationalFramework1997}
\bibinfo{author}{Schaff, J.}, \bibinfo{author}{Fink, C.},
  \bibinfo{author}{Slepchenko, B.}, \bibinfo{author}{Carson, J.} \&
  \bibinfo{author}{Loew, L.}
\newblock \bibinfo{journal}{\bibinfo{title}{A general computational framework
  for modeling cellular structure and function}}.
\newblock {\emph{\JournalTitle{Biophysical Journal}}}
  \textbf{\bibinfo{volume}{73}}, \bibinfo{pages}{1135--1146},
  \doiprefix\url{10.1016/S0006-3495(97)78146-3} (\bibinfo{year}{1997}).

\bibitem{cowanSpatialModelingCell2012}
\bibinfo{author}{Cowan, A.~E.}, \bibinfo{author}{Moraru, I.~I.},
  \bibinfo{author}{Schaff, J.~C.}, \bibinfo{author}{Slepchenko, B.~M.} \&
  \bibinfo{author}{Loew, L.~M.}
\newblock \bibinfo{title}{Spatial {{Modeling}} of {{Cell Signaling Networks}}}.
\newblock In \emph{\bibinfo{booktitle}{Methods in {{Cell Biology}}}}, vol.
  \bibinfo{volume}{110}, \bibinfo{pages}{195--221},
  \doiprefix\url{10.1016/B978-0-12-388403-9.00008-4}
  (\bibinfo{publisher}{Elsevier}, \bibinfo{year}{2012}).

\bibitem{kerrFastMonteCarlo2008}
\bibinfo{author}{Kerr, R.~A.} \emph{et~al.}
\newblock \bibinfo{journal}{\bibinfo{title}{Fast {{Monte Carlo Simulation
  Methods}} for {{Biological Reaction-Diffusion Systems}} in {{Solution}} and
  on {{Surfaces}}}}.
\newblock {\emph{\JournalTitle{SIAM Journal on Scientific Computing}}}
  \textbf{\bibinfo{volume}{30}}, \bibinfo{pages}{3126--3149},
  \doiprefix\url{10.1137/070692017} (\bibinfo{year}{2008}).

\bibitem{huckaSystemsBiologyMarkup2003}
\bibinfo{author}{Hucka, M.} \emph{et~al.}
\newblock \bibinfo{journal}{\bibinfo{title}{The systems biology markup language
  ({{SBML}}): A medium forrepresentation and exchange of biochemical network
  models}}.
\newblock {\emph{\JournalTitle{Bioinformatics}}} \textbf{\bibinfo{volume}{19}},
  \bibinfo{pages}{524--531}, \doiprefix\url{10.1093/bioinformatics/btg015}
  (\bibinfo{year}{2003}).

\bibitem{wuContactsEndoplasmicReticulum2017}
\bibinfo{author}{Wu, Y.} \emph{et~al.}
\newblock \bibinfo{journal}{\bibinfo{title}{Contacts between the endoplasmic
  reticulum and other membranes in neurons}}.
\newblock {\emph{\JournalTitle{Proceedings of the National Academy of Sciences
  of the United States of America}}} \textbf{\bibinfo{volume}{114}},
  \bibinfo{pages}{E4859--E4867}, \doiprefix\url{10.1073/pnas.1701078114}
  (\bibinfo{year}{2017}).

\bibitem{balayPETScPortableExtensible1998}
\bibinfo{author}{Balay, S.}, \bibinfo{author}{Gropp, W.},
  \bibinfo{author}{McInnes, L.~C.} \& \bibinfo{author}{Smith, B.~F.}
\newblock \bibinfo{title}{{{PETSc}}, the portable, extensible toolkit for
  scientific computation} (\bibinfo{year}{1998}).

\bibitem{sunComputationalModelYAP2016}
\bibinfo{author}{Sun, M.}, \bibinfo{author}{Spill, F.} \&
  \bibinfo{author}{Zaman, M.~H.}
\newblock \bibinfo{journal}{\bibinfo{title}{A {{Computational Model}} of
  {{YAP}}/{{TAZ Mechanosensing}}}}.
\newblock {\emph{\JournalTitle{Biophysical Journal}}}
  \textbf{\bibinfo{volume}{110}}, \bibinfo{pages}{2540--2550},
  \doiprefix\url{10.1016/j.bpj.2016.04.040} (\bibinfo{year}{2016}).

\bibitem{scottSpatialModelYAP2021}
\bibinfo{author}{Scott, K.~E.}, \bibinfo{author}{Fraley, S.~I.} \&
  \bibinfo{author}{Rangamani, P.}
\newblock \bibinfo{journal}{\bibinfo{title}{A spatial model of {{YAP}}/{{TAZ}}
  signaling reveals how stiffness, dimensionality, and shape contribute to
  emergent outcomes}}.
\newblock {\emph{\JournalTitle{Proceedings of the National Academy of
  Sciences}}} \textbf{\bibinfo{volume}{118}}, \bibinfo{pages}{e2021571118},
  \doiprefix\url{10.1073/pnas.2021571118} (\bibinfo{year}{2021}).

\bibitem{maduramSubcellularCurvaturePerimeter2008}
\bibinfo{author}{Maduram, J.~H.}, \bibinfo{author}{Goluch, E.},
  \bibinfo{author}{Hu, H.}, \bibinfo{author}{Liu, C.} \&
  \bibinfo{author}{Mrksich, M.}
\newblock \bibinfo{journal}{\bibinfo{title}{Subcellular curvature at the
  perimeter of micropatterned cells influences lamellipodial distribution and
  cell polarity}}.
\newblock {\emph{\JournalTitle{Cell motility and the cytoskeleton}}}
  \textbf{\bibinfo{volume}{65}}, \bibinfo{pages}{841--852},
  \doiprefix\url{10.1002/cm.20305} (\bibinfo{year}{2008}).

\bibitem{bellDendriticSpineGeometry2019a}
\bibinfo{author}{Bell, M.}, \bibinfo{author}{Bartol, T.},
  \bibinfo{author}{Sejnowski, T.} \& \bibinfo{author}{Rangamani, P.}
\newblock \bibinfo{journal}{\bibinfo{title}{Dendritic spine geometry and spine
  apparatus organization govern the spatiotemporal dynamics of calcium}}.
\newblock {\emph{\JournalTitle{Journal of General Physiology}}}
  \textbf{\bibinfo{volume}{151}}, \bibinfo{pages}{1017--1034},
  \doiprefix\url{10.1085/jgp.201812261} (\bibinfo{year}{2019}).

\bibitem{leungSystemsModelingPredicts2021b}
\bibinfo{author}{Leung, A.}, \bibinfo{author}{Ohadi, D.},
  \bibinfo{author}{Pekkurnaz, G.} \& \bibinfo{author}{Rangamani, P.}
\newblock \bibinfo{journal}{\bibinfo{title}{Systems modeling predicts that
  mitochondria {{ER}} contact sites regulate the postsynaptic energy
  landscape}}.
\newblock {\emph{\JournalTitle{NPJ Systems Biology and Applications}}}
  \textbf{\bibinfo{volume}{7}}, \bibinfo{pages}{26},
  \doiprefix\url{10.1038/s41540-021-00185-7} (\bibinfo{year}{2021}).

\bibitem{liOrganizationCa2Signaling2022}
\bibinfo{author}{Li, J.}, \bibinfo{author}{Richmond, B.} \&
  \bibinfo{author}{Hong, T.}
\newblock \bibinfo{title}{Organization of {{Ca2}}+ {{Signaling Microdomains}}
  in {{Cardiac Myocytes}}}.
\newblock In \bibinfo{editor}{Parinandi, N.~L.} \& \bibinfo{editor}{Hund,
  T.~J.} (eds.) \emph{\bibinfo{booktitle}{Cardiovascular {{Signaling}} in
  {{Health}} and {{Disease}}}} (\bibinfo{publisher}{Springer},
  \bibinfo{address}{Cham (CH)}, \bibinfo{year}{2022}).

\bibitem{hoshijimaCCDB3603MUS2004}
\bibinfo{author}{Hoshijima, M.} \emph{et~al.}
\newblock \bibinfo{title}{{{CCDB}}:3603, {{MUS MUSCULUS}}, {{T-tubules}},
  sarcoplasmic reticulum, myocyte}, \doiprefix\url{doi:10.7295/W9CCDB3603}
  (\bibinfo{year}{2004}).

\bibitem{mendelsohnMorphologicalPrinciplesNeuronal2022a}
\bibinfo{author}{Mendelsohn, R.} \emph{et~al.}
\newblock \bibinfo{journal}{\bibinfo{title}{Morphological principles of
  neuronal mitochondria}}.
\newblock {\emph{\JournalTitle{The Journal of Comparative Neurology}}}
  \textbf{\bibinfo{volume}{530}}, \bibinfo{pages}{886--902},
  \doiprefix\url{10.1002/cne.25254} (\bibinfo{year}{2022}).

\bibitem{garciaMitochondrialMorphologyGoverns2023}
\bibinfo{author}{Garcia, G.~C.}, \bibinfo{author}{Gupta, K.},
  \bibinfo{author}{Bartol, T.~M.}, \bibinfo{author}{Sejnowski, T.~J.} \&
  \bibinfo{author}{Rangamani, P.}
\newblock \bibinfo{journal}{\bibinfo{title}{Mitochondrial morphology governs
  {{ATP}} production rate}}.
\newblock {\emph{\JournalTitle{Journal of General Physiology}}}
  \textbf{\bibinfo{volume}{155}}, \doiprefix\url{10.1085/jgp.202213263}
  (\bibinfo{year}{2023}).

\bibitem{meyersPotentialControlSignaling2006}
\bibinfo{author}{Meyers, J.}, \bibinfo{author}{Craig, J.} \&
  \bibinfo{author}{Odde, D.~J.}
\newblock \bibinfo{journal}{\bibinfo{title}{Potential for {{Control}} of
  {{Signaling Pathways}} via {{Cell Size}} and {{Shape}}}}.
\newblock {\emph{\JournalTitle{Current Biology}}}
  \textbf{\bibinfo{volume}{16}}, \bibinfo{pages}{1685--1693},
  \doiprefix\url{10.1016/j.cub.2006.07.056} (\bibinfo{year}{2006}).

\bibitem{rangamaniDecodingInformationCell2013a}
\bibinfo{author}{Rangamani, P.} \emph{et~al.}
\newblock \bibinfo{journal}{\bibinfo{title}{Decoding {{Information}} in {{Cell
  Shape}}}}.
\newblock {\emph{\JournalTitle{Cell}}} \textbf{\bibinfo{volume}{154}},
  \bibinfo{pages}{1356--1369}, \doiprefix\url{10.1016/j.cell.2013.08.026}
  (\bibinfo{year}{2013}).

\bibitem{nevesCellShapeNegative2008}
\bibinfo{author}{Neves, S.~R.} \emph{et~al.}
\newblock \bibinfo{journal}{\bibinfo{title}{Cell {{Shape}} and {{Negative
  Links}} in {{Regulatory Motifs Together Control Spatial Information Flow}} in
  {{Signaling Networks}}}}.
\newblock {\emph{\JournalTitle{Cell}}} \textbf{\bibinfo{volume}{133}},
  \bibinfo{pages}{666--680}, \doiprefix\url{10.1016/j.cell.2008.04.025}
  (\bibinfo{year}{2008}).

\bibitem{calizoCellShapeRegulates2020}
\bibinfo{author}{Calizo, R.~C.} \emph{et~al.}
\newblock \bibinfo{journal}{\bibinfo{title}{Cell shape regulates subcellular
  organelle location to control early {{Ca2}}+ signal dynamics in vascular
  smooth muscle cells}}.
\newblock {\emph{\JournalTitle{Scientific Reports}}}
  \textbf{\bibinfo{volume}{10}}, \bibinfo{pages}{17866},
  \doiprefix\url{10.1038/s41598-020-74700-x} (\bibinfo{year}{2020}).

\bibitem{johnsonBuildingNextGeneration2023}
\bibinfo{author}{Johnson, G.~T.} \emph{et~al.}
\newblock \bibinfo{journal}{\bibinfo{title}{Building the next generation of
  virtual cells to understand cellular biology}}.
\newblock {\emph{\JournalTitle{Biophysical Journal}}}
  \textbf{\bibinfo{volume}{122}}, \bibinfo{pages}{3560--3569},
  \doiprefix\url{10.1016/j.bpj.2023.04.006} (\bibinfo{year}{2023}).

\bibitem{vianaIntegratedIntracellularOrganization2023}
\bibinfo{author}{Viana, M.~P.} \emph{et~al.}
\newblock \bibinfo{journal}{\bibinfo{title}{Integrated intracellular
  organization and its variations in human {{iPS}} cells}}.
\newblock {\emph{\JournalTitle{Nature}}} \textbf{\bibinfo{volume}{613}},
  \bibinfo{pages}{345--354}, \doiprefix\url{10.1038/s41586-022-05563-7}
  (\bibinfo{year}{2023}).

\bibitem{xuOpenaccessVolumeElectron2021}
\bibinfo{author}{Xu, C.~S.} \emph{et~al.}
\newblock \bibinfo{journal}{\bibinfo{title}{An open-access volume electron
  microscopy atlas of whole cells and tissues}}.
\newblock {\emph{\JournalTitle{Nature}}} \textbf{\bibinfo{volume}{599}},
  \bibinfo{pages}{147--151}, \doiprefix\url{10.1038/s41586-021-03992-4}
  (\bibinfo{year}{2021}).

\bibitem{edwardsVolRoverNEnhancingSurface2014}
\bibinfo{author}{Edwards, J.} \emph{et~al.}
\newblock \bibinfo{journal}{\bibinfo{title}{{{VolRoverN}}: {{Enhancing
  Surface}} and {{Volumetric Reconstruction}} for {{Realistic Dynamical
  Simulation}} of {{Cellular}} and {{Subcellular Function}}}}.
\newblock {\emph{\JournalTitle{Neuroinformatics}}}
  \textbf{\bibinfo{volume}{12}}, \bibinfo{pages}{277--289},
  \doiprefix\url{10.1007/s12021-013-9205-2} (\bibinfo{year}{2014}).

\bibitem{huFastTetrahedralMeshing2020}
\bibinfo{author}{Hu, Y.}, \bibinfo{author}{Schneider, T.},
  \bibinfo{author}{Wang, B.}, \bibinfo{author}{Zorin, D.} \&
  \bibinfo{author}{Panozzo, D.}
\newblock \bibinfo{journal}{\bibinfo{title}{Fast tetrahedral meshing in the
  wild}}.
\newblock {\emph{\JournalTitle{ACM Transactions on Graphics}}}
  \textbf{\bibinfo{volume}{39}}, \bibinfo{pages}{117:117:1--117:117:18},
  \doiprefix\url{10.1145/3386569.3392385} (\bibinfo{year}{2020}).

\bibitem{mituschDolfinadjoint2018Automated2019}
\bibinfo{author}{Mitusch, S.~K.}, \bibinfo{author}{Funke, S.~W.} \&
  \bibinfo{author}{Dokken, J.~S.}
\newblock \bibinfo{journal}{\bibinfo{title}{Dolfin-adjoint 2018.1: Automated
  adjoints for {{FEniCS}} and {{Firedrake}}}}.
\newblock {\emph{\JournalTitle{Journal of Open Source Software}}}
  \textbf{\bibinfo{volume}{4}}, \bibinfo{pages}{1292},
  \doiprefix\url{10.21105/joss.01292} (\bibinfo{year}{2019}).

\bibitem{novakDiffusionCurvedSurface2007}
\bibinfo{author}{Novak, I.~L.} \emph{et~al.}
\newblock \bibinfo{journal}{\bibinfo{title}{Diffusion on a {{Curved Surface
  Coupled}} to {{Diffusion}} in the {{Volume}}: {{Application}} to {{Cell
  Biology}}}}.
\newblock {\emph{\JournalTitle{Journal of computational physics}}}
  \textbf{\bibinfo{volume}{226}}, \bibinfo{pages}{1271--1290},
  \doiprefix\url{10.1016/j.jcp.2007.05.025} (\bibinfo{year}{2007}).

\bibitem{hepburnSTEPSEfficientSimulation2012}
\bibinfo{author}{Hepburn, I.}, \bibinfo{author}{Chen, W.},
  \bibinfo{author}{Wils, S.} \& \bibinfo{author}{De~Schutter, E.}
\newblock \bibinfo{journal}{\bibinfo{title}{{{STEPS}}: Efficient simulation of
  stochastic reaction--diffusion models in realistic morphologies}}.
\newblock {\emph{\JournalTitle{BMC Systems Biology}}}
  \textbf{\bibinfo{volume}{6}}, \bibinfo{pages}{36},
  \doiprefix\url{10.1186/1752-0509-6-36} (\bibinfo{year}{2012}).

\bibitem{hepburnVesicleReactiondiffusionHybrid2024}
\bibinfo{author}{Hepburn, I.} \emph{et~al.}
\newblock \bibinfo{journal}{\bibinfo{title}{Vesicle and reaction-diffusion
  hybrid modeling with {{STEPS}}}}.
\newblock {\emph{\JournalTitle{Communications Biology}}}
  \textbf{\bibinfo{volume}{7}}, \bibinfo{pages}{1--15},
  \doiprefix\url{10.1038/s42003-024-06276-5} (\bibinfo{year}{2024}).

\bibitem{robertsLatticeMicrobesHighperformance2013}
\bibinfo{author}{Roberts, E.}, \bibinfo{author}{Stone, J.~E.} \&
  \bibinfo{author}{{Luthey-Schulten}, Z.}
\newblock \bibinfo{journal}{\bibinfo{title}{Lattice microbes:
  {{High-performance}} stochastic simulation method for the reaction-diffusion
  master equation}}.
\newblock {\emph{\JournalTitle{Journal of Computational Chemistry}}}
  \textbf{\bibinfo{volume}{34}}, \bibinfo{pages}{245--255},
  \doiprefix\url{10.1002/jcc.23130} (\bibinfo{year}{2013}).

\bibitem{hoopsCOPASICOmplexPAthway2006}
\bibinfo{author}{Hoops, S.} \emph{et~al.}
\newblock \bibinfo{journal}{\bibinfo{title}{{{COPASI}}---a {{COmplex PAthway
  SImulator}}}}.
\newblock {\emph{\JournalTitle{Bioinformatics}}} \textbf{\bibinfo{volume}{22}},
  \bibinfo{pages}{3067--3074}, \doiprefix\url{10.1093/bioinformatics/btl485}
  (\bibinfo{year}{2006}).

\bibitem{clarkLogicbasedModelingBiological2024}
\bibinfo{author}{Clark, A.~P.}, \bibinfo{author}{Chowkwale, M.},
  \bibinfo{author}{Paap, A.}, \bibinfo{author}{Dang, S.} \&
  \bibinfo{author}{Saucerman, J.~J.}
\newblock \bibinfo{journal}{\bibinfo{title}{Logic-based modeling of biological
  networks with {{Netflux}}}}.
\newblock {\emph{\JournalTitle{bioRxiv}}} \bibinfo{pages}{2024.01.11.575227},
  \doiprefix\url{10.1101/2024.01.11.575227} (\bibinfo{year}{2024}).

\bibitem{geuzaineGmsh3DFinite2009}
\bibinfo{author}{Geuzaine, C.} \& \bibinfo{author}{Remacle, J.-F.}
\newblock \bibinfo{journal}{\bibinfo{title}{Gmsh: {{A}} 3-{{D}} finite element
  mesh generator with built-in pre- and post-processing facilities: {{THE GMSH
  PAPER}}}}.
\newblock {\emph{\JournalTitle{International Journal for Numerical Methods in
  Engineering}}} \textbf{\bibinfo{volume}{79}}, \bibinfo{pages}{1309--1331},
  \doiprefix\url{10.1002/nme.2579} (\bibinfo{year}{2009}).

\bibitem{francisBiologicalTestCases2024}
\bibinfo{author}{Francis, E.~A.}, \bibinfo{author}{Finsberg, H.~N.} \&
  \bibinfo{author}{Dokken, J.~S.}
\newblock \bibinfo{title}{Biological test cases implemented in {{SMART}}}.
\newblock \bibinfo{howpublished}{Zenodo},
  \doiprefix\url{10.5281/zenodo.11268945} (\bibinfo{year}{2024}).

\bibitem{alnaesUnifiedFormLanguage2014}
\bibinfo{author}{Aln{\ae}s, M.~S.}, \bibinfo{author}{Logg, A.},
  \bibinfo{author}{{\O}lgaard, K.~B.}, \bibinfo{author}{Rognes, M.~E.} \&
  \bibinfo{author}{Wells, G.~N.}
\newblock \bibinfo{journal}{\bibinfo{title}{Unified form language: {{A}}
  domain-specific language for weak formulations of partial differential
  equations}}.
\newblock {\emph{\JournalTitle{ACM Transactions on Mathematical Software}}}
  \textbf{\bibinfo{volume}{40}}, \bibinfo{pages}{9:1--9:37},
  \doiprefix\url{10.1145/2566630} (\bibinfo{year}{2014}).

\bibitem{greccoPintPhysicalQuantities}
\bibinfo{author}{Grecco, H.~E.}
\newblock \bibinfo{title}{Pint: {{Physical}} quantities module}.

\bibitem{dalcinParallelDistributedComputing2011a}
\bibinfo{author}{Dalcin, L.~D.}, \bibinfo{author}{Paz, R.~R.},
  \bibinfo{author}{Kler, P.~A.} \& \bibinfo{author}{Cosimo, A.}
\newblock \bibinfo{journal}{\bibinfo{title}{Parallel distributed computing
  using {{Python}}}}.
\newblock {\emph{\JournalTitle{Advances in Water Resources}}}
  \textbf{\bibinfo{volume}{34}}, \bibinfo{pages}{1124--1139},
  \doiprefix\url{10.1016/j.advwatres.2011.04.013} (\bibinfo{year}{2011}).

\bibitem{meurerSymPySymbolicComputing2017}
\bibinfo{author}{Meurer, A.} \emph{et~al.}
\newblock \bibinfo{journal}{\bibinfo{title}{{{SymPy}}: Symbolic computing in
  {{Python}}}}.
\newblock {\emph{\JournalTitle{PeerJ Computer Science}}}
  \textbf{\bibinfo{volume}{3}}, \bibinfo{pages}{e103},
  \doiprefix\url{10.7717/peerj-cs.103} (\bibinfo{year}{2017}).

\end{thebibliography}

\end{document}